\documentclass[fleqn,usenatbib]{mnras}
\usepackage[T1]{fontenc}
\usepackage{ae,aecompl}
\usepackage{graphicx}	
\usepackage{amsmath}	
\usepackage{amssymb}	
\usepackage{subfigure}
\usepackage{threeparttable}
\usepackage{lineno}
\usepackage{orcidlink}

\title[Research on transition redshift]{Constraints on transition redshift utilizing the latest H(z) measurements and comments on the Hubble tension}
\author[Jian-Ping, Hu et al.]{
J. P. Hu$^{\orcidlink{0000-0002-5819-5002}}$$^{1,2}$\thanks{E-mail: hjp1206@163.com}, X. D. Jia$^{\orcidlink{0009-0009-3583-552X}}$$^{2}$, D. H. Gao$^{\orcidlink{0000-0001-7176-8170}}$$^{2}$, J. Z. Gao$^{\orcidlink{0009-0004-8729-7544}}$$^{3}$, B. Q. Gao$^{4}$ and
F. Y. Wang$^{\orcidlink{0000-0003-4157-7714}}$$^{2,5}$\thanks{E-mail: fayinwang@nju.edu.cn}
\\
$^{1}$Ministry of Education Key Laboratory for Nonequilibrium Synthesis and Modulation of Condensed Matter, School of Physics,\\ Xi'an Jiaotong University, Xi'an 710049, China \\
$^{2}$School of Astronomy and Space Science, Nanjing University, Nanjing 210093, China\\
$^{3}$Institute of Theoretical Physics, School of Physics, Dalian University of Technology, Dalian 116024, China \\
$^{4}$Research Center for Astronomical Computing, Zhejiang Lab, Hangzhou, 311100, China \\
$^{5}$Key Laboratory of Modern Astronomy and Astrophysics (Nanjing University), Ministry of Education, Nanjing 210093, China \\
}

\date{Accepted XXX. Received YYY; in original form ZZZ}

\pubyear{2025}

\begin{document}
\label{firstpage}
\pagerange{\pageref{firstpage}--\pageref{lastpage}}
\maketitle

\begin{abstract}
The motivation of this paper is to obtain reliable constraints of transition redshift ($z_{ztr}$) and, in combination with the evolution of the Hubble constant ($H_{0}$) that could alleviate the Hubble tension, discuss the possible origin of the tension. Utilizing the latest H(z) measurements and different methods ($\Lambda$CDM model, Cosmography, and Gaussian process method), we investigated the impact of methodology and dataset on $z_{ztr}$ constraints, and find that the choice of method has a greater impact on $z_{tr}$ than the observations themselves. Through a statistical analysis of the $z_{ztr}$ constraints from 2004 to 2024, we find that total $z_{tr}$ constraints (2004$-$2024) can be well described by a Gaussian function with the mean value 0.65 and the standard deviation 0.16; that is, $\bar{z}_{tr}$(all) = 0.65 $\pm$ 0.16. And we confirmed that both dataset and methodology can indeed significantly affect the final constraints. The screened $z_{tr}$ constraints with free $H_{0}$ gives a new result $\bar{z}_{tr}$(free) = 0.64 $\pm$ 0.16. Coincidentally, the $z_{tr}$ results overlap with the initial moment of $H_{0}$ evolution ($H_{0}$ value starts to deviate from the Planck result). This may suggest that the Hubble tension might be closely related to this particular period in the evolution of the Universe.

\end{abstract}


\section{Introduction}\label{sec:intro}
Most observations indicate that the Universe is currently in a phase of accelerated expansion, including type Ia supernovae \citep[SNe Ia;][]{1998AJ....116.1009R,2007ApJ...659...98R,2018ApJ...859..101S,2022ApJ...938..110B,2023arXiv231112098R,2024ApJ...973L..14A}, Cosmic Microwave Background \citep[CMB;][]{2011ApJS..192...18K,2014A&A...571A..16P,2020AA...641A...6P}, Baryonic Acoustic Oscillations \citep[BAOs;][]{2005ApJ...633..560E,2012JMPh....3.1152R,2011AJ....142...72E,2013AJ....146...32S,2025JCAP...02..021A}, Hubble parameter (H(z)) measurements \citep{2013ApJ...764..138F,2013ApJ...766L...7F,2017ApJ...835...26F,2024PDU....4501522S,2024JHEAp..42...96K}, Gamma-ray bursts \citep[GRBs;][]{2015NewAR..67....1W,2021MNRAS.507..730H,2022MNRAS.510.2928C,2022ApJ...924...97W,2022ApJ...935....7L,2022MNRAS.516.2575J,2023MNRAS.518.2201D}, Dark energy survey \citep[DES;][]{2019PhRvL.122q1301A}, quasar lensing \citep{2020MNRAS.498.1420W} and so on. The so-called $\Lambda$CDM model is the simplest theoretical model that supports this acceleration phase, which consists of a cosmological constant $\Lambda$ term \citep{1989RvMP...61....1W,2003PhR...380..235P} and a cold dark matter component \citep{1985ApJ...292..371D,Bertone:2010zza}. In addition, there are several other models that can also explain the accelerated expansion, such as dark energy fluid \citep{2003RvMP...75..559P}, massive gravity theories \citep{2012PhRvD..86f1502V}, modifications of Newtonian theory \citep[MOND;][]{1983ApJ...270..365M}, f(R) gravity \citep{2011PhR...509..167C,2013PhRvD..88l4036G}, f(T) gravity \citep{2011EPJC...71.1752M,2016RPPh...79j6901C}, brane world models \citep{1999PhRvL..83.4245C,1999PhRvL..83.4690R}, string \citep{1994GReGr..26.1171D}, Kaluza-Klein theories \citep{1997PhR...283..303O}, and so on. Theoretically, the current accelerated expansion of the Universe is a smooth transition from a decelerating expansion. A model without a deceleration-acceleration transition fails on explaining current cosmological observations \citep{2020JCAP...04..053J}. 

Since the discovery of the accelerated expansion of the Universe \citep{2004ApJ...607..665R}, there have been many attempts, using different type observations and methods, to precisely constrain the parameter that determines such transition from deceleration to an accelerated phase, that is transition redshift $z_{tr}$ \citep{2005MNRAS.360L...1F,2005ApJ...633..611L,2009JCAP...10..010G,2015JCAP...12..045R,2018ApJ...856....3Y,2022MNRAS.509.5399C,2023GrCo...29..177R}. Various approaches have been proposed and can be roughly divided into two categories: (i) model-dependent and (ii) model-independent. The former (i) gives the constraints of $z_{tr}$ based on commonly cosmological models, for example $\Lambda$CDM \citep{2005MNRAS.360L...1F}, $w_{0}$CDM \citep{2005MNRAS.360L...1F}, o$\Lambda$CDM \citep{2016JCAP...05..014M}, Chevallier-Polarski-Linder (CPL) model\citep{2021CQGra..38r4001K}, Chameleon model \citep{2022EPJC...82.1165S}. The latter (ii) usually adopts a model-independent method to constrain the parameter $z_{tr}$, for instance Cosmograhpy \citep{2004ApJ...607..665R,2007ApJ...659...98R,2009JCAP...10..010G}, linear parametrization \citep{2008MNRAS.390..210C,2015JCAP...12..045R,2018JCAP...05..073J,2023MNRAS.523.4938M}, Gaussian process \citep[GP;][]{2012JCAP...06..036S,2018ApJ...856....3Y,2023IJMPD..3250039K}, piecewise linear fit \citep{2016JCAP...05..014M}, and Bezier curve \citep{2023MNRAS.523.4938M}. In theory, for the given cosmic evolution history, consistent constraints should be given on the transition redshift $z_{tr}$ when using different methods and different type of observations. For the determination of transition redshift $z_{tr}$, it is important for understanding the late-time evolution history of the Universe. Until now, the local observations can not provide accurate constraints on $z_{tr}$ like the CMB and the their distribution is relatively scattered, most of which fall within the redshift range (0.3, 1.0). In order to accurately estimate the transition redshift, much high-quality data at high enough redshift above the transition redshift are needed. At the same time, different methods and observations are also necessary to facilitate cross-verification. 

Recently, there have been some hints that the Hubble constant $H_{0}(z)$ might evolve with the redshift of the data used. They come from a combination of different observations and methods, and the main observations used include quasar lens \citep{2020MNRAS.498.1420W,2020A&A...639A.101M}, SNe Ia \citep{2021ApJ...912..150D,2022A&A...668A..34H,2024EPJC...84..317M,2024PDU....4401464O,2025arXiv250111772D}, H(z) measurements \citep{2022MNRAS.517..576H}, and composite observations \citep{2020PhRvD.102j3525K,2022Galax..10...24D,2022PhRvD.106d1301O,2023A&A...674A..45J,2024MNRAS.530.5091X,2025ApJ...979L..34J,2025JCAP...03..026L}. The evolutionary behavior of $H_{0}(z)$ is one of the proposed solutions that can alleviate the Hubble tension which is currently one of the most serious challenge to the standard $\Lambda$CDM model. This tension arises from the 5$\sigma$ discrepancy between the values of $H_{0}$ measured from the local distance ladder \citep[SH0ES;][]{2022ApJ...934L...7R} and the Planck cosmic microwave background \citep[CMB;][]{2020A&A...641A...6P}. More information about the Hubble tension can be obtained from the review literature \citep{2021CQGra..38o3001D,2022JHEAp..34...49A,2022NewAR..9501659P,2023Univ....9...94H,2023eppg.confE.231K,2023Univ....9..393V,2024Univ...10..140C,2024IAUS..376...15R,2024MNRAS.527.7692W}. It is noteworthy that the evolutionary behavior occurs at redshifts between 0.30 and 0.70 \citep{2022MNRAS.517..576H,2023A&A...674A..45J,2025ApJ...979L..34J}, which seems to coincide with the possible redshift range of the transition redshift $z_{tr}$. This may not be a coincidence. There are reasons to suspect that the Hubble tension might be related to this special period in the evolution of the Universe \citep{2004ApJ...607..665R,2007ApJ...659...98R,2022MNRAS.517..576H}. Therefore, accurately constraining $z_{tr}$ might help us better understand the evolution of the Hubble constant and explore the nature of the Hubble tension.

In this present work, our primary goal is to constrain the transition redshift $z_{tr}$ utilizing the different combinations of methods and observations. Afterwards, we collected the results from studies from 2004 to 2024 and constructed a statistical analysis. Finally, based the final results and recent research on the evolution behavior of the Hubble constant that can alleviate the Hubble tension, we gave some comments on the tension. Our research encompasses three methods ($\Lambda$CDM model, Cosmography and Gaussian process (GP) method) and three kinds of H(z) measurements including cosmic chronometers (CCs), BAOs, and Dark Energy Spectroscopic Instrument (DESI) BAOs \citep{2025JCAP...02..021A}. The outline of this paper is as follows. Section \ref{S2} provides basic information about the datasets used. Section \ref{S3} presents the methods for deriving estimates of $z_{tr}$ in detail. Section \ref{S4} contains the presentation of our principal results. A comprehensive analysis and discussion can be found in Sect. \ref{S5}. Finally, a brief conclusion is presented in Sect. \ref{S6}.

\section{Data sample} \label{S2}
There are 60 H(z) measurements used in this paper. It covers redshift range (0.07, 2.40), consists 35 CCs \citep{2012JCAP...08..006M,2022ApJ...928L...4B,2023ApJS..265...48J,2023JCAP...11..047J,2023A&A...679A..96T} and 25 BAOs \citep{2024ApJ...975L..36H,2024SciBu..69.2698Y}. The former is derived by comparing relative ages of galaxies at different redshifts \citep{2002ApJ...573...37J}. The latter depends on the sound horizon at the drag epoch $r_{d}$ \citep{2025JCAP...02..021A}. More detailed information can be found in Tables \ref{tab:CC} and \ref{tab:BAO}. These two datasets have been widely used in cosmological researches, such as cosmological constraints \citep{2018ApJ...856....3Y,2020CQGra..38e5007B,2020EPJC...80..562V,2022MNRAS.512..439C,2024arXiv241218493L}, cosmic anisotropy \citep{2024PhRvD.109l3533B,2024A&A...689A.215H,2024arXiv241110838S}, calibration of the GRB correlations \citep{2015NewAR..67....1W,2021MNRAS.507..730H,2022ApJ...941...84L,2022ApJ...924...97W,2023MNRAS.521.4406L,2024JHEAp..44..323F}, constraints of the transition redshift \citep{2016JCAP...05..014M,2020JCAP...04..053J}, cosmographic constraints \citep{2020MNRAS.491.4960L,2021PhRvD.103f3537A,2022BrJPh..52..115V} and alleviation of the Hubble tension \citep{2022MNRAS.517..576H,2024arXiv241105744D,2024ApJ...975L..36H,2024PhRvD.110b1304L,2024MNRAS.533..244L,2024MNRAS.530.5091X,2025JCAP...03..066B,2025JCAP...03..026L}. 

Uncertainty in the H(z) measurements consists of two components: statistical and systematic errors. The former will decrease with the advancement of observation technology and the increase of sample size, but the latter will not necessarily decrease accordingly. Systematic error can arise from a variety of factors, including instrumentation, environment, and observational methods. For CCs measurements, the systematic errors primarily stem from the assumption of stellar population synthesis (SPS) model, error in the CC star formation history (SFH), error in the CC metallicity estimate and rejuvenation effect \citep{2022LRR....25....6M}. For BAOs measurements, the systematic errors primarily stem from theoretical systematics, modelling choices and observational systematics \citep{2024MNRAS.534..544C}. For the treatments of systematic of CCs and BAOs, see references \citet{2020ApJ...898...82M} and \citet{2024PhRvD.109b3525G}, respectively. In this paper, we not only considered statistical errors in the research process, but also analyzed and discussed the possible impact of systematic errors with examples.

\section{Methodology} \label{S3}
In this section, we will show how to reconstruct the deceleration parameter, q(z), from H(z) measurements, and then find the transition redshift from the condition q($z_{tr}$) = 0. In order to perform this reconstruction, we must know how to obtain q(z) from H(z) measurements utilizing different approaches including the $\Lambda$CDM model, Cosmography, and GP method. Assuming a Friedman-Lemaitre-Robertson-Walker (FLRW) metric, the definition of the deceleration parameter q(z) and its relation to H(z) can be given by: 
\begin{equation}
	q(z) \equiv -\frac{1}{H^{2}}\frac{\ddot{a}}{a} =  \frac{1+z}{H} \frac{d H}{d z} - 1,
	\label{qz}
\end{equation}
where $a$ is the scale factor, and H is Hubble parameter H(z). As one may see, the deceleration parameter q(z) can be obtained from H(z) and from its first derivative, H$^{'}$(z). 

\subsection{$\Lambda$CDM model}
For the flat $\Lambda$CDM model, H(z) can be given by 
\begin{equation}
	H(z) = H_{0} E(z) = H_{0}\sqrt{\Omega_{m}(1+z)^{3}+\Omega_{\Lambda}}.
	\label{HzL}
\end{equation}
Here $H_{0}$ is the Hubble constant, which characterizes the current expansion rate of the Universe. Parameters $\Omega_{m}$ and $\Omega_{\Lambda}$ represent the matter density and the dark energy density, respectively. In the flat $\Lambda$CDM model, the sum of these two parts equals 1.00, that is $\Omega_{m} + \Omega_{\Lambda} = 1$. Substituting Eqs. (\ref{HzL}) into (\ref{qz}), we get 
\begin{equation}
	q(z) = - \frac{3\Omega_{m}(1+z)^{3}}{2[\Omega_{m}(1+z)^{3}+1-\Omega_{m}]} - 1.
	\label{qzn}
\end{equation}
Utilizing Eq. (\ref{qzn}), we can get the expression of transition redshift as follows
\begin{equation}
	z_{tr} = [\frac{2(1-\Omega_{m})}{\Omega_{m}}]^{1/3} - 1,
	\label{zt}
\end{equation}
and its corresponding 1$\sigma$ error is given by
\begin{equation}
	\sigma_{z_{tr}} = \frac{2}{3}[\frac{2(1-\Omega_{m})}{\Omega_{m}^{4}}]^{-\frac{2}{3}}\times \sigma_{\Omega_{m}},
	\label{zts}
\end{equation}
where $\sigma_{\Omega_{m}}$ represents the 1$\sigma$ error of the best fits of $\Omega_{m}$. Finally, the estimation of the transition redshift $z_{tr}$ can be given by combining Eqs. (\ref{zt}) and (\ref{zts}) with the best fitting results of $\Omega_{m}$. Parameter fitting is performed employing the MCMC method, which introduced in detail in section \ref{mcmethod}.

\subsection{Cosmography} 
Cosmography has been widely used in cosmological data processing to stint the state of the kinematics of our Universe in a model-independent way \citep{2015CQGra..32m5007V,2016IJGMM..1330002D,2018MNRAS.476.3924C,2019MNRAS.484.4484C,2019IJMPD..2830016C,2019A&A...628L...4L,2021A&A...649A..65B,2024MNRAS.527.7861G,2024JCAP...09..069K,2025JCAP...02..076K,2025PDU....4701759M,2025MNRAS.537..436P}. It relies only on the assumption of a homogeneous and isotropy universe as described by the FLRW metric \citep{1972gcpa.book.....W}. Its methodology is essentially based on expanding measurable cosmological quantities as Taylor series with respect to the current time. In this framework, the evolution of universe can be described by some cosmographic parameters, such as Hubble parameter $H$, deceleration $q$, jerk $j$, snap $s$, and lerk $l$ parameters. The corresponding definitions of these parameters can be expressed as follows:
	\begin{eqnarray}
		\label{eq:q0j0}
		H = \frac{\dot{a}}{a}, q = -\frac{1}{H^{2}}\frac{\ddot{a}}{a}, j=\frac{1}{H^3}\frac{\dot{\ddot{a}}}{a},
		s=\frac{1}{H^4}\frac{\ddot{\ddot{a}}}{a}, l=\frac{1}{H^5}\frac{\dot{\ddot{\ddot{a}}}}{a}.
	\end{eqnarray} 
 
The $z$-redshift method is the earliest Taylor series used in Cosmography. The corresponding luminosity distance can be conveniently expressed as \citep{2007CQGra..24.5985C,2011PhRvD..84l4061C}
\begin{eqnarray}
        \label{eq:dlz}
        d_{L}(z) &=& \frac{c}{H_{0}}[z + \frac{1}{2} (1-q_{0})z^{2} -\frac{1}{6}(1-q_{0}-3q_{0}^{2}+j_{0})z^{3} \nonumber \\
        &+& \frac{1}{24}(2-2q_{0}-15q_{0}^{2}-15q_{0}^{3} +5j_{0}+10q_{0}j_{0}+s_{0})z^{4} \nonumber \\
        &+& \frac{1}{120}(-6+6q_{0}+81q_{0}^{2}+165q_{0}^{3}+105q_{0}^{4}+10j_{0}^{2} \nonumber \\
        &-&27j_{0}-110q_{0}j_{0}-105q_{0}^{2}j_{0}-15q_{0}s_{0}-11s_{0}-l_{0})z^{5} \nonumber \\
        &+&\textit{O}(z^6)], \nonumber \\
\end{eqnarray}
where $H_{0}$, $q_{0}$, $j_{0}$, $s_{0}$, and $l_{0}$ represent the current values. The first two terms above are Weinberg's version of the Hubble law which can be found from Equation (14.6.8) in the book \citep{1972gcpa.book.....W}. The third and fourth terms are equivalent to that obtained by \citet{1999MNRAS.306..696N} and \citet{2004CQGra..21.2603V}, respectively. Meanwhile, the corresponding expression of H(z) can be written as \citep{2020MNRAS.494.2576C}
\begin{eqnarray}
        \label{eq:hz}
        H(z) &=& H_{0}[1 + (1+q_{0})z + \frac{1}{2} (j_{0} - q_{0}^{2})z^{2} - \frac{1}{6}(-3q_{0}^{2} - 3q_{0}^{3} \nonumber \\ 
        &+& j_{0}(3 + 4q_{0}) + s_{0})z^{3} + \frac{1}{24} (-4j_{0} + l_{0} - 12q_{0}^{2} - 24q_{0}^{3} \nonumber \\ 
        &-& 15q_{0}^{4} + j0(12 + 32q_{0} + 25q_{0}^{2}) + 8s_{0} + 7q_{0}s_{0})z^{4} \nonumber \\ 
        &+&\textit{O}(z^5)].
\end{eqnarray}
Substituting Eqs. (\ref{eq:hz}) in (\ref{qz}), we get 
\begin{equation}
	q(z) =  \frac{p_{1}+2p_{2}z+(3p_{3}+p_{2})z^{2} + (4p_{4}+2p_{3})z^{3} + 3p_{4}z^{4}}{1+p_{1}z+p_{2}z^{2}+p_{3}z^{3}+p_{4}z^{4}},
	\label{qzn1}
\end{equation}
where 
\begin{eqnarray}
    p_{1} &=& 1+q_{0},  \nonumber \\
    p_{2} &=& \frac{1}{2} (j_{0} - q_{0}^{2}),  \nonumber \\
    p_{3} &=& - \frac{1}{6}(-3q_{0}^{2} - 3q_{0}^{3} + j_{0}(3 + 4q_{0}) + s_{0}), \nonumber \\
    p_{4} &=& \frac{1}{24} (-4j_{0} + l_{0} - 12q_{0}^{2} - 24q_{0}^{3} - 15q_{0}^{4} + j0(12 + 32q_{0}\nonumber \\ 
    &+& 25q_{0}^{2}) + 8s_{0} + 7q_{0}s_{0})z^{4}.
	\label{pp}
\end{eqnarray}
Considering the small amount of data, in order to avoid overfitting and high uncertainty, we only give constraints on first two parameters $q_0$ and $j_{0}$ \citep{2022A&A...661A..71H}. With the simplified Eqs. (\ref{eq:hz}) and (\ref{qzn1}), we can derive the transition redshift according to the condition q($z_{tr}$) = 0 in the following form 
\begin{equation}
	z_{tr} = -1 + \sqrt{1+\frac{-2q_{0}}{j_{0} - q_{0}^{2}}}.
	\label{zt1}
\end{equation}
The corresponding 1$\sigma$ error can be given by 
\begin{eqnarray}
	\sigma_{z_{tr}} &=& \frac{1}{2}(1+\frac{-2q_{0}}{j_{0} - 2q_{0}^{2}})^{-1/2} \times [(\frac{-2j_{0}-2q_{0}^{2}}{(j_{0}-q_{0}^{2})^{2}})^{2}\sigma_{q_{0}}^{2} \nonumber \\ 
 &+& (\frac{-2q_{0}}{(j_{0}-q_{0}^{2})^{2}})^{2}\sigma_{j_{0}}^{2}]^{1/2},
	\label{zt1s}
\end{eqnarray}
where $\sigma_{q_{0}}$ and $\sigma_{j_{0}}$ are the 1$\sigma$ errors corresponding to the best fits.

\subsection{Gaussian process method}
Gaussian process has been extensively used for cosmological applications, such as constraint on cosmological parameters \citep{2018ApJ...856....3Y,2018JCAP...04..051G,2019ApJ...886L..23L,2020CQGra..38e5007B}, calibration of high redshift cosmic probe \citep{2021MNRAS.507..730H,2022ApJ...924...97W} and comparison of cosmological models \citep{2018JCAP...02..034M}. Here, we only give a brief introduction to the GP method. A more detailed explanation can be discovered from the literature \citep{2006gpml.book.....R,SCHULZ20181}. In this work, the GP regression is implemented by employing the package \emph{GaPP3} \footnote{https://github.com/lighink/GaPP3} \citep{2011JMLR...12.2825P} in the Python environment. It will reconstruct a continuous function $h(x)$ that is the best representative of a discrete set of measurements $h(x_{i})\pm\sigma_{i}$ at $x_{i}$, where $i$ = 1,2,..., $N$ and $\sigma_{i}$ is the 1$\sigma$ error. The GP method assumes that the value of $h(x_{i})$ at any position $x_{i}$ is random that follows a Gaussian distribution with expectation $\mu(x)$ and standard deviation $\sigma(x)$. They can be determined from observations through a defined covariance function $k(x,x_{i})$ or kernel function
\begin{eqnarray}
	\label{eq:gp_mu}
	\mu(x) &=& \sum_{i,j = 1}^{N} k(x, x_{i})(M^{-1})_{ij}h(x_{j}),
\end{eqnarray}
and
\begin{eqnarray}
	\label{eq:gp_sigma}
	\sigma(x) &=& k(x, x_{i}) - \sum_{i,j = 1}^{N} k(x, x_{i})(M^{-1})_{ij}k(x_{j},x),
\end{eqnarray}
where the matrix $M_{ij} = k(x_{i}, x_{j}) + c_{ij}$ and $c_{ij}$ is the covariance matrix of the observations. For uncorrelated data, it can be simplified as $diag(\sigma^{2}_{i})$. Equations (\ref{eq:gp_mu}) and (\ref{eq:gp_sigma}) specify the posterior distribution of the extrapolated points. For given data sets ($x_{i}, y_{i}$), considering a suitable GP kernel, it is straightforward to derive the continuous function $h(x)$ which used to get the value of $H_{0}$, that is $h(0)$. 

In this work, we consider a common kernel function, the Radial Basis Function kernel (RBF) kernel, also known as the Gaussian kernel, which can be written as
\begin{equation}\label{eq:RBF}
	k(x,\tilde{x})=\sigma_h^2\exp{\left[-\frac{(x-\tilde{x})^2}{2l^2}\right]},
\end{equation}
where, parameters $\sigma_{h}$ and $l$ control the strength of the correlation of the function value and the coherence length of the correlation in $x$, respectively. More detailed information about covariance functions for the Gaussian process can be found in chapter 4 of the book \citep{2006gpml.book.....R}. Parameters $\sigma_h$ and $l$ are optimized for the observations, $h(x_i)\pm\sigma_i$, by minimizing the log marginal likelihood function \citep{2012JCAP...06..036S}
\begin{eqnarray}\label{likelihood}
	\ln\mathcal{L} &=& -\frac{1}{2}\sum_{i,j=1}^N[h(x_i)-\mu(x_i)](M^{-1})_{ij}[h(x_j)-\mu(x_j)] \nonumber \\
	&-&\frac{1}{2}
	\ln|M|-\frac{1}{2}N\ln{2\pi},
\end{eqnarray}
where $|M|$ is the determinant of $M_{ij}$.

The GP method, as performed by GaPP3, furnish not only the H(z) reconstruction from H(z) measurements, but also its derivatives up to fourth order and their estimated uncertainties and covariances. More details can be found from \citet{2012JCAP...06..036S}. Therefore, it just remains to estimate the uncertainty of q(z), $\sigma_{q}$, via the error propagating Eq. (\ref{qz}). We get 
\begin{equation}\label{sigmaqz}
	q(z) =  \frac{1+z}{H} H^{'} - 1,
\end{equation}
and 
\begin{equation}\label{sigmaq}
	(\frac{\sigma_{q}}{1 + q})^{2} = (\frac{\sigma_{H^{'}}}{H^{'}})^{2} + (\frac{\sigma_{H}}{H})^{2} -\frac{2\sigma_{HH^{'}}}{HH^{'}},
\end{equation}
where, $H^{'}$ represents the derivative of H(z) with respect to redshift z.

\subsection{MCMC method} \label{mcmethod}
Constraints of free parameters $P_{i}$ (including $\Omega_{m}$, $H_{0}$, $q_{0}$ and $j_{0}$) for a fixed cosmological model can be given by minimizing the chi-square ($\chi^{2}$) function. When only statistical error is considered, the corresponding format of $\chi^{2}$ is as follows:
\begin{eqnarray}
	\label{eq:chi}
	\chi^{2} = \sum_{i=1}^{N}\frac{(\Delta{H})^{2}}{\sigma_{i}^{2}}.
\end{eqnarray}
Here, $\sigma_{i}$ is the corresponding 1$\sigma$ statistical error. $\Delta{H}$ represents the difference between the observational Hubble parameter $H_{\rm obs}(z)$ and the theoretical prediction $H_{\rm th}(P_{i}, z_{i})$:
\begin{eqnarray}
	\label{eq:deltaH}
	\Delta{H} = H_{\rm obs}(z) - H_{\rm th}(P_{i}, z_{i}).
\end{eqnarray}
Equations (\ref{HzL}) and (\ref{eq:hz}) can provide the $H_{\rm th}(P_{i},z_{i})$ for the flat $\Lambda$CDM model and the Cosmography method, respectively. When both statistical error ($C_{\rm stat}$) and systematic error ($C_{\rm syst}$) are considered, the corresponding format of $\chi^{2}$ is as follows:
\begin{eqnarray}
	\label{eq:chi1}
	\chi^{2} = \Delta{H}C^{-1}_{\rm stat+syst}\Delta{H^{T}}.
\end{eqnarray}
At this time, the statistical ($C_{stat}$) and systematic ($C_{syst}$) covariance matrices are combined and adopted to constrain the cosmological parameters:
\begin{eqnarray}
	\label{eq:CC_ss}
	C_{\rm stat+syst} = C_{\rm syst} + C_{\rm stat}.
\end{eqnarray}
In this work, the minimization is performed employing a Bayesian Markov chain Monte Carlo (MCMC) \citep{2013PASP..125..306F} method with the $emcee$ package\footnote{https://emcee.readthedocs.io/en/stable/}. All the fittings in this paper are obtained adopting this python package.

\section{Results} \label{S4}
\begin{figure}
	\centering
	\includegraphics[width=0.32\textwidth]{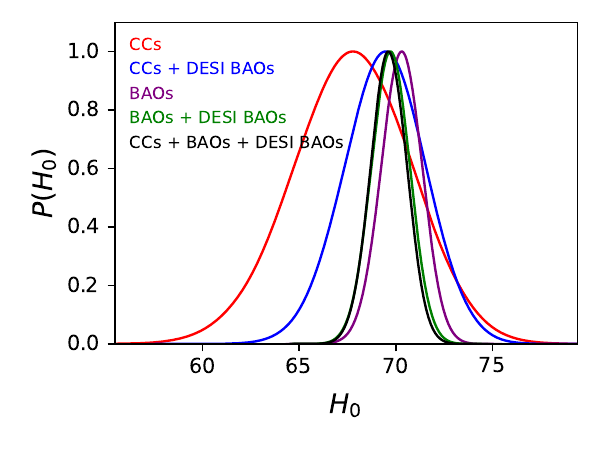} \\
    \includegraphics[width=0.32\textwidth]{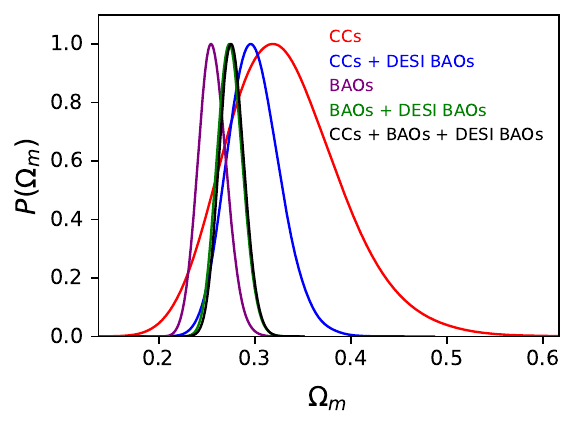}
	\caption{Cosmological constraints of the flat $\Lambda$CDM model from a diverse combination of observations which consists CCs, BAOs and DESI BAOs.}
	\label{omh0}       
\end{figure}

First, we gave the cosmological constraints of the $\Lambda$CDM model from a diverse combination of observations which including CCs, BAOs and DESI BAOs. For the CCs + BAOs + DESI BAOs combination, the corresponding constraints are $\Omega_{m}$ = $0.28 \pm 0.01$ and $H_{0}$ = $69.63 \pm 0.94$ km/s/Mpc. They are in line with ones obtained from other combinations within 1$\sigma$ level. All the marginalized likelihood distributions are shown in Fig. \ref{omh0} and the corresponding best fits are summarized in Table \ref{T1}. According to the cosmological constraints, we derived the corresponding $z_{tr}$ estimations utilizing the Eqs. (\ref{zt}) and (\ref{zts}). Results derived from various combinations are 0.60$\pm$0.13 (CCs), 0.67$\pm$0.08 (CCs + DESI BAOs), 0.79$\pm$0.05 (BAOs), 0.76$\pm$0.03 (BAOs + DESI BAOs) and 0.73$\pm$0.03 (CCs + BAOs + DESI BAOs). 

\begin{figure}
	\centering
	\includegraphics[width=0.40\textwidth]{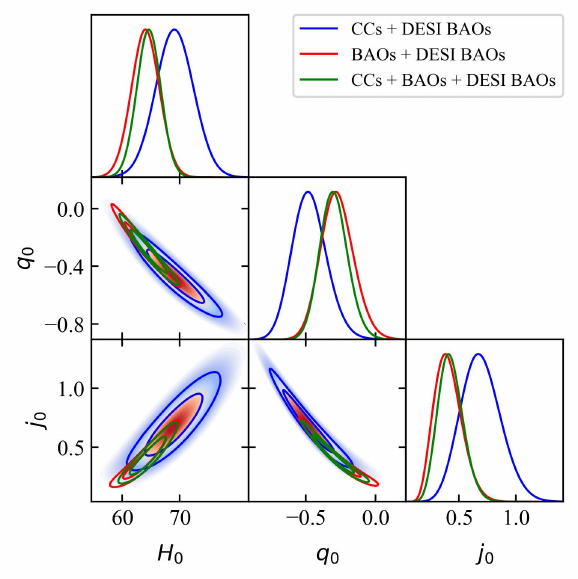}
	\caption{Confidence contours (1$\sigma$ and 2$\sigma$) for the parameters space ($H_{0}$, $q_{0}$, $j_{0}$) utilizing the Cosmography method. }
	\label{fitall}       
\end{figure}

For the estimation of $z_{tr}$ based on the Cosmography, we still need to first give the cosmographic constraints on different observational combinations. Subsequently, by inserting the best fits into Eqs. (\ref{zt1}) and (\ref{zt1s}), the estimations of $z_{tr}$ can be achieved. Individual CCs or BAOs measurements do not give tight constraints here. Thus, we only considered three combination scenarios: (1) CCs + DESI BAOs, (2) BAOs + DESI BAOs and (3) CCs + BAOs + DESI BAOs. The best fitting results are ($H_{0}$ = 69.11$_{-3.28}^{+3.31}$ km/s/Mpc, $q_{0}$ = -0.48$_{-0.12}^{+0.13}$, $j_{0}$ =0.69$_{-0.17}^{+0.18}$), ($H_{0}$ = 64.03$_{-2.43}^{+2.44}$ km/s/Mpc, $q_{0}$ = -0.28$_{-0.11}^{+0.12}$, $j_{0}$ = 0.40$_{-0.11}^{+0.13}$) and ($H_{0}$ = 64.61$\pm$2.08 km/s/Mpc, $q_{0}$ = -0.30$\pm$0.10, $j_{0}$ = 0.42$_{-0.10}^{+0.12}$) for the observational combinations CCs + DESI BAOs, BAOs + DESI BAOs and CCs + BAOs + DESI BAOs, respectively. The corresponding confidence contours and best fits are shown in Fig. \ref{fitall} and Table \ref{T1}, respectively. According the Cosmographic constraints, we gave the final estimations of $z_{tr}$: (1) 0.45$\pm$0.38, (2) 0.32$\pm$0.38 and (3) 0.35$\pm$0.33. 

\begin{figure}
	\centering
	\includegraphics[width=0.235\textwidth]{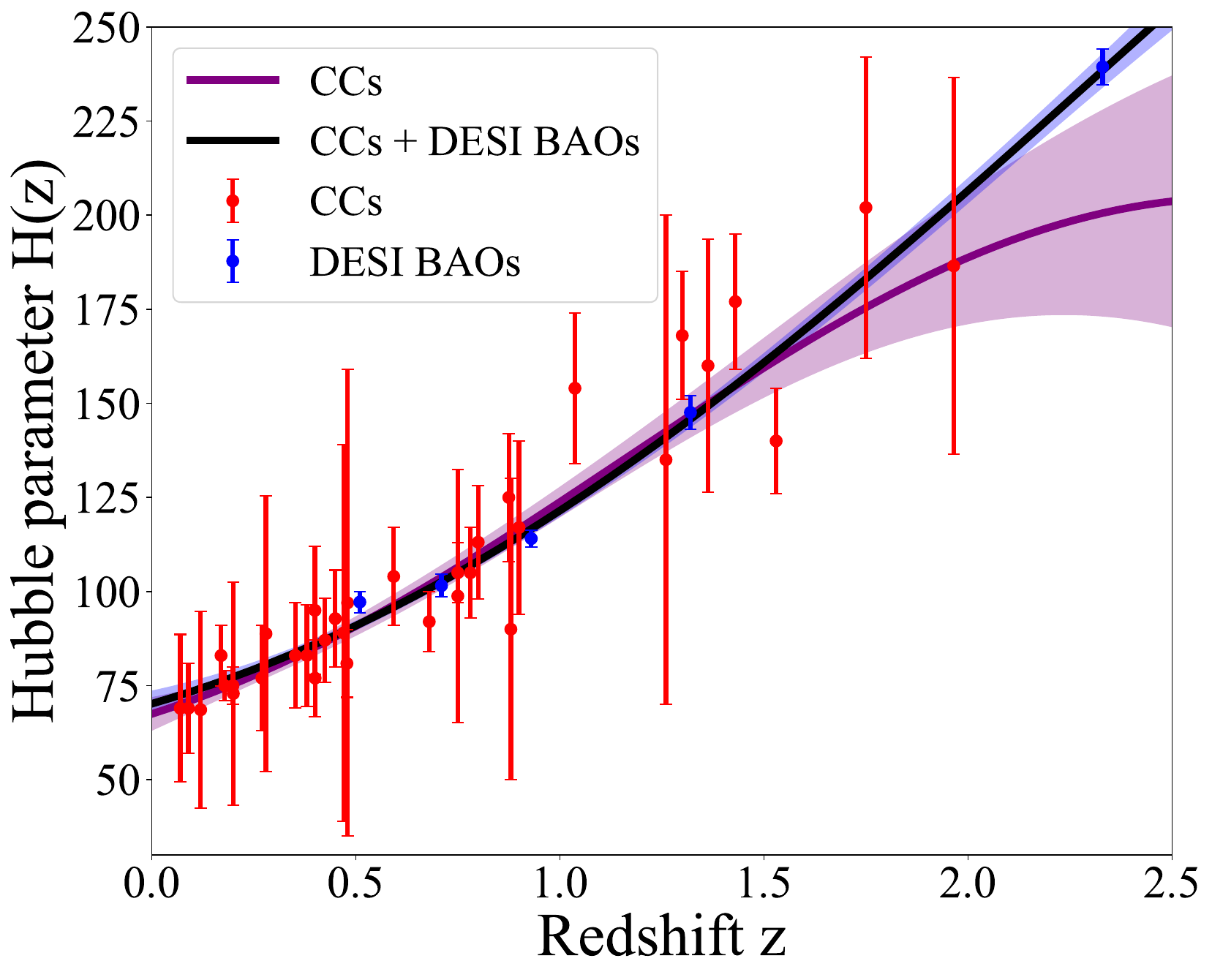}
        \includegraphics[width=0.235\textwidth]{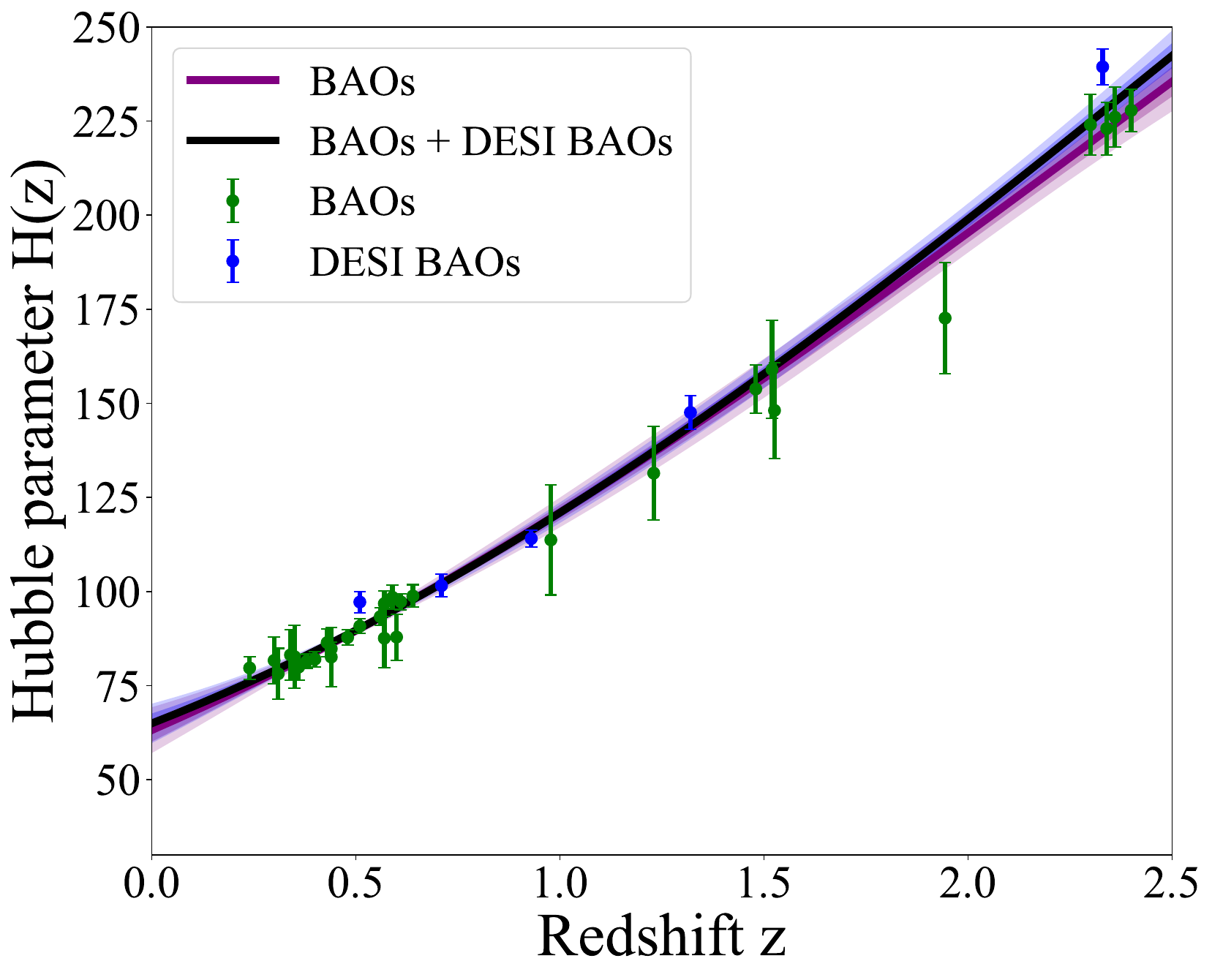}
        \includegraphics[width=0.235\textwidth]{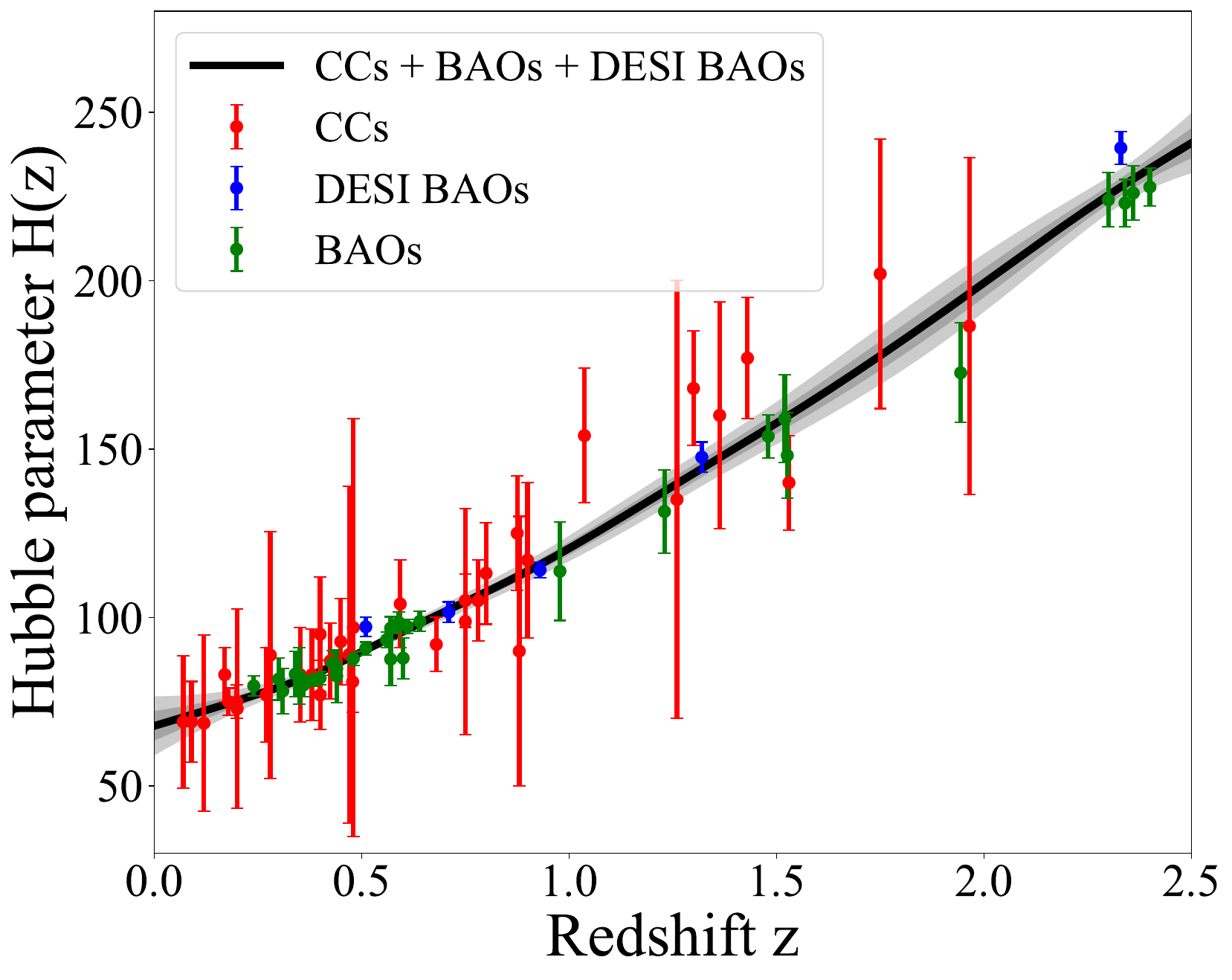}
	\caption{Reconstruction of the Hubble parameter H(z) adopting the GP method with different combinations of CCs, BAOs, DESI BAOs datasets.}
	\label{Fhz}       
\end{figure}

\begin{figure}
	\centering
	\includegraphics[width=0.235\textwidth]{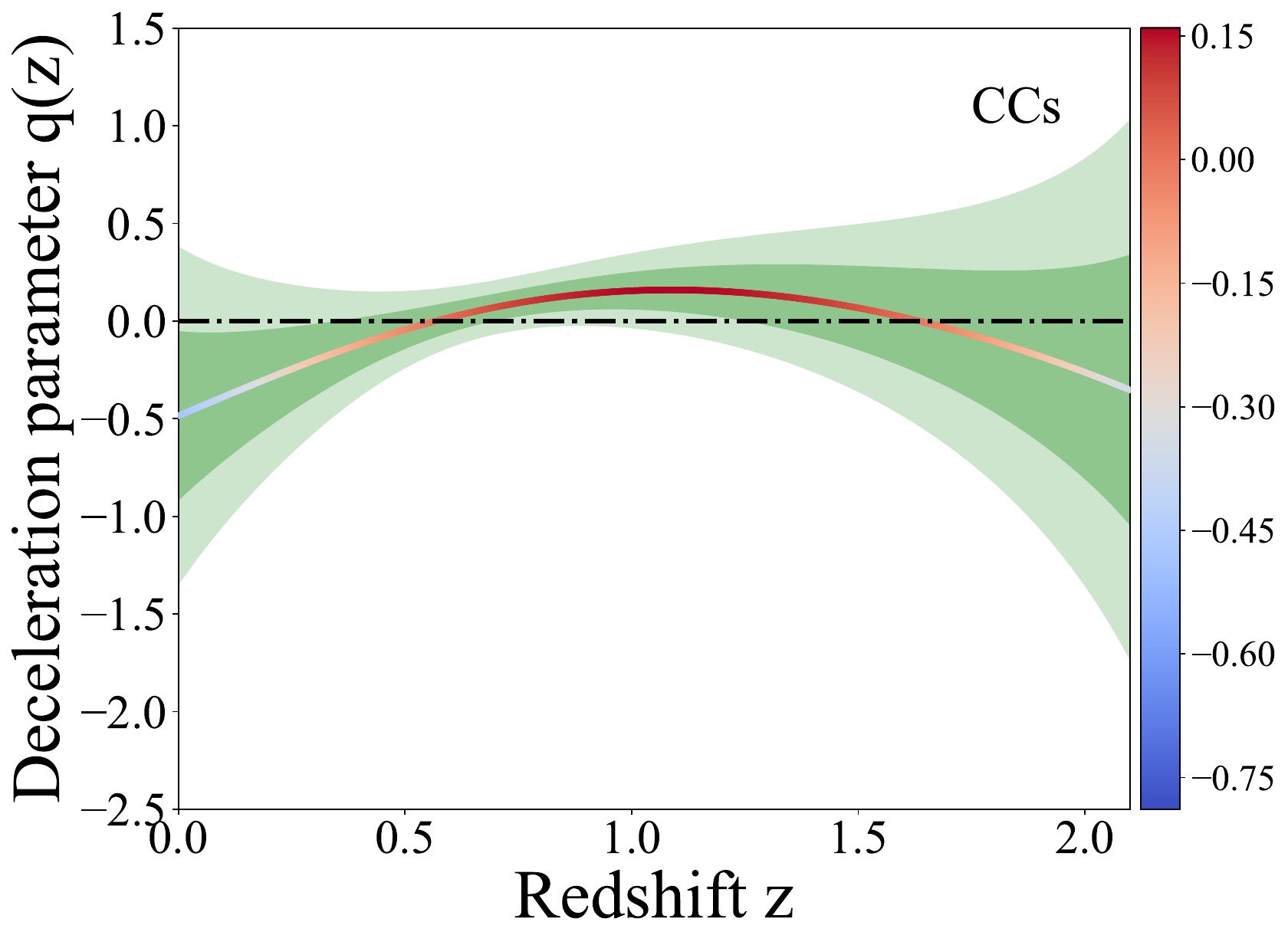}
    \includegraphics[width=0.235\textwidth]{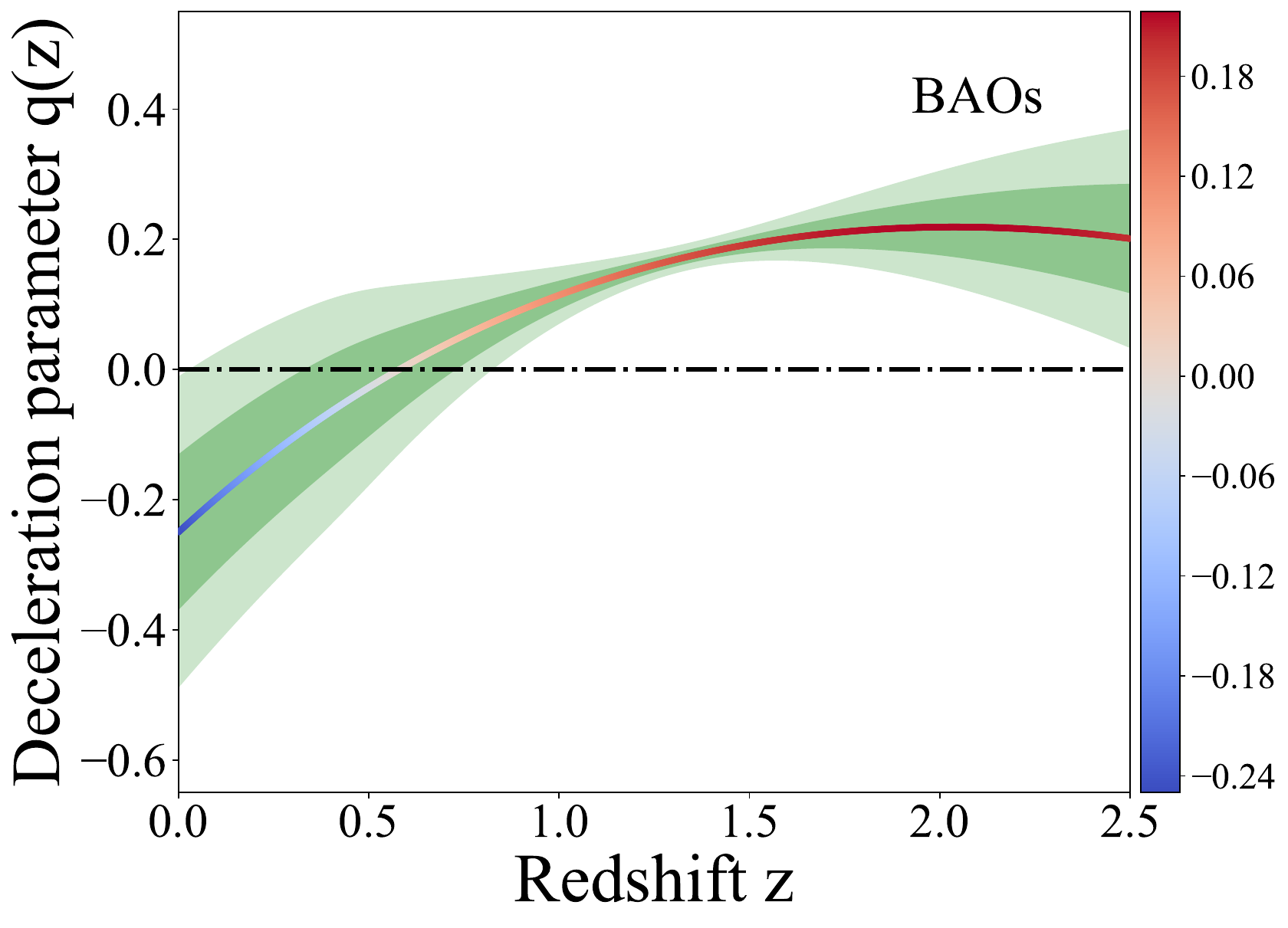} \\
    \includegraphics[width=0.235\textwidth]{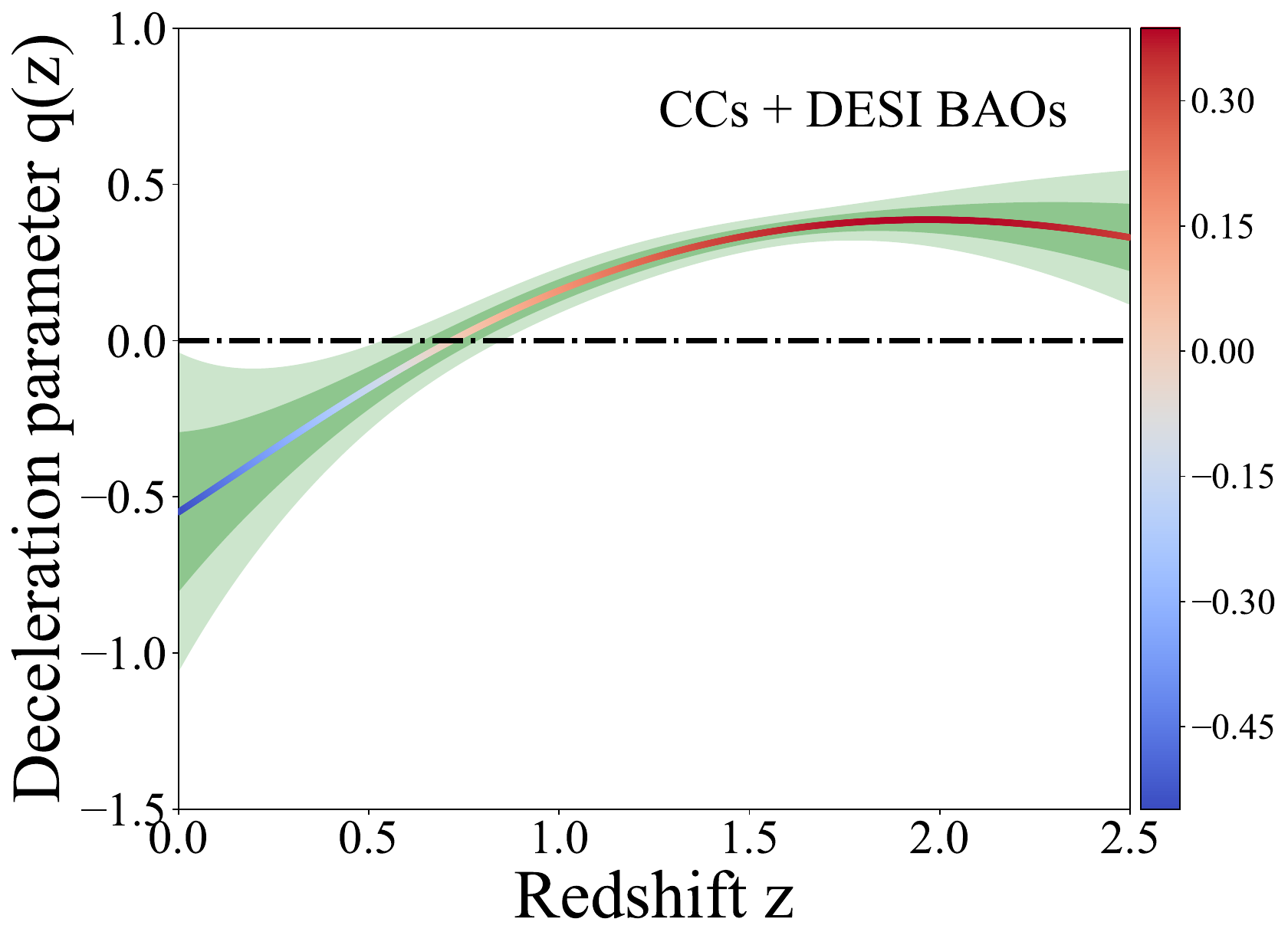} 
    \includegraphics[width=0.235\textwidth]{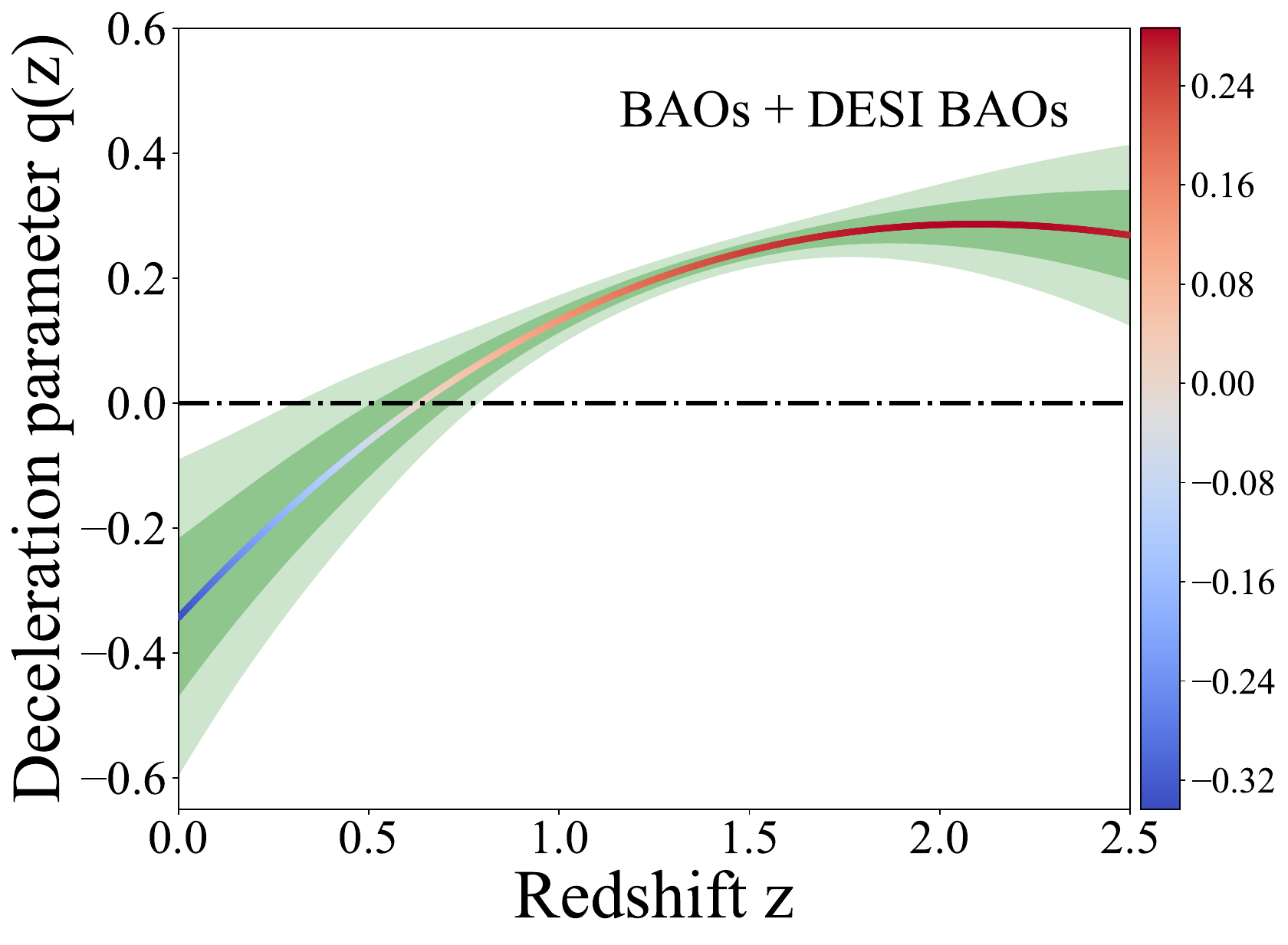} \\
    \includegraphics[width=0.235\textwidth]{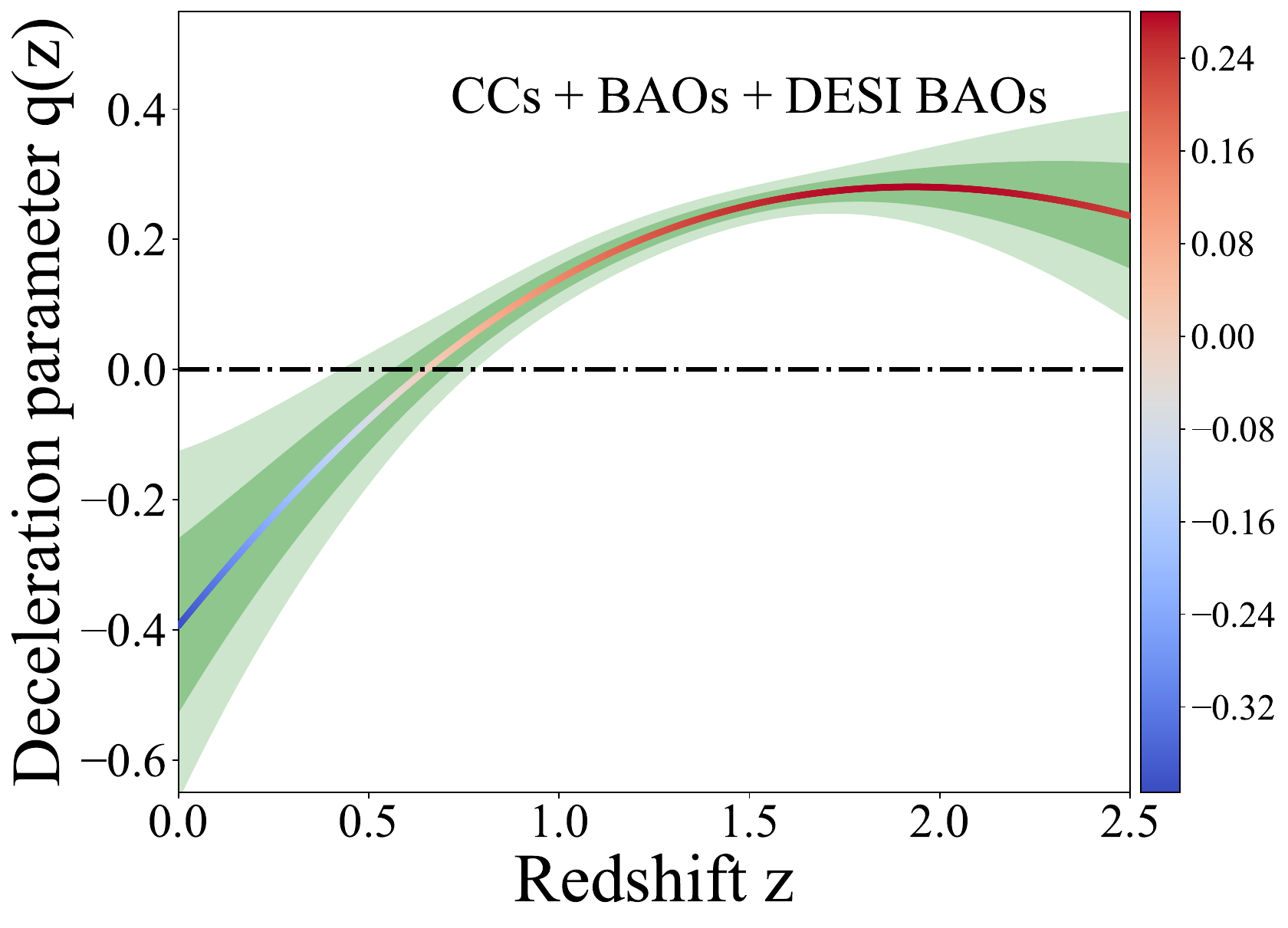}
	\caption{Reconstruction of the deceleration parameter q(z) adopting the GP method with different combinations of CCs, BAOs, DESI BAOs datasets.} 
	\label{Fqz}       
\end{figure}

For the GP method, we first drew the reconstructed H(z) function with 1$\sigma$ error based on the different observational combinations, as shown in Fig. \ref{Fhz}. Utilizing the reconstruction function of H(z), it is convenient to obtain the estimates of $H_{0}$. The $H_{0}$ constraints are overall biased towards Planck results \citep{2020AA...641A...6P}. The numerical results are summarized in Table \ref{T1}. Combining the reconstructed functions of H(z) and $H^{'}$(z) with Eqs. (\ref{sigmaqz}) and (\ref{sigmaq}), we can reconstruct the q(z) function. From the reconstructed results in Fig. \ref{Fqz}, it is easy to get the transition redshift employing the condition q($z_{tr}$) = 0, and the estimation of $q_{0}$. The estimations of $q_{0}$ from CCs, CCs + DESI BAOs, BAOs, BAOs + DESI BAOs, CCs + BAOs + DESI BAOs datasets are -0.49$\pm$0.44, -0.55$\pm$0.26, -0.25$\pm$0.12, -0.34$\pm$0.13, and -0.39$\pm$0.14, respectively. These $q_{0}$ values are consistent with the result of standard $\Lambda$CDM model with $\Omega_{m}$ = 0.30. The results of the transition redshift $z_{tr}$ are 0.57$_{-0.23}^{+0.13}$ (CCs), 0.72$_{-0.08}^{+0.07}$ (CCs + DESI BAOs), 0.58$_{-0.26}^{+0.15}$ (BAOs), 0.63$_{-0.13}^{+0.09}$ (BAOs + DESI BAOs) and 0.65$_{-0.09}^{+0.07}$ (CCs + BAOs + DESI BAOs). 

In addition, we also gave the results considering the systematic error to discuss the influence of the systematic error on the $z_{tr}$ constraints. Taking CCs + DESI BAOs sample as an example, we gave the constraints obtained by different methods when considering the CCs systematic and statistical errors, as shown in Table \ref{T2}. For the $\Lambda$CDM model, Cosmography and GP method, the estimations of $z_{tr}$ are 0.69$\pm$0.08, 0.50$\pm$0.44 and 0.77$^{+0.07}_{-0.09}$, respectively. In here, the system covariance matrix of CCs is calculated utilizing the method proposed by \citet{2020ApJ...898...82M}, which combines statistical and systematic errors. These covariance matrices have been widely used in cosmological researches \citep{2024PhRvD.109b3525G,2025EPJC...85..124O}.   

\begin{table*}\footnotesize
\centering
	\caption{Cosmological constraints ($\Omega_{m}$, $H_{0}$, $q_{0}$ and $j_{0}$) and derived estimations ($z_{tr}$) utilizing different combinations of methods and datasets. Parameter $H_{0}$ in units of $\textrm{km}~\textrm{s}^{-1} \textrm{Mpc}^{-1}$.\label{T1}}
		\begin{tabular}{ccccccc}
			\hline\hline
			Method & Parameter & Dataset (A) & Dataset (B)  & Dataset (C) & Dataset (D) & Dataset (E) \\ \hline	
			flat $\Lambda$CDM  & $H_{0}$   & 67.76$_{-3.09}^{+3.04}$  &  69.44$_{-2.13}^{+2.10}$  & 70.29$_{-1.06}^{+1.05}$  & 69.72$_{-1.00}^{+0.99}$ &  69.63$\pm$0.94 \\
                   & $\Omega_{m}$ & 0.33$_{-0.05}^{+0.06}$ &  0.30$\pm$0.03 & 0.26$_{-0.01}^{+0.02}$ & 0.27$\pm$0.01 & 0.28$\pm$0.01 \\
                   & $z_{tr}$ & 0.60$\pm$0.13  &  0.67$\pm$0.08 & 0.79$\pm$0.05  & 0.76$\pm$0.03 & 0.73$\pm$0.03\\ \hline	
            Cosmography & $H_{0}$  &  --   & 69.11$^{+3.31}_{-3.28}$  & -- & 64.03$^{+2.44}_{-2.43}$ & 64.61$\pm$2.08 \\
                   & $q_{0}$   & --  & -0.48$^{+0.13}_{-0.12}$  & -- & -0.28$^{+0.12}_{-0.11}$ & -0.30$\pm$0.01 \\
                   & $j_{0}$   &  --  & 0.69$^{+0.18}_{-0.17}$   & -- & 0.40$^{+0.13}_{-0.11}$ &  0.42$^{+0.12}_{-0.10}$ \\
                   & $z_{tr}$ &  --  & 0.45$\pm$0.38  & -- & 0.32$\pm$0.38 & 0.35$\pm$0.33 \\  \hline	
            GP & $H_{0}$ & 67.48 $\pm$ 4.58  & 70.19$\pm$3.58 & 63.17$\pm$3.11  & 64.97$\pm$2.68 & 67.79$\pm$4.48 \\
			     & $q_{0}$  & -0.49$\pm$0.44 &  -0.55$\pm$0.26 &  -0.25$\pm$0.12  & -0.34$\pm$0.13 & -0.39$\pm$0.14 \\
			   & $z_{tr}$  & 0.57$_{-0.23}^{+0.13}$  &  0.72$_{-0.08}^{+0.07}$ & 0.58$_{-0.26}^{+0.15}$  &  0.63$_{-0.13}^{+0.09}$ &  0.65$_{-0.09}^{+0.07}$ \\
			\hline\hline
		\end{tabular}
		\begin{itemize}	
			\tiny
			\item[Note:] Dataset combination: (A) CCs; (B) CCs + DESI BAOs; (C) BAOs; (D) BAOs + DESI BAOs; (E) CCs + BAOs + DESI BAOs.
		\end{itemize}
\end{table*}

\begin{table}\footnotesize
\centering
	\caption{Cosmological constraints ($\Omega_{m}$, $H_{0}$, $q_{0}$ and $j_{0}$) and derived estimations ($z_{tr}$) obtained from CCs + DESI BAOs utilizing different methods, taking into account both CCs systematic and statistical errors. Parameter $H_{0}$ in units of $\textrm{km}~\textrm{s}^{-1} \textrm{Mpc}^{-1}$.\label{T2}}
		\begin{tabular}{ccc}
			\hline\hline
			Method & Parameter  & Dataset (B)  \\ \hline	
			flat $\Lambda$CDM  & $H_{0}$   &  70.31$_{-2.53}^{+2.48}$   \\
                   & $\Omega_{m}$ &  0.29$\pm$0.03 \\
                   & $z_{tr}$ &  0.69$\pm$0.08 \\ \hline	
            Cosmography & $H_{0}$  & 70.98$^{+4.06}_{-4.08}$  \\
                   & $q_{0}$  & -0.53$^{+0.14}_{-0.13}$  \\
                   & $j_{0}$  & 0.75$^{+0.20}_{-0.19}$  \\
                   & $z_{tr}$ & 0.50$\pm$0.44   \\  \hline	
            GP & $H_{0}$  & 72.54$\pm$4.56  \\
			     & $q_{0}$  &  -0.61$\pm$0.35  \\
			   & $z_{tr}$ &  0.77$_{-0.09}^{+0.07}$ \\
			\hline\hline
		\end{tabular}
\end{table}

\begin{figure}
	\centering
    \includegraphics[width=0.235\textwidth]{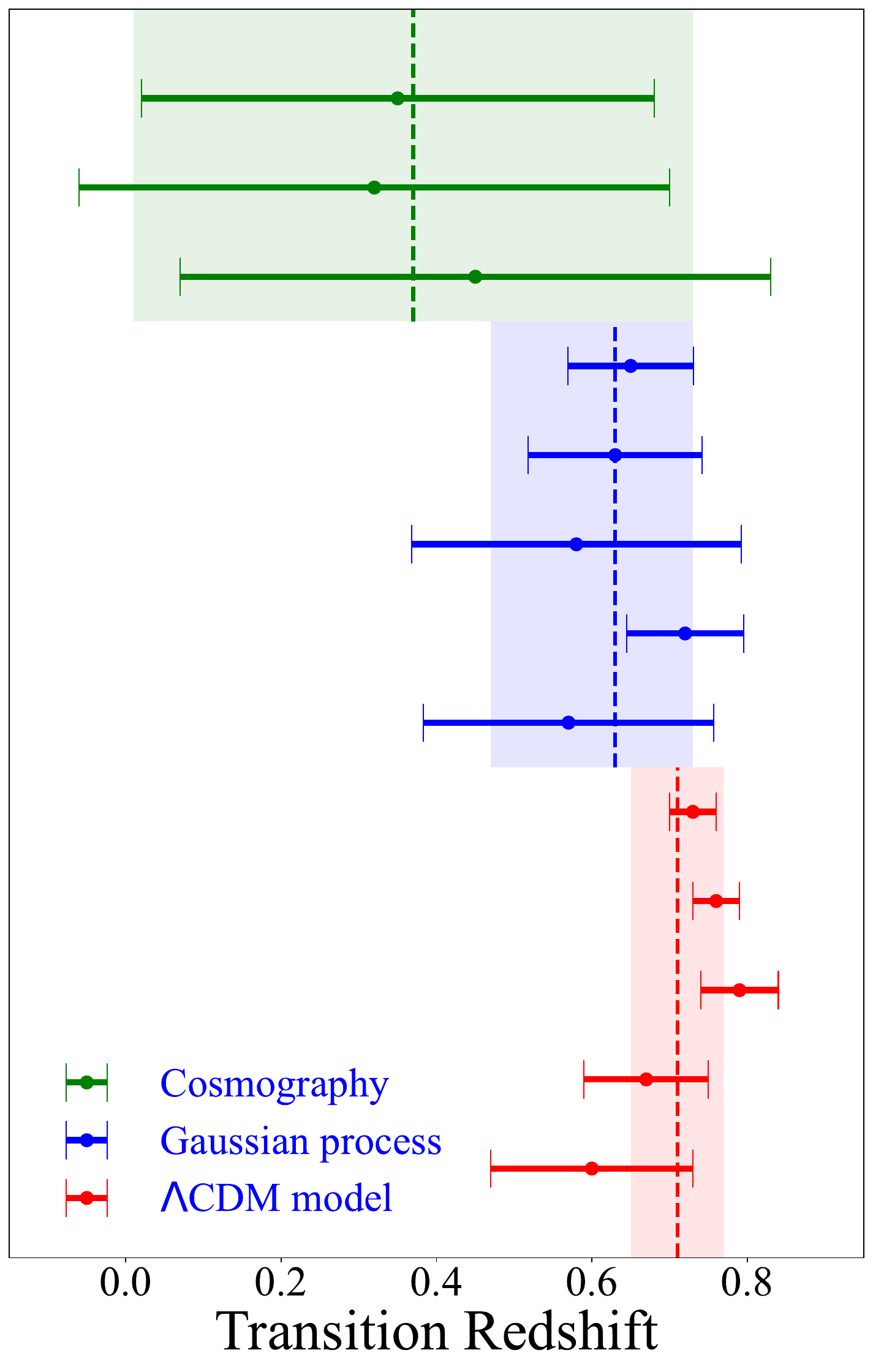} 
    \includegraphics[width=0.235\textwidth]{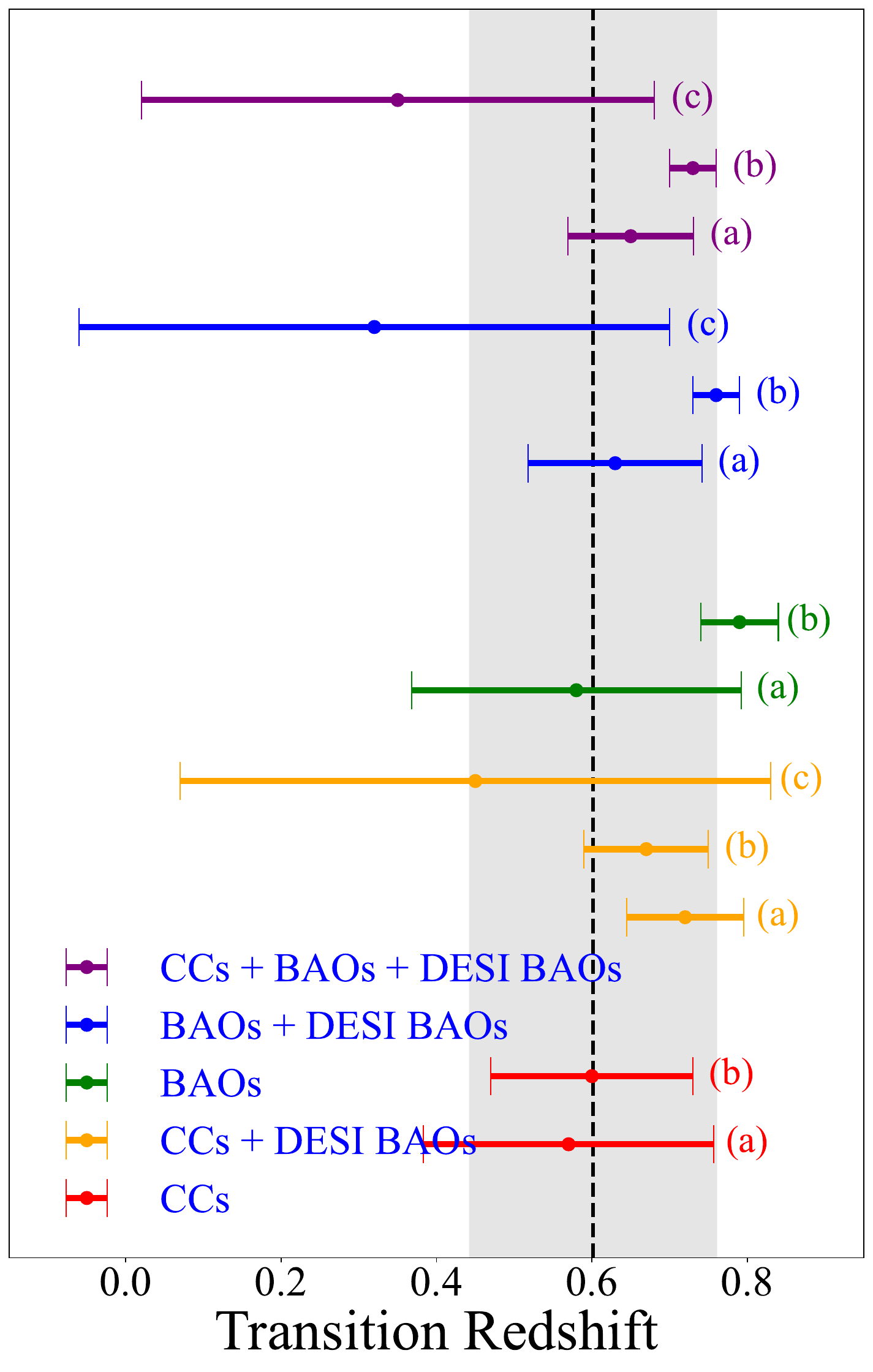}
	\caption{Estimations of the transition redshift $z_{tr}$ adopting the different methods with diverse combinations of of CCs, BAOs, DESI BAOs datasets. Left panel shows all the results classified by the approaches used. Dotted lines and shadow regions represent the mean value and corresponding 1$\sigma$ range calculated by the results from different methods. For the flat $\Lambda$CDM model, Cosmography and GP method, the calculations are  $\tilde{z}_{tr}$ = 0.71$\pm$0.06, 0.37$\pm$0.36 and 0.63$_{-0.16}^{+0.10}$, respectively. Right panel shows all the results classified according to the dataset used. Labels (a), (b) and (c) represent the GP method, the flat $\Lambda$CDM model, and the Cosmography method, respectively. Dotted line and shadow regions represent the mean value and corresponding 1$\sigma$ range calculated by all our results, that is $\tilde{z}_{tr}$ = 0.60 $\pm$ 0.16.} 
	\label{tm}       
\end{figure}

\section{Analysis and discussion} \label{S5}
Full numerical results from our analysis are displayed in Table \ref{T1} and Fig. \ref{tm}. The left panel of Fig. \ref{tm} is classified according to methodology, which helps to understand the discrepancy between different methods. The right panel shows the classification according to used samples. From Fig. \ref{tm}, we can find that the $z_{tr}$ constraints obtained from different datasets employing the same approach are consistent with each other within 1$\sigma$ level. Afterwards, we calculated the mean values $\tilde{z}_{tr}$ of the same method. For the flat $\Lambda$CDM, the result is $\tilde{z}_{tr}$ = 0.71 $\pm$ 0.06 ($\tilde{\sigma}_{z_{tr}}$). The ones from the Cosmography and GP method are 0.37 $\pm$ 0.36 and 0.63$_{-0.16}^{+0.10}$, respectively. It can be found that $\tilde{z}_{tr}$(Cosmography) $<$ $\tilde{z}_{tr}$(GP method) $<$ $\tilde{z}_{tr}$($\Lambda$CDM model) and $\tilde{\sigma}_{z_{tr}}$($\Lambda$CDM model) $<$ $\tilde{\sigma}_{z_{tr}}$(GP method) $<$ $\tilde{\sigma}_{z_{tr}}$(Cosmography). The results obtained by the model-independent method are significantly smaller than those of the $\Lambda$CDM model. Different methods have different abilities to constrain the transition redshift $z_{tr}$. The flat $\Lambda$CDM model and GP method give smaller 1$\sigma$ error of $z_{tr}$ constraints than that from the Cosmography method. Cosmography method has more free parameters, which will make the constraints worse and increase the 1$\sigma$ error of $z_{tr}$. From the right panel of Fig. \ref{tm}, we can see that the impact of the dataset on the results is weaker than that of the methodology. Afterwards, we calculated the statistical mean of our results and its 1$\sigma$ error; that is $\tilde{z}_{tr}$ = 0.60 and $\tilde{\sigma}_{{z}_{tr}}$ = 0.16. At the same time, we also investigated the impact of systematic errors on the results using CCs + DESI BAOs as an example. By comparing the results before and after considering the systematic error, as shown in Tables \ref{T1} and \ref{T2}, we find that considering the systematic error can cause the best fits of the cosmological parameters ($\Omega_{m}$, $H_{0}$, $q_{0}$ and $j_{0}$) to change insignificantly (less than 1$\sigma$), and the corresponding 1$\sigma$ error increases slightly. Changes in constraints on cosmological parameters ($\Omega_{m}$, $q_{0}$ and $j_{0}$) lead to a slight increase (less than 1$\sigma$) in the estimated value of $z_{tr}$. It is worth noting that the Hubble constant $H_{0}$ is the most affected, with a maximum increase of 2.35 km/s/Mpc, but it is not directly related to the estimation of $z_{tr}$. This finding shows that systematic errors can significantly affect the value of $H_{0}$, suggesting that the current Hubble tension might also be caused by unknown systematic errors. We should be cautious about systematic errors.

\begin{figure*}
	\centering
	\includegraphics[width=0.475\textwidth]{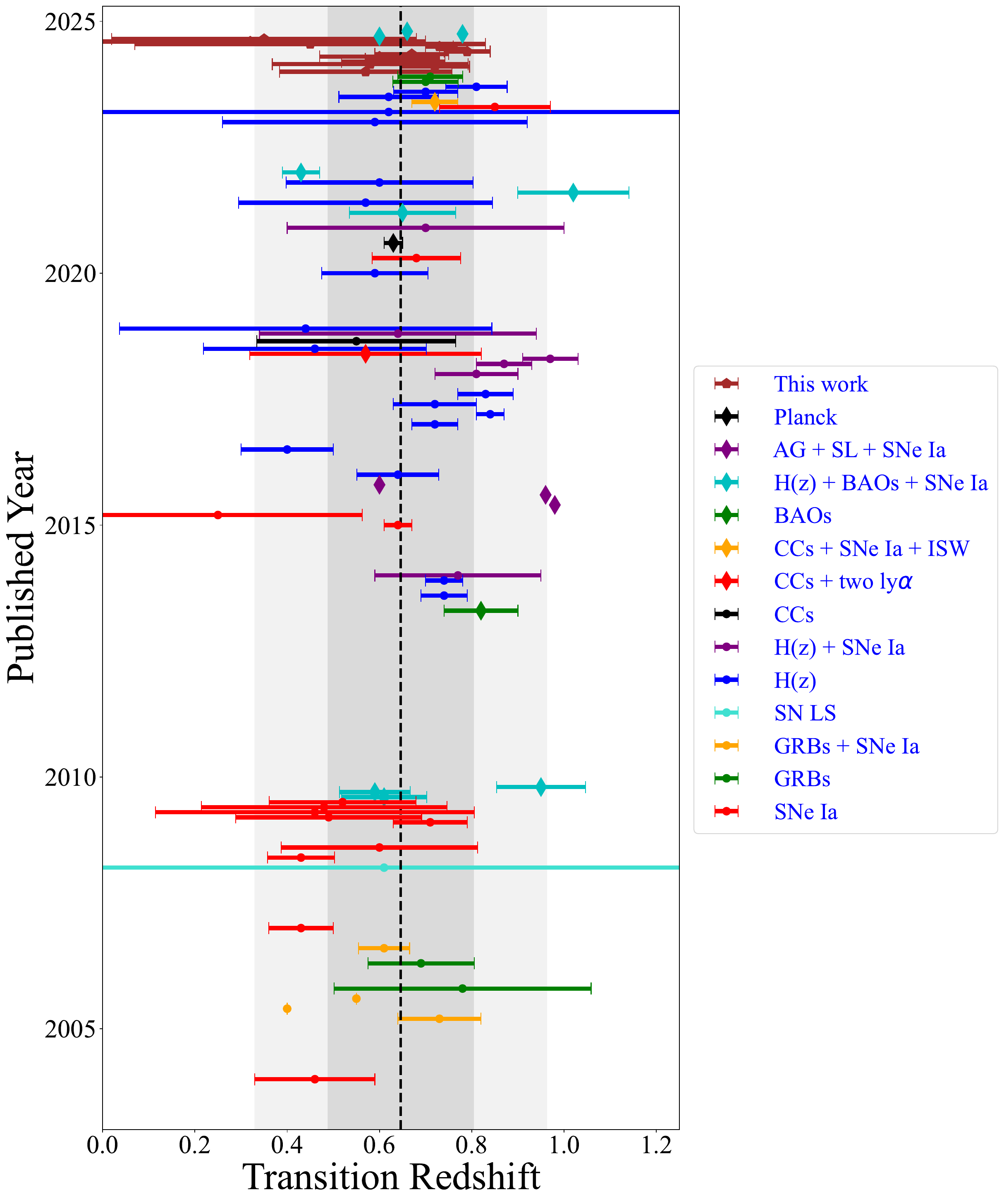}
    \includegraphics[width=0.45\textwidth]{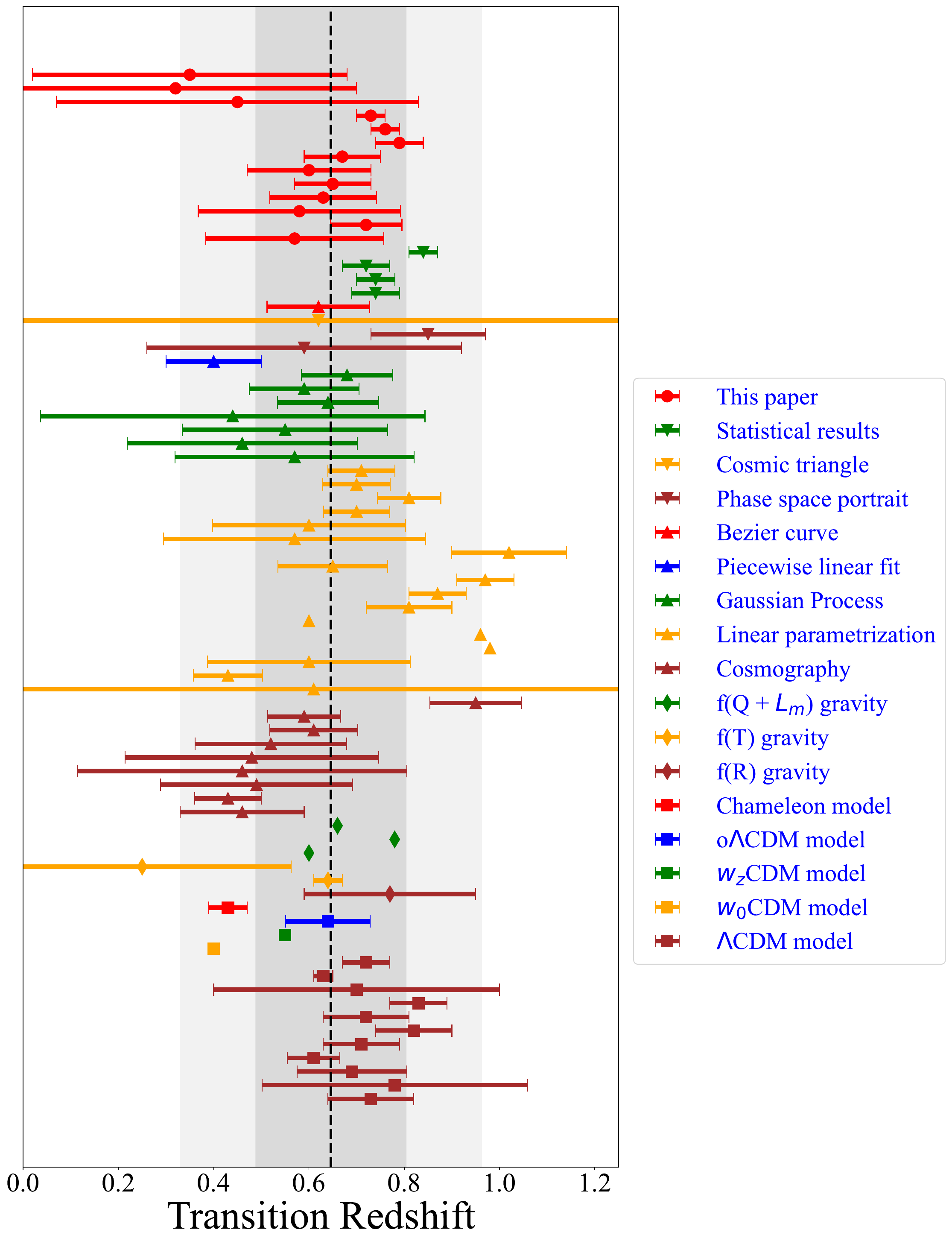}
	\caption{Estimations of transition redshift $z_{tr}$ obtained from different combinations of observations and methods. In the left panel, all $z_{tr}$ results are sorted by submitted year and the datasets used are marked. The right panel shows the result classified according to the analysis method. Black dotted line represents the mean value $\bar{z}_{tr}$ = 0.65. The shaded regions represent the 1$\sigma$ and 2$\sigma$ ranges. }
	\label{ztr}       
\end{figure*}

\begin{figure}
	\centering
	\includegraphics[width=0.30 \textwidth]{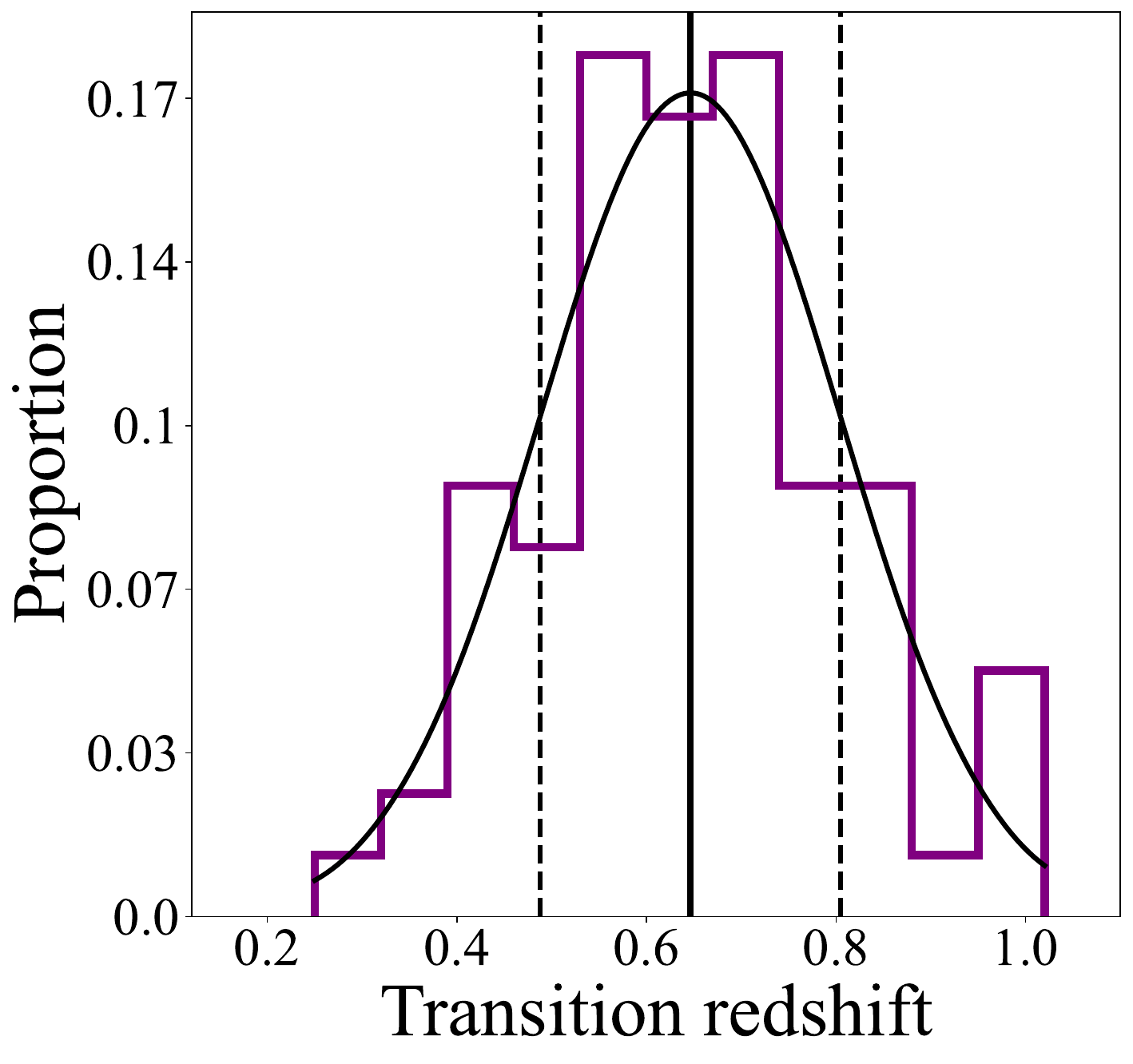}
	\caption{Statistical result of transition redshift $z_{tr}$ which obtained between 2004 and 2024. The $z_{tr}$ distribution well described by a Gaussian function with the mean value 0.65 (solid line) and the standard deviation 0.16 (dotted line).}
	\label{ztr1}       
\end{figure}

\begin{figure}
	\centering
	\includegraphics[width=0.30 \textwidth]{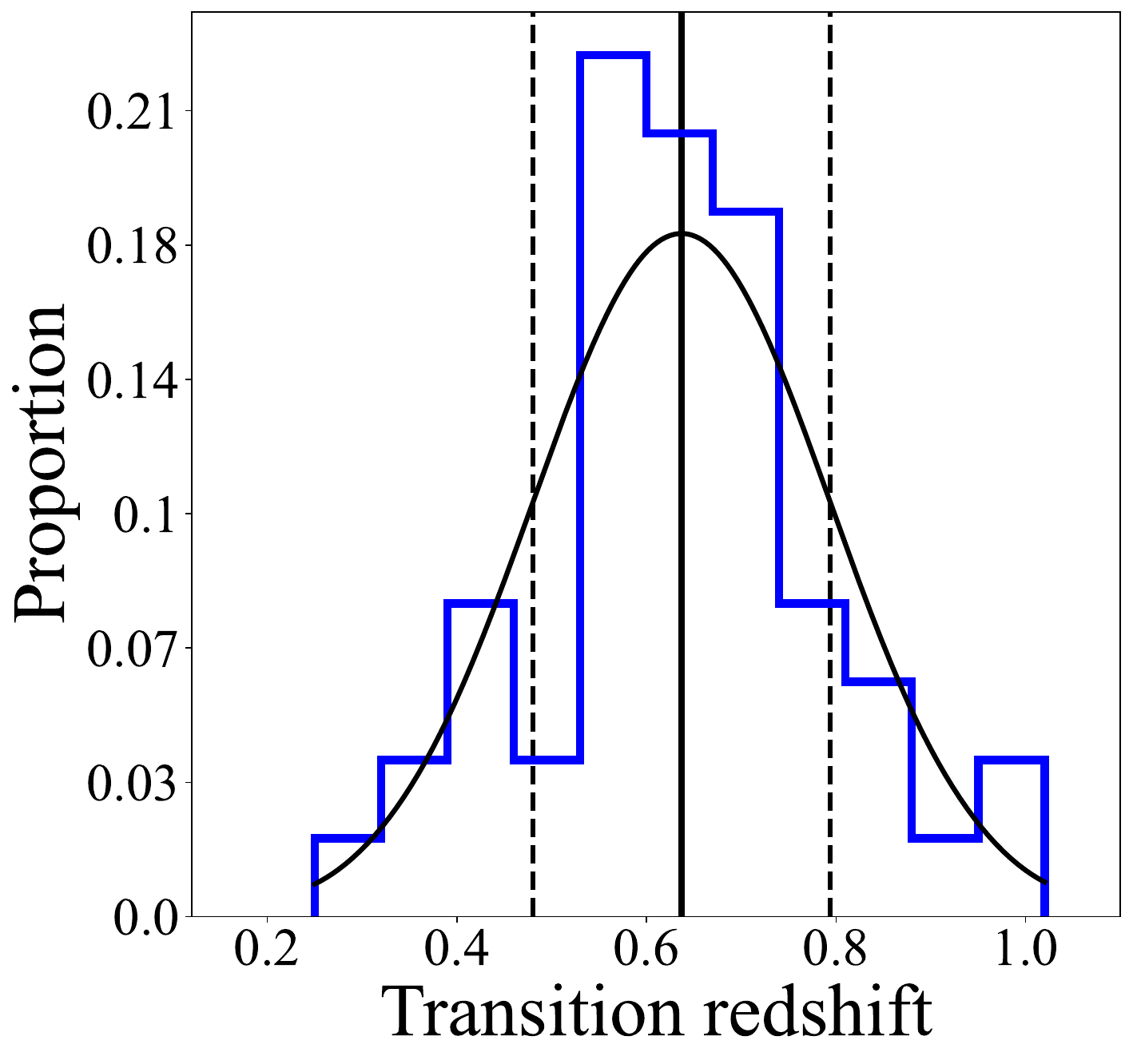}
	\caption{Statistical result of transition redshift $z_{tr}$ which obtained between 2004 and 2024 with free $H_{0}$. The $z_{tr}$ distribution well described by a Gaussian function with the mean value 0.64 (solid line) and the standard deviation 0.16 (dotted line).}
	\label{ztr2}       
\end{figure}

Since the discovery of the accelerated expansion of the Universe, there have been many researches on the transition redshift $z_{tr}$. Many $z_{tr}$ related researches (2004-2024) which utilizing different types of observations and approaches are showed in Table \ref{A3}. According to the type of dataset and the methodology, we categorized the collected results, as shown in Fig. \ref{ztr}. The left panel shows the classification results according to the dataset, and the right panel shows the classification results according to the methodology. From Fig. \ref{ztr}, it is easy to find that our results are in line with with most of the results, with only a few being inconsistent \citep{2015JCAP...12..045R,2018JCAP...05..073J,2022MNRAS.509.5399C}. Overall, 
current constraints on $z_{tr}$ are quite diffuse, covering a wide redshift range (0.25, 1.02). It can be seen that both the dataset and methodology can both affect the final result, with the latter having a stronger impact. Therefore, the current results of the constraints of $z_{tr}$ are still inconclusive. Through statistical analysis, we find that all $z_{tr}$ constraints (2004$-$2024) can be well described by a Gaussian function with the mean value 0.65 ($\bar{z}_{tr}$(all)) and the standard deviation 0.16 ($\sigma_{\bar{z}_{tr}}$(all)), as shown in Fig. \ref{ztr1}.

From the left panel of Fig. \ref{ztr}, taking $\bar{z}_{tr}$(all) = 0.65 as reference, we can find that $z_{tr}$ constraints obtained from different type datasets have different preferences. Some datasets prefer smaller value, such as SNe Ia \citep{2004ApJ...607..665R,2007ApJ...659...98R,2008MNRAS.390..210C,2009JCAP...10..010G,2015PhRvD..91l4037C}, CCs \citep{2018ApJ...856....3Y}, CCs + two ly$\alpha$ \citep{2018ApJ...856....3Y}. Some prefer larger value, for instant, GRBs \citep{2005ApJ...633..611L,2006MNRAS.368..371W,2023MNRAS.523.4938M}, H(z) + SNe Ia \citep{2018JCAP...05..073J,2020EPJC...80..562V}, CCs + SNe Ia + ISW \citep{2023GrCo...29..177R}, BAOs \citep{2013AA...552A..96B}. The rest show no clear preference \citep{2005MNRAS.360L...1F,2006MNRAS.368..371W,2008MNRAS.390..210C,2009JCAP...07..031X,2020AA...641A...6P}. Compared with the dataset type, the preferences for different methods are more obvious. From right panel of Fig. \ref{ztr}, we can see the tendencies of different methods. The $z_{tr}$ constraints utilizing the flat $\Lambda$CDM model \citep{2005MNRAS.360L...1F,2005ApJ...633..611L,2006MNRAS.368..371W,2009JCAP...10..010G,2017ApJ...835...26F,2020AA...641A...6P,2020EPJC...80..562V,2023GrCo...29..177R}, f(R) gravity \citep{2014PhRvD..90d4016C}, linear parametrization \citep{2008MNRAS.390..210C,2015JCAP...12..045R,2018JCAP...05..073J,2022MNRAS.509.5399C,2023MNRAS.523.4938M}, and statistical analysis \citep{2013ApJ...766L...7F,2013PhLB..726...72F,2017ApJ...835...26F} favor a larger value. Ones from the $w_{0}$CDM model \citep{2005MNRAS.360L...1F}, $w_{z}$CDM model \citep{2005MNRAS.360L...1F}, Chameleon model \citep{2022EPJC...82.1165S}, f(T) gravity \citep{2015PhRvD..91l4037C}, GP method \citep{2018ApJ...856....3Y,2018JCAP...10..015H,2020JCAP...04..053J}, and Piecewise linear fit \citep{2016JCAP...05..014M} prefer a smaller value. In general, there are certain preferences for the $z_{tr}$ constraints given by different methods and different datasets. In addition to methods and observations, different treatments of nuisance parameter ($H_{0}$) can also affect the constraints of $z_{tr}$. In past studies, there are three common treatments: (a) fixed \citep{2005MNRAS.360L...1F,2005ApJ...633..611L,2006MNRAS.368..371W,2013ApJ...766L...7F,2013PhLB..726...72F,2017ApJ...835...26F}, (b) marginalized \citep{2008MNRAS.390..210C,2016JCAP...05..014M,2023IJMPD..3250039K}, and (c) free \citep{2004ApJ...607..665R,2009JCAP...07..031X,2018JCAP...05..073J,2018ApJ...856....3Y,2018JCAP...10..015H,2020JCAP...04..053J,2023MNRAS.523.4938M,2024PDU....4601614M}. Specifically, category (a) usually fixes $H_{0}$ to the measurements of SH0ES \citep{2022ApJ...934L...7R} or CMB \citep{2020AA...641A...6P}; category (b) typically assumes a Gaussian prior on $H_{0}$ based on the SH0ES \citep{2022ApJ...934L...7R} or CMB \citep{2020AA...641A...6P} results; and category (c) treats $H_{0}$ as a free parameter. Categories (a) and (c) are relatively easy to understand. For category (b), we give an example to help readers understand, for instance, \citet{2016JCAP...05..014M} assumed a gaussian prior on the Hubble constant $H_{0}$ = 73$\pm$2.4 km/s/Mpc \citep{2011ApJ...730..119R} to obtain the $z_{tr}$ constraints. When $H_{0}$ is fixed, $z_{tr}$ is positively correlated with it, and the larger $H_{0}$ is, the larger the $z_{tr}$ is \citep{2013ApJ...766L...7F,2017ApJ...835...26F}. Considering that the value of the nuisance parameter $H_{0}$ can affect the estimation of $z_{tr}$, we screened out $z_{tr}$ with $H_{0}$ as a free parameter from the total $z_{tr}$ set (2004 $-$ 2024), and then show it in Fig. \ref{ztr2}. For convenience, we defined the $z_{tr}$ constraints selected by the free parameter $H_{0}$ as the screening condition as "screened $z_{tr}$ constraints". From Fig. \ref{ztr2}, it can be found that the screened $z_{tr}$ constraints can be well described by a Gaussian function with the mean value 0.64 ($\bar{z}_{tr}$(free)) and the standard deviation 0.16 ($\sigma_{\bar{z}_{tr}}$(free)). Compared with the previous results ($\bar{z}_{tr}$(free) = 0.65 $\pm$ 0.16), there is no significant change. All in all, the $z_{tr}$ constraints are closely related to many factors, such as data type, research method, redshift range and assumption. To clarify the details, more precisely measurements are needed for further research.    

Recently, a possible correlation between the lens redshift and the inferred of $H_{0}(z)$ was found using time-delay cosmography of lensed quasars by the H0LiCOW collaboration \citep{2020MNRAS.498.1420W}. The statistical significance was approximately 1.9$\sigma$. Subsequently, the addition of a new H0LiCOW lens (DES J0408-5354) by \citet{2020A&A...639A.101M} slightly weakened the trend to 1.7$\sigma$. Inspired by this, \citet{2020PhRvD.102j3525K} obtained $H_{0}$ constraints for different redshift ranges ($z <$ 0.7) by binning a joint dataset consisting of megamasers, CCs, SNe Ia and BAO according to redshift. They also found a similar $H_0$ descending trend with low significance in different cosmological models \citep{2020PhRvD.102j3525K}. To date, the same $H_0$ descending trend has been verified by multiple methods and datasets \citep{2021ApJ...912..150D,2022A&A...668A..34H,2022MNRAS.517..576H,2022MNRAS.515L...1W,2022PhRvD.106d1301O,2023A&A...674A..45J,2025ApJ...979L..34J,2025PDU....4801847M,2025A&A...698A.215G,2025PDU....4801943D}. These researches hint that the Hubble constant $H_{0}$ might evolve with redshift. What caused the evolution is still unclear. There are many speculations about the possible reasons for the evolution of $H_{0}$, for instance the local void \citep{2008GReGr..40..451E,2008JCAP...04..003G,2013ApJ...775...62K,2013NatPh...9..465W,2025MNRAS.536.3232M,2025MNRAS.540..545B}, cosmic anisotropy \citep{2023PDU....4201365Y,2024A&A...681A..88H,2024PDU....4601626Y,2024ApJ...975L..36H}, modified gravity \citep{2020FoPh...50..893C,2021MNRAS.500.1795B,2024PDU....4601652E}, dynamic dark energy \citep{2024arXiv241212905C,2024JCAP...10..035G,2025JCAP...02..021A,2025PDU....4801847M,2025arXiv250103480P}, dark energy - dark matter interaction \citep{2016RPPh...79i6901W,2025PDU....4801848M} and so on. But there is no conclusion yet. 

According to the research on $H_{0}$ redshift evolution, we know that starting from a certain moment in the late universe, the $H_{0}$ constraint began to deviate from the CMB result \citep{2020AA...641A...6P}, gradually evolving into the current Hubble tension \citep{2020PhRvD.102j3525K,2022MNRAS.517..576H,2023A&A...674A..45J,2025ApJ...979L..34J,2025MNRAS.536.3232M}. From the study of \citet{2020PhRvD.102j3525K}, it can be found that the divergence of $H_{0}$ begins around redshift 0.50. \citet{2022MNRAS.517..576H} reported a late-time transition of $H_{0}$; that is, $H_{0}$ changes from a low value to a high one from early to late cosmic time by combining the GP method and $H(z)$ measurements. The $H_{0}$ transition occurs around redshift 0.49, which is also where the divergence of $H_{0}$ begins. In addition, some studies have shown that the divergence of $H_{0}$ begins around 0.70 \citep{2023A&A...674A..45J,2025ApJ...979L..34J}. Thus, these findings suggest that the Hubble tension might begin near the redshift interval of 0.49 to 0.70. Coincidentally, we find that the transition redshift ($\bar{z}_{tr}$(all) = 0.65$\pm$ 0.16 and $\bar{z}_{tr}$(free) = 0.64$\pm$ 0.16) obtained based on the statistical analysis from 2004 to 2024 are exactly within this range. Therefore, we suspect that Hubble tension might be related to this special period of cosmic evolution, when the Universe smoothly transitions from a decelerating expansion to an accelerating expansion. Previously, we have always believed that the Hubble tension was just a contradiction between the early-time and late-time universe. But the redshift evolution of the Hubble constant seems to tell us that the Hubble tension may also be a contradiction between different redshift intervals in the late-time universe. And combined with the history of cosmic evolution, we suspect that the Hubble tension might be closely related to this special period of cosmic evolution. Near the transition redshift $z_{tr}$, the proportion of dark energy increases slightly, causing the Hubble constant to evolve with redshift. There are many physical mechanisms that cause the increase in the dark energy percentage, such as dynamical dark energy \citep{2024arXiv241212905C,2024JCAP...10..035G,2025JCAP...02..021A,2025PDU....4801847M,2025arXiv250103480P,2025arXiv250420664S} and dark energy - dark matter interaction \citep{2016RPPh...79i6901W,2025PDU....4801848M}. From Planck + DR2 + DESY5 dataset, \citet{2025arXiv250420664S} found a transition from a phantom-like regime ($w$ $<$ -1) at early times to a quintessence-like regime ($w$ $>$ -1) today, driven by the negative values of the deviation parameter $\Delta$. The transition occurs at around redshift 0.50. Detailed information can be found in Figure 5 in literature \citep{2025arXiv250420664S}. In addition, an unknown physical mechanism might have occurred during the special period, exacerbating the accelerated expansion of the Universe and causing the current Hubble tension. The unknown mechanism might also be related to the special properties of dark matter and dark energy. Further research not only requires more precise observations to accurately constrain the transition redshift $z_{tr}$ and the redshift evolution behavior of Hubble constant, but also requires a deeper understanding of dark matter and dark energy. In summary, precise positioning of $z_{tr}$ might help to find the physical nature of the redshift evolution of the Hubble constant and the Hubble tension.

\section{Conclusion}  \label{S6}
Transition redshift $z_{tr}$ is a special period in the history of cosmic evolution, when the Universe smoothly transitions from decelerating expansion to accelerating expansion. Determining the exact moment in which accelerated phase began is an interesting question. In the present paper, we constrain the transition redshift $z_{tr}$ utilizing the latest H(z) measurements (35 CCs, 20 BAOs and 5 DESI BAOs) and three kinds of methods ($\Lambda$CDM model, Cosmography, and GP method). By comparing the constraints given by different methods and datasets, we find that dataset and methodology can both affect the final constraints, with the latter having a stronger impact. Compared with the other two model-independent methods, the $\Lambda$CDM model prefers a larger $z_{tr}$ constraint. Different methods have different abilities to constrain the transition redshift $z_{tr}$. Through statistical analyses of the research results from 2004 to 2024, we find that all $z_{tr}$ constraints and the screened $z_{tr}$ constraints with free $H_{0}$ can be well described by Gaussian functions; that is $\bar{z}_{tr}$(all) = 0.65 $\pm$ 0.16 and $\bar{z}_{tr}$(free) = 0.64 $\pm$ 0.16. Coincidentally, these results overlap with the initial position of the redshift evolution of the Hubble constant. Therefore, we speculate that the origin of the Hubble tension might be associated with the special period. However, a reasonable physical mechanism to explain this coincidence and firmly link the two is lacking. In future research, we will pay more attention to studying the late-time evolution of the Universe, especially around redshift range (0.50, 0.70), to find a suitable physical explanation.

In summary, the determination of the transition redshift $z_{tr}$ is similar importance in modern cosmology as the Hubble constant $H_{0}$, deceleration parameter $q_{0}$, and cosmic constant $\Lambda$. Accurately constraining the redshift is crucial to understanding the evolutionary history of the late-time Universe and will provide new perspectives for exploring the redshift evolution of the Hubble constant and the physical nature of the Hubble tension.

\section*{Acknowledgements}
Anonymous reviewer provided many constructive comments, for which we are deeply grateful. This work was supported by the National Natural Science Foundation of China (grant No. 12273009, No. 12373026), Leading Innovation and Entrepreneurship Team of Zhejiang Province of China (grant No. 2023R01008), and Key R\&D Program of Zhejiang, China (grant No. 2024SSYS0012).

\section*{DATA AVAILABILITY}
The data used in the paper are publicly available in Tables \ref{tab:CC} and \ref{tab:BAO}.

\bibliographystyle{mnras}
\bibliography{mnras_mg} 

\begin{thebibliography}{}
\makeatletter
\relax
\def\mn@urlcharsother{\let\do\@makeother \do\$\do\&\do\#\do\^\do\_\do\%\do\~}
\def\mn@doi{\begingroup\mn@urlcharsother \@ifnextchar [ {\mn@doi@}
  {\mn@doi@[]}}
\def\mn@doi@[#1]#2{\def\@tempa{#1}\ifx\@tempa\@empty \href
  {http://dx.doi.org/#2} {doi:#2}\else \href {http://dx.doi.org/#2} {#1}\fi
  \endgroup}
\def\mn@eprint#1#2{\mn@eprint@#1:#2::\@nil}
\def\mn@eprint@arXiv#1{\href {http://arxiv.org/abs/#1} {{\tt arXiv:#1}}}
\def\mn@eprint@dblp#1{\href {http://dblp.uni-trier.de/rec/bibtex/#1.xml}
  {dblp:#1}}
\def\mn@eprint@#1:#2:#3:#4\@nil{\def\@tempa {#1}\def\@tempb {#2}\def\@tempc
  {#3}\ifx \@tempc \@empty \let \@tempc \@tempb \let \@tempb \@tempa \fi \ifx
  \@tempb \@empty \def\@tempb {arXiv}\fi \@ifundefined
  {mn@eprint@\@tempb}{\@tempb:\@tempc}{\expandafter \expandafter \csname
  mn@eprint@\@tempb\endcsname \expandafter{\@tempc}}}

\bibitem[\protect\citeauthoryear{{Abbott} et~al.,}{{Abbott}
  et~al.}{2019}]{2019PhRvL.122q1301A}
{Abbott} T.~M.~C.,  et~al., 2019, \mn@doi [\prl]
  {10.1103/PhysRevLett.122.171301}, \href
  {https://ui.adsabs.harvard.edu/abs/2019PhRvL.122q1301A} {122, 171301}

\bibitem[\protect\citeauthoryear{{Abbott} et~al.,}{{Abbott}
  et~al.}{2024}]{2024ApJ...973L..14A}
{Abbott} T.~M.~C.,  et~al., 2024, \mn@doi [\apjl] {10.3847/2041-8213/ad6f9f},
  \href {https://ui.adsabs.harvard.edu/abs/2024ApJ...973L..14A} {973, L14}

\bibitem[\protect\citeauthoryear{{Abdalla} et~al.,}{{Abdalla}
  et~al.}{2022}]{2022JHEAp..34...49A}
{Abdalla} E.,  et~al., 2022, \mn@doi [Journal of High Energy Astrophysics]
  {10.1016/j.jheap.2022.04.002}, \href
  {https://ui.adsabs.harvard.edu/abs/2022JHEAp..34...49A} {34, 49}

\bibitem[\protect\citeauthoryear{{Adame} et~al.,}{{Adame}
  et~al.}{2025a}]{2025JCAP...01..124A}
{Adame} A.~G.,  et~al., 2025a, \mn@doi [\jcap] {10.1088/1475-7516/2025/01/124},
  \href {https://ui.adsabs.harvard.edu/abs/2025JCAP...01..124A} {2025, 124}

\bibitem[\protect\citeauthoryear{{Adame} et~al.,}{{Adame}
  et~al.}{2025b}]{2025JCAP...02..021A}
{Adame} A.~G.,  et~al., 2025b, \mn@doi [\jcap] {10.1088/1475-7516/2025/02/021},
  \href {https://ui.adsabs.harvard.edu/abs/2025JCAP...02..021A} {2025, 021}

\bibitem[\protect\citeauthoryear{{Adame} et~al.,}{{Adame}
  et~al.}{2025c}]{2025JCAP...04..012A}
{Adame} A.~G.,  et~al., 2025c, \mn@doi [\jcap] {10.1088/1475-7516/2025/04/012},
  \href {https://ui.adsabs.harvard.edu/abs/2025JCAP...04..012A} {2025, 012}

\bibitem[\protect\citeauthoryear{{Alam} et~al.,}{{Alam}
  et~al.}{2017}]{2017MNRAS.470.2617A}
{Alam} S.,  et~al., 2017, \mn@doi [\mnras] {10.1093/mnras/stx721}, \href
  {https://ui.adsabs.harvard.edu/abs/2017MNRAS.470.2617A} {470, 2617}

\bibitem[\protect\citeauthoryear{{Anderson} et~al.,}{{Anderson}
  et~al.}{2014}]{2014MNRAS.441...24A}
{Anderson} L.,  et~al., 2014, \mn@doi [\mnras] {10.1093/mnras/stu523}, \href
  {https://ui.adsabs.harvard.edu/abs/2014MNRAS.441...24A} {441, 24}

\bibitem[\protect\citeauthoryear{{Arjona} \& {Nesseris}}{{Arjona} \&
  {Nesseris}}{2021}]{2021PhRvD.103f3537A}
{Arjona} R.,  {Nesseris} S.,  2021, \mn@doi [\prd]
  {10.1103/PhysRevD.103.063537}, \href
  {https://ui.adsabs.harvard.edu/abs/2021PhRvD.103f3537A} {103, 063537}

\bibitem[\protect\citeauthoryear{{Banik} \& {Kalaitzidis}}{{Banik} \&
  {Kalaitzidis}}{2025}]{2025MNRAS.540..545B}
{Banik} I.,  {Kalaitzidis} V.,  2025, \mn@doi [\mnras] {10.1093/mnras/staf781},
  \href {https://ui.adsabs.harvard.edu/abs/2025MNRAS.540..545B} {540, 545}

\bibitem[\protect\citeauthoryear{{Bargiacchi}, {Risaliti}, {Benetti},
  {Capozziello}, {Lusso}, {Saccardi}  \& {Signorini}}{{Bargiacchi}
  et~al.}{2021}]{2021A&A...649A..65B}
{Bargiacchi} G.,  {Risaliti} G.,  {Benetti} M.,  {Capozziello} S.,  {Lusso} E.,
   {Saccardi} A.,   {Signorini} M.,  2021, \mn@doi [\aap]
  {10.1051/0004-6361/202140386}, \href
  {https://ui.adsabs.harvard.edu/abs/2021A&A...649A..65B} {649, A65}

\bibitem[\protect\citeauthoryear{{Benetti}, {Capozziello}  \&
  {Lambiase}}{{Benetti} et~al.}{2021}]{2021MNRAS.500.1795B}
{Benetti} M.,  {Capozziello} S.,   {Lambiase} G.,  2021, \mn@doi [\mnras]
  {10.1093/mnras/staa3368}, \href
  {https://ui.adsabs.harvard.edu/abs/2021MNRAS.500.1795B} {500, 1795}

\bibitem[\protect\citeauthoryear{{Bengaly}, {Alcaniz}  \& {Pigozzo}}{{Bengaly}
  et~al.}{2024}]{2024PhRvD.109l3533B}
{Bengaly} C. A.~P.,  {Alcaniz} J.~S.,   {Pigozzo} C.,  2024, \mn@doi [\prd]
  {10.1103/PhysRevD.109.123533}, \href
  {https://ui.adsabs.harvard.edu/abs/2024PhRvD.109l3533B} {109, 123533}

\bibitem[\protect\citeauthoryear{{Blake} et~al.,}{{Blake}
  et~al.}{2012}]{2012MNRAS.425..405B}
{Blake} C.,  et~al., 2012, \mn@doi [\mnras] {10.1111/j.1365-2966.2012.21473.x},
  \href {https://ui.adsabs.harvard.edu/abs/2012MNRAS.425..405B} {425, 405}

\bibitem[\protect\citeauthoryear{{Borghi}, {Moresco}  \& {Cimatti}}{{Borghi}
  et~al.}{2022}]{2022ApJ...928L...4B}
{Borghi} N.,  {Moresco} M.,   {Cimatti} A.,  2022, \mn@doi [\apjl]
  {10.3847/2041-8213/ac3fb2}, \href
  {https://ui.adsabs.harvard.edu/abs/2022ApJ...928L...4B} {928, L4}

\bibitem[\protect\citeauthoryear{{Boubel}, {Colless}, {Said}  \&
  {Staveley-Smith}}{{Boubel} et~al.}{2025}]{2025JCAP...03..066B}
{Boubel} P.,  {Colless} M.,  {Said} K.,   {Staveley-Smith} L.,  2025, \mn@doi
  [\jcap] {10.1088/1475-7516/2025/03/066}, \href
  {https://ui.adsabs.harvard.edu/abs/2025JCAP...03..066B} {2025, 066}

\bibitem[\protect\citeauthoryear{{Briffa}, {Capozziello}, {Said}, {Mifsud}  \&
  {Saridakis}}{{Briffa} et~al.}{2020}]{2020CQGra..38e5007B}
{Briffa} R.,  {Capozziello} S.,  {Said} J.~L.,  {Mifsud} J.,   {Saridakis}
  E.~N.,  2020, \mn@doi [Classical and Quantum Gravity]
  {10.1088/1361-6382/abd4f5}, \href
  {https://ui.adsabs.harvard.edu/abs/2020CQGra..38e5007B} {38, 055007}

\bibitem[\protect\citeauthoryear{{Brout} et~al.,}{{Brout}
  et~al.}{2022}]{2022ApJ...938..110B}
{Brout} D.,  et~al., 2022, \mn@doi [\apj] {10.3847/1538-4357/ac8e04}, \href
  {https://ui.adsabs.harvard.edu/abs/2022ApJ...938..110B} {938, 110}

\bibitem[\protect\citeauthoryear{{Busca} et~al.,}{{Busca}
  et~al.}{2013}]{2013AA...552A..96B}
{Busca} N.~G.,  et~al., 2013, \mn@doi [\aap] {10.1051/0004-6361/201220724},
  \href {https://ui.adsabs.harvard.edu/abs/2013A&A...552A..96B} {552, A96}

\bibitem[\protect\citeauthoryear{{Cai}, {Capozziello}, {De Laurentis}  \&
  {Saridakis}}{{Cai} et~al.}{2016}]{2016RPPh...79j6901C}
{Cai} Y.-F.,  {Capozziello} S.,  {De Laurentis} M.,   {Saridakis} E.~N.,  2016,
  \mn@doi [Reports on Progress in Physics] {10.1088/0034-4885/79/10/106901},
  \href {https://ui.adsabs.harvard.edu/abs/2016RPPh...79j6901C} {79, 106901}

\bibitem[\protect\citeauthoryear{{Cao}, {Khadka}  \& {Ratra}}{{Cao}
  et~al.}{2022a}]{2022MNRAS.510.2928C}
{Cao} S.,  {Khadka} N.,   {Ratra} B.,  2022a, \mn@doi [\mnras]
  {10.1093/mnras/stab3559}, \href
  {https://ui.adsabs.harvard.edu/abs/2022MNRAS.510.2928C} {510, 2928}

\bibitem[\protect\citeauthoryear{{Cao}, {Dainotti}  \& {Ratra}}{{Cao}
  et~al.}{2022b}]{2022MNRAS.512..439C}
{Cao} S.,  {Dainotti} M.,   {Ratra} B.,  2022b, \mn@doi [\mnras]
  {10.1093/mnras/stac517}, \href
  {https://ui.adsabs.harvard.edu/abs/2022MNRAS.512..439C} {512, 439}

\bibitem[\protect\citeauthoryear{{Capozziello} \& {Ruchika}}{{Capozziello} \&
  {Ruchika}}{2019}]{2019MNRAS.484.4484C}
{Capozziello} S.,  {Ruchika} Sen A.~A.,  2019, \mn@doi [\mnras]
  {10.1093/mnras/stz176}, \href
  {https://ui.adsabs.harvard.edu/abs/2019MNRAS.484.4484C} {484, 4484}

\bibitem[\protect\citeauthoryear{{Capozziello} \& {de Laurentis}}{{Capozziello}
  \& {de Laurentis}}{2011}]{2011PhR...509..167C}
{Capozziello} S.,  {de Laurentis} M.,  2011, \mn@doi [\physrep]
  {10.1016/j.physrep.2011.09.003}, \href
  {https://ui.adsabs.harvard.edu/abs/2011PhR...509..167C} {509, 167}

\bibitem[\protect\citeauthoryear{{Capozziello}, {Lazkoz}  \&
  {Salzano}}{{Capozziello} et~al.}{2011}]{2011PhRvD..84l4061C}
{Capozziello} S.,  {Lazkoz} R.,   {Salzano} V.,  2011, \mn@doi [\prd]
  {10.1103/PhysRevD.84.124061}, \href
  {https://ui.adsabs.harvard.edu/abs/2011PhRvD..84l4061C} {84, 124061}

\bibitem[\protect\citeauthoryear{{Capozziello}, {Farooq}, {Luongo}  \&
  {Ratra}}{{Capozziello} et~al.}{2014}]{2014PhRvD..90d4016C}
{Capozziello} S.,  {Farooq} O.,  {Luongo} O.,   {Ratra} B.,  2014, \mn@doi
  [\prd] {10.1103/PhysRevD.90.044016}, \href
  {https://ui.adsabs.harvard.edu/abs/2014PhRvD..90d4016C} {90, 044016}

\bibitem[\protect\citeauthoryear{{Capozziello}, {Luongo}  \&
  {Saridakis}}{{Capozziello} et~al.}{2015}]{2015PhRvD..91l4037C}
{Capozziello} S.,  {Luongo} O.,   {Saridakis} E.~N.,  2015, \mn@doi [\prd]
  {10.1103/PhysRevD.91.124037}, \href
  {https://ui.adsabs.harvard.edu/abs/2015PhRvD..91l4037C} {91, 124037}

\bibitem[\protect\citeauthoryear{{Capozziello}, {D'Agostino}  \&
  {Luongo}}{{Capozziello} et~al.}{2018}]{2018MNRAS.476.3924C}
{Capozziello} S.,  {D'Agostino} R.,   {Luongo} O.,  2018, \mn@doi [\mnras]
  {10.1093/mnras/sty422}, \href
  {https://ui.adsabs.harvard.edu/abs/2018MNRAS.476.3924C} {476, 3924}

\bibitem[\protect\citeauthoryear{{Capozziello}, {D'Agostino}  \&
  {Luongo}}{{Capozziello} et~al.}{2019}]{2019IJMPD..2830016C}
{Capozziello} S.,  {D'Agostino} R.,   {Luongo} O.,  2019, \mn@doi
  [International Journal of Modern Physics D] {10.1142/S0218271819300167},
  \href {https://ui.adsabs.harvard.edu/abs/2019IJMPD..2830016C} {28, 1930016}

\bibitem[\protect\citeauthoryear{{Capozziello}, {Benetti}  \&
  {Spallicci}}{{Capozziello} et~al.}{2020a}]{2020FoPh...50..893C}
{Capozziello} S.,  {Benetti} M.,   {Spallicci} A. D.~A.~M.,  2020a, \mn@doi
  [Foundations of Physics] {10.1007/s10701-020-00356-2}, \href
  {https://ui.adsabs.harvard.edu/abs/2020FoPh...50..893C} {50, 893}

\bibitem[\protect\citeauthoryear{{Capozziello}, {D'Agostino}  \&
  {Luongo}}{{Capozziello} et~al.}{2020b}]{2020MNRAS.494.2576C}
{Capozziello} S.,  {D'Agostino} R.,   {Luongo} O.,  2020b, \mn@doi [\mnras]
  {10.1093/mnras/staa871}, \href
  {https://ui.adsabs.harvard.edu/abs/2020MNRAS.494.2576C} {494, 2576}

\bibitem[\protect\citeauthoryear{{Capozziello}, {Dunsby}  \&
  {Luongo}}{{Capozziello} et~al.}{2022}]{2022MNRAS.509.5399C}
{Capozziello} S.,  {Dunsby} P. K.~S.,   {Luongo} O.,  2022, \mn@doi [\mnras]
  {10.1093/mnras/stab3187}, \href
  {https://ui.adsabs.harvard.edu/abs/2022MNRAS.509.5399C} {509, 5399}

\bibitem[\protect\citeauthoryear{{Capozziello}, {Sarracino}  \& {De
  Somma}}{{Capozziello} et~al.}{2024}]{2024Univ...10..140C}
{Capozziello} S.,  {Sarracino} G.,   {De Somma} G.,  2024, \mn@doi [Universe]
  {10.3390/universe10030140}, \href
  {https://ui.adsabs.harvard.edu/abs/2024Univ...10..140C} {10, 140}

\bibitem[\protect\citeauthoryear{{Catto{\"e}n} \& {Visser}}{{Catto{\"e}n} \&
  {Visser}}{2007}]{2007CQGra..24.5985C}
{Catto{\"e}n} C.,  {Visser} M.,  2007, \mn@doi [Classical and Quantum Gravity]
  {10.1088/0264-9381/24/23/018}, \href
  {https://ui.adsabs.harvard.edu/abs/2007CQGra..24.5985C} {24, 5985}

\bibitem[\protect\citeauthoryear{{Chen} et~al.,}{{Chen}
  et~al.}{2024}]{2024MNRAS.534..544C}
{Chen} S.~F.,  et~al., 2024, \mn@doi [\mnras] {10.1093/mnras/stae2090}, \href
  {https://ui.adsabs.harvard.edu/abs/2024MNRAS.534..544C} {534, 544}

\bibitem[\protect\citeauthoryear{{Chuang} et~al.,}{{Chuang}
  et~al.}{2013}]{2013MNRAS.433.3559C}
{Chuang} C.-H.,  et~al., 2013, \mn@doi [\mnras] {10.1093/mnras/stt988}, \href
  {https://ui.adsabs.harvard.edu/abs/2013MNRAS.433.3559C} {433, 3559}

\bibitem[\protect\citeauthoryear{{Cline}, {Grojean}  \& {Servant}}{{Cline}
  et~al.}{1999}]{1999PhRvL..83.4245C}
{Cline} J.~M.,  {Grojean} C.,   {Servant} G.,  1999, \mn@doi [\prl]
  {10.1103/PhysRevLett.83.4245}, \href
  {https://ui.adsabs.harvard.edu/abs/1999PhRvL..83.4245C} {83, 4245}

\bibitem[\protect\citeauthoryear{{Colg{\'a}in} \&
  {Sheikh-Jabbari}}{{Colg{\'a}in} \&
  {Sheikh-Jabbari}}{2024}]{2024arXiv241212905C}
{Colg{\'a}in} E.~{\'O}.,  {Sheikh-Jabbari} M.~M.,  2024, \mn@doi [arXiv
  e-prints] {10.48550/arXiv.2412.12905}, \href
  {https://ui.adsabs.harvard.edu/abs/2024arXiv241212905C} {p. arXiv:2412.12905}

\bibitem[\protect\citeauthoryear{{Cunha} \& {Lima}}{{Cunha} \&
  {Lima}}{2008}]{2008MNRAS.390..210C}
{Cunha} J.~V.,  {Lima} J.~A.~S.,  2008, \mn@doi [\mnras]
  {10.1111/j.1365-2966.2008.13640.x}, \href
  {https://ui.adsabs.harvard.edu/abs/2008MNRAS.390..210C} {390, 210}

\bibitem[\protect\citeauthoryear{{Dainotti}, {De Simone}, {Schiavone},
  {Montani}, {Rinaldi}  \& {Lambiase}}{{Dainotti}
  et~al.}{2021}]{2021ApJ...912..150D}
{Dainotti} M.~G.,  {De Simone} B.,  {Schiavone} T.,  {Montani} G.,  {Rinaldi}
  E.,   {Lambiase} G.,  2021, \mn@doi [\apj] {10.3847/1538-4357/abeb73}, \href
  {https://ui.adsabs.harvard.edu/abs/2021ApJ...912..150D} {912, 150}

\bibitem[\protect\citeauthoryear{{Dainotti}, {De Simone}, {Schiavone},
  {Montani}, {Rinaldi}, {Lambiase}, {Bogdan}  \& {Ugale}}{{Dainotti}
  et~al.}{2022}]{2022Galax..10...24D}
{Dainotti} M.~G.,  {De Simone} B.~D.,  {Schiavone} T.,  {Montani} G.,
  {Rinaldi} E.,  {Lambiase} G.,  {Bogdan} M.,   {Ugale} S.,  2022, \mn@doi
  [Galaxies] {10.3390/galaxies10010024}, \href
  {https://ui.adsabs.harvard.edu/abs/2022Galax..10...24D} {10, 24}

\bibitem[\protect\citeauthoryear{{Dainotti}, {Lenart}, {Chraya}, {Sarracino},
  {Nagataki}, {Fraija}, {Capozziello}  \& {Bogdan}}{{Dainotti}
  et~al.}{2023}]{2023MNRAS.518.2201D}
{Dainotti} M.~G.,  {Lenart} A.~{\L}.,  {Chraya} A.,  {Sarracino} G.,
  {Nagataki} S.,  {Fraija} N.,  {Capozziello} S.,   {Bogdan} M.,  2023, \mn@doi
  [\mnras] {10.1093/mnras/stac2752}, \href
  {https://ui.adsabs.harvard.edu/abs/2023MNRAS.518.2201D} {518, 2201}

\bibitem[\protect\citeauthoryear{{Dainotti} et~al.,}{{Dainotti}
  et~al.}{2025}]{2025arXiv250111772D}
{Dainotti} M.~G.,  et~al., 2025, \mn@doi [arXiv e-prints]
  {10.48550/arXiv.2501.11772}, \href
  {https://ui.adsabs.harvard.edu/abs/2025arXiv250111772D} {p. arXiv:2501.11772}

\bibitem[\protect\citeauthoryear{{Damour} \& {Polyakov}}{{Damour} \&
  {Polyakov}}{1994}]{1994GReGr..26.1171D}
{Damour} T.,  {Polyakov} A.~M.,  1994, \mn@doi [General Relativity and
  Gravitation] {10.1007/BF02106709}, \href
  {https://ui.adsabs.harvard.edu/abs/1994GReGr..26.1171D} {26, 1171}

\bibitem[\protect\citeauthoryear{{Davis}, {Efstathiou}, {Frenk}  \&
  {White}}{{Davis} et~al.}{1985}]{1985ApJ...292..371D}
{Davis} M.,  {Efstathiou} G.,  {Frenk} C.~S.,   {White} S.~D.~M.,  1985,
  \mn@doi [\apj] {10.1086/163168}, \href
  {https://ui.adsabs.harvard.edu/abs/1985ApJ...292..371D} {292, 371}

\bibitem[\protect\citeauthoryear{{De Simone}, {van Putten}, {Dainotti}  \&
  {Lambiase}}{{De Simone} et~al.}{2024}]{2024arXiv241105744D}
{De Simone} B.,  {van Putten} M.~H.~P.~M.,  {Dainotti} M.~G.,   {Lambiase} G.,
  2024, \mn@doi [arXiv e-prints] {10.48550/arXiv.2411.05744}, \href
  {https://ui.adsabs.harvard.edu/abs/2024arXiv241105744D} {p. arXiv:2411.05744}

\bibitem[\protect\citeauthoryear{{Delubac} et~al.,}{{Delubac}
  et~al.}{2015}]{2015AA...574A..59D}
{Delubac} T.,  et~al., 2015, \mn@doi [\aap] {10.1051/0004-6361/201423969},
  \href {https://ui.adsabs.harvard.edu/abs/2015A&A...574A..59D} {574, A59}

\bibitem[\protect\citeauthoryear{{Di Valentino} et~al.,}{{Di Valentino}
  et~al.}{2021}]{2021CQGra..38o3001D}
{Di Valentino} E.,  et~al., 2021, \mn@doi [Classical and Quantum Gravity]
  {10.1088/1361-6382/ac086d}, \href
  {https://ui.adsabs.harvard.edu/abs/2021CQGra..38o3001D} {38, 153001}

\bibitem[\protect\citeauthoryear{{Dilsiz}, {Deliduman}  \& {Binici}}{{Dilsiz}
  et~al.}{2025}]{2025PDU....4801943D}
{Dilsiz} F.~{\c{S}}.,  {Deliduman} C.,   {Binici} S.~S.,  2025, \mn@doi
  [Physics of the Dark Universe] {10.1016/j.dark.2025.101943}, \href
  {https://ui.adsabs.harvard.edu/abs/2025PDU....4801943D} {48, 101943}

\bibitem[\protect\citeauthoryear{{Dunsby} \& {Luongo}}{{Dunsby} \&
  {Luongo}}{2016}]{2016IJGMM..1330002D}
{Dunsby} P. K.~S.,  {Luongo} O.,  2016, \mn@doi [International Journal of
  Geometric Methods in Modern Physics] {10.1142/S0219887816300026}, \href
  {https://ui.adsabs.harvard.edu/abs/2016IJGMM..1330002D} {13, 1630002}

\bibitem[\protect\citeauthoryear{{Eisenstein} et~al.,}{{Eisenstein}
  et~al.}{2005}]{2005ApJ...633..560E}
{Eisenstein} D.~J.,  et~al., 2005, \mn@doi [\apj] {10.1086/466512}, \href
  {https://ui.adsabs.harvard.edu/abs/2005ApJ...633..560E} {633, 560}

\bibitem[\protect\citeauthoryear{{Eisenstein} et~al.,}{{Eisenstein}
  et~al.}{2011}]{2011AJ....142...72E}
{Eisenstein} D.~J.,  et~al., 2011, \mn@doi [\aj] {10.1088/0004-6256/142/3/72},
  \href {https://ui.adsabs.harvard.edu/abs/2011AJ....142...72E} {142, 72}

\bibitem[\protect\citeauthoryear{{Enqvist}}{{Enqvist}}{2008}]{2008GReGr..40..451E}
{Enqvist} K.,  2008, \mn@doi [General Relativity and Gravitation]
  {10.1007/s10714-007-0553-9}, \href
  {https://ui.adsabs.harvard.edu/abs/2008GReGr..40..451E} {40, 451}

\bibitem[\protect\citeauthoryear{{Escamilla}, {Fiorucci}, {Montani}  \& {Di
  Valentino}}{{Escamilla} et~al.}{2024}]{2024PDU....4601652E}
{Escamilla} L.~A.,  {Fiorucci} D.,  {Montani} G.,   {Di Valentino} E.,  2024,
  \mn@doi [Physics of the Dark Universe] {10.1016/j.dark.2024.101652}, \href
  {https://ui.adsabs.harvard.edu/abs/2024PDU....4601652E} {46, 101652}

\bibitem[\protect\citeauthoryear{{Farooq} \& {Ratra}}{{Farooq} \&
  {Ratra}}{2013}]{2013ApJ...766L...7F}
{Farooq} O.,  {Ratra} B.,  2013, \mn@doi [\apjl] {10.1088/2041-8205/766/1/L7},
  \href {https://ui.adsabs.harvard.edu/abs/2013ApJ...766L...7F} {766, L7}

\bibitem[\protect\citeauthoryear{{Farooq}, {Crandall}  \& {Ratra}}{{Farooq}
  et~al.}{2013a}]{2013PhLB..726...72F}
{Farooq} O.,  {Crandall} S.,   {Ratra} B.,  2013a, \mn@doi [Physics Letters B]
  {10.1016/j.physletb.2013.08.078}, \href
  {https://ui.adsabs.harvard.edu/abs/2013PhLB..726...72F} {726, 72}

\bibitem[\protect\citeauthoryear{{Farooq}, {Mania}  \& {Ratra}}{{Farooq}
  et~al.}{2013b}]{2013ApJ...764..138F}
{Farooq} O.,  {Mania} D.,   {Ratra} B.,  2013b, \mn@doi [\apj]
  {10.1088/0004-637X/764/2/138}, \href
  {https://ui.adsabs.harvard.edu/abs/2013ApJ...764..138F} {764, 138}

\bibitem[\protect\citeauthoryear{{Farooq}, {Ranjeet Madiyar}, {Crandall}  \&
  {Ratra}}{{Farooq} et~al.}{2017}]{2017ApJ...835...26F}
{Farooq} O.,  {Ranjeet Madiyar} F.,  {Crandall} S.,   {Ratra} B.,  2017,
  \mn@doi [\apj] {10.3847/1538-4357/835/1/26}, \href
  {https://ui.adsabs.harvard.edu/abs/2017ApJ...835...26F} {835, 26}

\bibitem[\protect\citeauthoryear{{Favale}, {Dainotti}, {G{\'o}mez-Valent}  \&
  {Migliaccio}}{{Favale} et~al.}{2024}]{2024JHEAp..44..323F}
{Favale} A.,  {Dainotti} M.~G.,  {G{\'o}mez-Valent} A.,   {Migliaccio} M.,
  2024, \mn@doi [Journal of High Energy Astrophysics]
  {10.1016/j.jheap.2024.10.010}, \href
  {https://ui.adsabs.harvard.edu/abs/2024JHEAp..44..323F} {44, 323}

\bibitem[\protect\citeauthoryear{{Firmani}, {Ghisellini}, {Ghirlanda}  \&
  {Avila-Reese}}{{Firmani} et~al.}{2005}]{2005MNRAS.360L...1F}
{Firmani} C.,  {Ghisellini} G.,  {Ghirlanda} G.,   {Avila-Reese} V.,  2005,
  \mn@doi [\mnras] {10.1111/j.1745-3933.2005.00023.x}, \href
  {https://ui.adsabs.harvard.edu/abs/2005MNRAS.360L...1F} {360, L1}

\bibitem[\protect\citeauthoryear{{Font-Ribera} et~al.,}{{Font-Ribera}
  et~al.}{2014}]{2014JCAP...05..027F}
{Font-Ribera} A.,  et~al., 2014, \mn@doi [\jcap]
  {10.1088/1475-7516/2014/05/027}, \href
  {https://ui.adsabs.harvard.edu/abs/2014JCAP...05..027F} {2014, 027}

\bibitem[\protect\citeauthoryear{{Foreman-Mackey}, {Hogg}, {Lang}  \&
  {Goodman}}{{Foreman-Mackey} et~al.}{2013}]{2013PASP..125..306F}
{Foreman-Mackey} D.,  {Hogg} D.~W.,  {Lang} D.,   {Goodman} J.,  2013, \mn@doi
  [\pasp] {10.1086/670067}, \href
  {https://ui.adsabs.harvard.edu/abs/2013PASP..125..306F} {125, 306}

\bibitem[\protect\citeauthoryear{{Gao}, {Zhou}, {Du}, {Zou}, {Hu}  \&
  {Xu}}{{Gao} et~al.}{2024}]{2024MNRAS.527.7861G}
{Gao} J.,  {Zhou} Z.,  {Du} M.,  {Zou} R.,  {Hu} J.,   {Xu} L.,  2024, \mn@doi
  [\mnras] {10.1093/mnras/stad3708}, \href
  {https://ui.adsabs.harvard.edu/abs/2024MNRAS.527.7861G} {527, 7861}

\bibitem[\protect\citeauthoryear{{Gao}, {Wu}, {Hu}, {Yi}, {Zhou}, {Wang}  \&
  {Dai}}{{Gao} et~al.}{2025}]{2025A&A...698A.215G}
{Gao} D.~H.,  {Wu} Q.,  {Hu} J.~P.,  {Yi} S.~X.,  {Zhou} X.,  {Wang} F.~Y.,
  {Dai} Z.~G.,  2025, \mn@doi [\aap] {10.1051/0004-6361/202453006}, \href
  {https://ui.adsabs.harvard.edu/abs/2025A&A...698A.215G} {698, A215}

\bibitem[\protect\citeauthoryear{{Garcia-Bellido} \&
  {Haugb{\o}lle}}{{Garcia-Bellido} \&
  {Haugb{\o}lle}}{2008}]{2008JCAP...04..003G}
{Garcia-Bellido} J.,  {Haugb{\o}lle} T.,  2008, \mn@doi [\jcap]
  {10.1088/1475-7516/2008/04/003}, \href
  {https://ui.adsabs.harvard.edu/abs/2008JCAP...04..003G} {2008, 003}

\bibitem[\protect\citeauthoryear{{Gazta{\~n}aga}, {Cabr{\'e}}  \&
  {Hui}}{{Gazta{\~n}aga} et~al.}{2009}]{2009MNRAS.399.1663G}
{Gazta{\~n}aga} E.,  {Cabr{\'e}} A.,   {Hui} L.,  2009, \mn@doi [\mnras]
  {10.1111/j.1365-2966.2009.15405.x}, \href
  {https://ui.adsabs.harvard.edu/abs/2009MNRAS.399.1663G} {399, 1663}

\bibitem[\protect\citeauthoryear{{Giar{\`e}}, {Najafi}, {Pan}, {Di Valentino}
  \& {Firouzjaee}}{{Giar{\`e}} et~al.}{2024}]{2024JCAP...10..035G}
{Giar{\`e}} W.,  {Najafi} M.,  {Pan} S.,  {Di Valentino} E.,   {Firouzjaee}
  J.~T.,  2024, \mn@doi [\jcap] {10.1088/1475-7516/2024/10/035}, \href
  {https://ui.adsabs.harvard.edu/abs/2024JCAP...10..035G} {2024, 035}

\bibitem[\protect\citeauthoryear{{G{\'o}mez-Valent} \&
  {Amendola}}{{G{\'o}mez-Valent} \& {Amendola}}{2018}]{2018JCAP...04..051G}
{G{\'o}mez-Valent} A.,  {Amendola} L.,  2018, \mn@doi [\jcap]
  {10.1088/1475-7516/2018/04/051}, \href
  {https://ui.adsabs.harvard.edu/abs/2018JCAP...04..051G} {2018, 051}

\bibitem[\protect\citeauthoryear{{G{\'o}mez-Valent}, {Favale}, {Migliaccio}  \&
  {Sen}}{{G{\'o}mez-Valent} et~al.}{2024}]{2024PhRvD.109b3525G}
{G{\'o}mez-Valent} A.,  {Favale} A.,  {Migliaccio} M.,   {Sen} A.~A.,  2024,
  \mn@doi [\prd] {10.1103/PhysRevD.109.023525}, \href
  {https://ui.adsabs.harvard.edu/abs/2024PhRvD.109b3525G} {109, 023525}

\bibitem[\protect\citeauthoryear{{Guimar{\~a}es}, {Cunha}  \&
  {Lima}}{{Guimar{\~a}es} et~al.}{2009}]{2009JCAP...10..010G}
{Guimar{\~a}es} A.~C.~C.,  {Cunha} J.~V.,   {Lima} J.~A.~S.,  2009, \mn@doi
  [\jcap] {10.1088/1475-7516/2009/10/010}, \href
  {https://ui.adsabs.harvard.edu/abs/2009JCAP...10..010G} {2009, 010}

\bibitem[\protect\citeauthoryear{{Guo} \& {Frolov}}{{Guo} \&
  {Frolov}}{2013}]{2013PhRvD..88l4036G}
{Guo} J.-Q.,  {Frolov} A.~V.,  2013, \mn@doi [\prd]
  {10.1103/PhysRevD.88.124036}, \href
  {https://ui.adsabs.harvard.edu/abs/2013PhRvD..88l4036G} {88, 124036}

\bibitem[\protect\citeauthoryear{{Haridasu}, {Lukovi{\'c}}, {Moresco}  \&
  {Vittorio}}{{Haridasu} et~al.}{2018}]{2018JCAP...10..015H}
{Haridasu} B.~S.,  {Lukovi{\'c}} V.~V.,  {Moresco} M.,   {Vittorio} N.,  2018,
  \mn@doi [\jcap] {10.1088/1475-7516/2018/10/015}, \href
  {https://ui.adsabs.harvard.edu/abs/2018JCAP...10..015H} {2018, 015}

\bibitem[\protect\citeauthoryear{{Horstmann}, {Pietschke}  \&
  {Schwarz}}{{Horstmann} et~al.}{2022}]{2022A&A...668A..34H}
{Horstmann} N.,  {Pietschke} Y.,   {Schwarz} D.~J.,  2022, \mn@doi [\aap]
  {10.1051/0004-6361/202142640}, \href
  {https://ui.adsabs.harvard.edu/abs/2022A&A...668A..34H} {668, A34}

\bibitem[\protect\citeauthoryear{{Hu} \& {Wang}}{{Hu} \&
  {Wang}}{2022a}]{2022MNRAS.517..576H}
{Hu} J.~P.,  {Wang} F.~Y.,  2022a, \mn@doi [\mnras] {10.1093/mnras/stac2728},
  \href {https://ui.adsabs.harvard.edu/abs/2022MNRAS.517..576H} {517, 576}

\bibitem[\protect\citeauthoryear{{Hu} \& {Wang}}{{Hu} \&
  {Wang}}{2022b}]{2022A&A...661A..71H}
{Hu} J.~P.,  {Wang} F.~Y.,  2022b, \mn@doi [\aap]
  {10.1051/0004-6361/202142162}, \href
  {https://ui.adsabs.harvard.edu/abs/2022A&A...661A..71H} {661, A71}

\bibitem[\protect\citeauthoryear{{Hu} \& {Wang}}{{Hu} \&
  {Wang}}{2023}]{2023Univ....9...94H}
{Hu} J.-P.,  {Wang} F.-Y.,  2023, \mn@doi [Universe] {10.3390/universe9020094},
  \href {https://ui.adsabs.harvard.edu/abs/2023Univ....9...94H} {9, 94}

\bibitem[\protect\citeauthoryear{{Hu}, {Wang}  \& {Dai}}{{Hu}
  et~al.}{2021}]{2021MNRAS.507..730H}
{Hu} J.~P.,  {Wang} F.~Y.,   {Dai} Z.~G.,  2021, \mn@doi [\mnras]
  {10.1093/mnras/stab2180}, \href
  {https://ui.adsabs.harvard.edu/abs/2021MNRAS.507..730H} {507, 730}

\bibitem[\protect\citeauthoryear{{Hu}, {Wang}, {Hu}  \& {Wang}}{{Hu}
  et~al.}{2024a}]{2024A&A...681A..88H}
{Hu} J.~P.,  {Wang} Y.~Y.,  {Hu} J.,   {Wang} F.~Y.,  2024a, \mn@doi [\aap]
  {10.1051/0004-6361/202347121}, \href
  {https://ui.adsabs.harvard.edu/abs/2024A&A...681A..88H} {681, A88}

\bibitem[\protect\citeauthoryear{{Hu}, {Hu}, {Jia}, {Gao}  \& {Wang}}{{Hu}
  et~al.}{2024b}]{2024A&A...689A.215H}
{Hu} J.~P.,  {Hu} J.,  {Jia} X.~D.,  {Gao} B.~Q.,   {Wang} F.~Y.,  2024b,
  \mn@doi [\aap] {10.1051/0004-6361/202450342}, \href
  {https://ui.adsabs.harvard.edu/abs/2024A&A...689A.215H} {689, A215}

\bibitem[\protect\citeauthoryear{{Hu}, {Jia}, {Hu}  \& {Wang}}{{Hu}
  et~al.}{2024c}]{2024ApJ...975L..36H}
{Hu} J.~P.,  {Jia} X.~D.,  {Hu} J.,   {Wang} F.~Y.,  2024c, \mn@doi [\apjl]
  {10.3847/2041-8213/ad85cf}, \href
  {https://ui.adsabs.harvard.edu/abs/2024ApJ...975L..36H} {975, L36}

\bibitem[\protect\citeauthoryear{{Jesus}, {Holanda}  \& {Pereira}}{{Jesus}
  et~al.}{2018}]{2018JCAP...05..073J}
{Jesus} J.~F.,  {Holanda} R.~F.~L.,   {Pereira} S.~H.,  2018, \mn@doi [\jcap]
  {10.1088/1475-7516/2018/05/073}, \href
  {https://ui.adsabs.harvard.edu/abs/2018JCAP...05..073J} {2018, 073}

\bibitem[\protect\citeauthoryear{{Jesus}, {Valentim}, {Escobal}  \&
  {Pereira}}{{Jesus} et~al.}{2020}]{2020JCAP...04..053J}
{Jesus} J.~F.,  {Valentim} R.,  {Escobal} A.~A.,   {Pereira} S.~H.,  2020,
  \mn@doi [\jcap] {10.1088/1475-7516/2020/04/053}, \href
  {https://ui.adsabs.harvard.edu/abs/2020JCAP...04..053J} {2020, 053}

\bibitem[\protect\citeauthoryear{{Jia}, {Hu}, {Yang}, {Zhang}  \& {Wang}}{{Jia}
  et~al.}{2022}]{2022MNRAS.516.2575J}
{Jia} X.~D.,  {Hu} J.~P.,  {Yang} J.,  {Zhang} B.~B.,   {Wang} F.~Y.,  2022,
  \mn@doi [\mnras] {10.1093/mnras/stac2356}, \href
  {https://ui.adsabs.harvard.edu/abs/2022MNRAS.516.2575J} {516, 2575}

\bibitem[\protect\citeauthoryear{{Jia}, {Hu}  \& {Wang}}{{Jia}
  et~al.}{2023}]{2023A&A...674A..45J}
{Jia} X.~D.,  {Hu} J.~P.,   {Wang} F.~Y.,  2023, \mn@doi [\aap]
  {10.1051/0004-6361/202346356}, \href
  {https://ui.adsabs.harvard.edu/abs/2023A&A...674A..45J} {674, A45}

\bibitem[\protect\citeauthoryear{{Jia}, {Hu}, {Yi}  \& {Wang}}{{Jia}
  et~al.}{2025}]{2025ApJ...979L..34J}
{Jia} X.~D.,  {Hu} J.~P.,  {Yi} S.~X.,   {Wang} F.~Y.,  2025, \mn@doi [\apjl]
  {10.3847/2041-8213/ada94d}, \href
  {https://ui.adsabs.harvard.edu/abs/2025ApJ...979L..34J} {979, L34}

\bibitem[\protect\citeauthoryear{{Jiao}, {Borghi}, {Moresco}  \&
  {Zhang}}{{Jiao} et~al.}{2023}]{2023ApJS..265...48J}
{Jiao} K.,  {Borghi} N.,  {Moresco} M.,   {Zhang} T.-J.,  2023, \mn@doi [\apjs]
  {10.3847/1538-4365/acbc77}, \href
  {https://ui.adsabs.harvard.edu/abs/2023ApJS..265...48J} {265, 48}

\bibitem[\protect\citeauthoryear{{Jimenez} \& {Loeb}}{{Jimenez} \&
  {Loeb}}{2002}]{2002ApJ...573...37J}
{Jimenez} R.,  {Loeb} A.,  2002, \mn@doi [\apj] {10.1086/340549}, \href
  {https://ui.adsabs.harvard.edu/abs/2002ApJ...573...37J} {573, 37}

\bibitem[\protect\citeauthoryear{{Jimenez}, {Moresco}, {Verde}  \&
  {Wandelt}}{{Jimenez} et~al.}{2023}]{2023JCAP...11..047J}
{Jimenez} R.,  {Moresco} M.,  {Verde} L.,   {Wandelt} B.~D.,  2023, \mn@doi
  [\jcap] {10.1088/1475-7516/2023/11/047}, \href
  {https://ui.adsabs.harvard.edu/abs/2023JCAP...11..047J} {2023, 047}

\bibitem[\protect\citeauthoryear{{Kalbouneh}, {Marinoni}  \&
  {Maartens}}{{Kalbouneh} et~al.}{2024}]{2024JCAP...09..069K}
{Kalbouneh} B.,  {Marinoni} C.,   {Maartens} R.,  2024, \mn@doi [\jcap]
  {10.1088/1475-7516/2024/09/069}, \href
  {https://ui.adsabs.harvard.edu/abs/2024JCAP...09..069K} {2024, 069}

\bibitem[\protect\citeauthoryear{{Kalbouneh}, {Santiago}, {Marinoni},
  {Maartens}, {Clarkson}  \& {Sarma}}{{Kalbouneh}
  et~al.}{2025}]{2025JCAP...02..076K}
{Kalbouneh} B.,  {Santiago} J.,  {Marinoni} C.,  {Maartens} R.,  {Clarkson} C.,
    {Sarma} M.,  2025, \mn@doi [\jcap] {10.1088/1475-7516/2025/02/076}, \href
  {https://ui.adsabs.harvard.edu/abs/2025JCAP...02..076K} {2025, 076}

\bibitem[\protect\citeauthoryear{{Keenan}, {Barger}  \& {Cowie}}{{Keenan}
  et~al.}{2013}]{2013ApJ...775...62K}
{Keenan} R.~C.,  {Barger} A.~J.,   {Cowie} L.~L.,  2013, \mn@doi [\apj]
  {10.1088/0004-637X/775/1/62}, \href
  {https://ui.adsabs.harvard.edu/abs/2013ApJ...775...62K} {775, 62}

\bibitem[\protect\citeauthoryear{{Komatsu} et~al.,}{{Komatsu}
  et~al.}{2011}]{2011ApJS..192...18K}
{Komatsu} E.,  et~al., 2011, \mn@doi [\apjs] {10.1088/0067-0049/192/2/18},
  \href {https://ui.adsabs.harvard.edu/abs/2011ApJS..192...18K} {192, 18}

\bibitem[\protect\citeauthoryear{{Koussour}, {Myrzakulov}  \& {Ali}}{{Koussour}
  et~al.}{2024}]{2024JHEAp..42...96K}
{Koussour} M.,  {Myrzakulov} N.,   {Ali} M.~K.~M.,  2024, \mn@doi [Journal of
  High Energy Astrophysics] {10.1016/j.jheap.2024.04.003}, \href
  {https://ui.adsabs.harvard.edu/abs/2024JHEAp..42...96K} {42, 96}

\bibitem[\protect\citeauthoryear{{Krishnan}, {Colg{\'a}in}, {Ruchika},
  {Sheikh-Jabbari}  \& {Yang}}{{Krishnan} et~al.}{2020}]{2020PhRvD.102j3525K}
{Krishnan} C.,  {Colg{\'a}in} E.~{\'O}.,  {Ruchika} A.~A. S.,  {Sheikh-Jabbari}
  M.~M.,   {Yang} T.,  2020, \mn@doi [\prd] {10.1103/PhysRevD.102.103525},
  \href {https://ui.adsabs.harvard.edu/abs/2020PhRvD.102j3525K} {102, 103525}

\bibitem[\protect\citeauthoryear{{Krishnan}, {Mohayaee}, {Colg{\'a}in},
  {Sheikh-Jabbari}  \& {Yin}}{{Krishnan} et~al.}{2021}]{2021CQGra..38r4001K}
{Krishnan} C.,  {Mohayaee} R.,  {Colg{\'a}in} E.~{\'O}.,  {Sheikh-Jabbari}
  M.~M.,   {Yin} L.,  2021, \mn@doi [Classical and Quantum Gravity]
  {10.1088/1361-6382/ac1a81}, \href
  {https://ui.adsabs.harvard.edu/abs/2021CQGra..38r4001K} {38, 184001}

\bibitem[\protect\citeauthoryear{{Kroupa} et~al.,}{{Kroupa}
  et~al.}{2023}]{2023eppg.confE.231K}
{Kroupa} P.,  et~al., 2023, in School and Workshops on Elementary Particle
  Physics and Gravity. p.~231 (\mn@eprint {arXiv} {2309.11552}),
  \mn@doi{10.48550/arXiv.2309.11552}

\bibitem[\protect\citeauthoryear{{Kumar}, {Jain}, {Mahajan}, {Mukherjee}  \&
  {Rana}}{{Kumar} et~al.}{2023}]{2023IJMPD..3250039K}
{Kumar} D.,  {Jain} D.,  {Mahajan} S.,  {Mukherjee} A.,   {Rana} A.,  2023,
  \mn@doi [International Journal of Modern Physics D]
  {10.1142/S0218271823500396}, \href
  {https://ui.adsabs.harvard.edu/abs/2023IJMPD..3250039K} {32, 2350039}

\bibitem[\protect\citeauthoryear{{Li}, {Du}  \& {Xu}}{{Li}
  et~al.}{2020}]{2020MNRAS.491.4960L}
{Li} E.-K.,  {Du} M.,   {Xu} L.,  2020, \mn@doi [\mnras]
  {10.1093/mnras/stz3308}, \href
  {https://ui.adsabs.harvard.edu/abs/2020MNRAS.491.4960L} {491, 4960}

\bibitem[\protect\citeauthoryear{{Li}, {Zhang}  \& {Liang}}{{Li}
  et~al.}{2023}]{2023MNRAS.521.4406L}
{Li} Z.,  {Zhang} B.,   {Liang} N.,  2023, \mn@doi [\mnras]
  {10.1093/mnras/stad838}, \href
  {https://ui.adsabs.harvard.edu/abs/2023MNRAS.521.4406L} {521, 4406}

\bibitem[\protect\citeauthoryear{{Liang} \& {Zhang}}{{Liang} \&
  {Zhang}}{2005}]{2005ApJ...633..611L}
{Liang} E.,  {Zhang} B.,  2005, \mn@doi [\apj] {10.1086/491594}, \href
  {https://ui.adsabs.harvard.edu/abs/2005ApJ...633..611L} {633, 611}

\bibitem[\protect\citeauthoryear{{Liang}, {Li}, {Xie}  \& {Wu}}{{Liang}
  et~al.}{2022}]{2022ApJ...941...84L}
{Liang} N.,  {Li} Z.,  {Xie} X.,   {Wu} P.,  2022, \mn@doi [\apj]
  {10.3847/1538-4357/aca08a}, \href
  {https://ui.adsabs.harvard.edu/abs/2022ApJ...941...84L} {941, 84}

\bibitem[\protect\citeauthoryear{{Liao}, {Shafieloo}, {Keeley}  \&
  {Linder}}{{Liao} et~al.}{2019}]{2019ApJ...886L..23L}
{Liao} K.,  {Shafieloo} A.,  {Keeley} R.~E.,   {Linder} E.~V.,  2019, \mn@doi
  [\apjl] {10.3847/2041-8213/ab5308}, \href
  {https://ui.adsabs.harvard.edu/abs/2019ApJ...886L..23L} {886, L23}

\bibitem[\protect\citeauthoryear{{Liu}, {Liang}, {Xie}, {Yuan}, {Yu}  \&
  {Wu}}{{Liu} et~al.}{2022}]{2022ApJ...935....7L}
{Liu} Y.,  {Liang} N.,  {Xie} X.,  {Yuan} Z.,  {Yu} H.,   {Wu} P.,  2022,
  \mn@doi [\apj] {10.3847/1538-4357/ac7de5}, \href
  {https://ui.adsabs.harvard.edu/abs/2022ApJ...935....7L} {935, 7}

\bibitem[\protect\citeauthoryear{{Liu}, {Yu}  \& {Wu}}{{Liu}
  et~al.}{2024a}]{2024PhRvD.110b1304L}
{Liu} Y.,  {Yu} H.,   {Wu} P.,  2024a, \mn@doi [\prd]
  {10.1103/PhysRevD.110.L021304}, \href
  {https://ui.adsabs.harvard.edu/abs/2024PhRvD.110b1304L} {110, L021304}

\bibitem[\protect\citeauthoryear{{Liu}, {Wang}, {Yu}  \& {Wu}}{{Liu}
  et~al.}{2024b}]{2024MNRAS.533..244L}
{Liu} Y.,  {Wang} B.,  {Yu} H.,   {Wu} P.,  2024b, \mn@doi [\mnras]
  {10.1093/mnras/stae1808}, \href
  {https://ui.adsabs.harvard.edu/abs/2024MNRAS.533..244L} {533, 244}

\bibitem[\protect\citeauthoryear{{Lopez-Hernandez} \&
  {De-Santiago}}{{Lopez-Hernandez} \&
  {De-Santiago}}{2025}]{2025JCAP...03..026L}
{Lopez-Hernandez} M.,  {De-Santiago} J.,  2025, \mn@doi [\jcap]
  {10.1088/1475-7516/2025/03/026}, \href
  {https://ui.adsabs.harvard.edu/abs/2025JCAP...03..026L} {2025, 026}

\bibitem[\protect\citeauthoryear{{Luongo} \& {Muccino}}{{Luongo} \&
  {Muccino}}{2024}]{2024arXiv241218493L}
{Luongo} O.,  {Muccino} M.,  2024, \mn@doi [arXiv e-prints]
  {10.48550/arXiv.2412.18493}, \href
  {https://ui.adsabs.harvard.edu/abs/2024arXiv241218493L} {p. arXiv:2412.18493}

\bibitem[\protect\citeauthoryear{{Lusso}, {Piedipalumbo}, {Risaliti},
  {Paolillo}, {Bisogni}, {Nardini}  \& {Amati}}{{Lusso}
  et~al.}{2019}]{2019A&A...628L...4L}
{Lusso} E.,  {Piedipalumbo} E.,  {Risaliti} G.,  {Paolillo} M.,  {Bisogni} S.,
  {Nardini} E.,   {Amati} L.,  2019, \mn@doi [\aap]
  {10.1051/0004-6361/201936223}, \href
  {https://ui.adsabs.harvard.edu/abs/2019A&A...628L...4L} {628, L4}

\bibitem[\protect\citeauthoryear{{Malekjani}, {Mc Conville}, {{\'O}
  Colg{\'a}in}, {Pourojaghi}  \& {Sheikh-Jabbari}}{{Malekjani}
  et~al.}{2024}]{2024EPJC...84..317M}
{Malekjani} M.,  {Mc Conville} R.,  {{\'O} Colg{\'a}in} E.,  {Pourojaghi} S.,
  {Sheikh-Jabbari} M.~M.,  2024, \mn@doi [European Physical Journal C]
  {10.1140/epjc/s10052-024-12667-z}, \href
  {https://ui.adsabs.harvard.edu/abs/2024EPJC...84..317M} {84, 317}

\bibitem[\protect\citeauthoryear{{Mazurenko}, {Banik}  \& {Kroupa}}{{Mazurenko}
  et~al.}{2025}]{2025MNRAS.536.3232M}
{Mazurenko} S.,  {Banik} I.,   {Kroupa} P.,  2025, \mn@doi [\mnras]
  {10.1093/mnras/stae2758}, \href
  {https://ui.adsabs.harvard.edu/abs/2025MNRAS.536.3232M} {536, 3232}

\bibitem[\protect\citeauthoryear{{Melia} \& {Yennapureddy}}{{Melia} \&
  {Yennapureddy}}{2018}]{2018JCAP...02..034M}
{Melia} F.,  {Yennapureddy} M.~K.,  2018, \mn@doi [\jcap]
  {10.1088/1475-7516/2018/02/034}, \href
  {https://ui.adsabs.harvard.edu/abs/2018JCAP...02..034M} {2018, 034}

\bibitem[\protect\citeauthoryear{{Milgrom}}{{Milgrom}}{1983}]{1983ApJ...270..365M}
{Milgrom} M.,  1983, \mn@doi [\apj] {10.1086/161130}, \href
  {https://ui.adsabs.harvard.edu/abs/1983ApJ...270..365M} {270, 365}

\bibitem[\protect\citeauthoryear{{Millon} et~al.,}{{Millon}
  et~al.}{2020}]{2020A&A...639A.101M}
{Millon} M.,  et~al., 2020, \mn@doi [\aap] {10.1051/0004-6361/201937351}, \href
  {https://ui.adsabs.harvard.edu/abs/2020A&A...639A.101M} {639, A101}

\bibitem[\protect\citeauthoryear{{Mishra}, {Kavya}, {Sahoo}  \&
  {Venkatesha}}{{Mishra} et~al.}{2025}]{2025PDU....4701759M}
{Mishra} S.~S.,  {Kavya} N.~S.,  {Sahoo} P.~K.,   {Venkatesha} V.,  2025,
  \mn@doi [Physics of the Dark Universe] {10.1016/j.dark.2024.101759}, \href
  {https://ui.adsabs.harvard.edu/abs/2025PDU....4701759M} {47, 101759}

\bibitem[\protect\citeauthoryear{{Montani}, {Carlevaro}  \&
  {Dainotti}}{{Montani} et~al.}{2025a}]{2025PDU....4801847M}
{Montani} G.,  {Carlevaro} N.,   {Dainotti} M.~G.,  2025a, \mn@doi [Physics of
  the Dark Universe] {10.1016/j.dark.2025.101847}, \href
  {https://ui.adsabs.harvard.edu/abs/2025PDU....4801847M} {48, 101847}

\bibitem[\protect\citeauthoryear{{Montani}, {Carlevaro}, {Escamilla}  \& {Di
  Valentino}}{{Montani} et~al.}{2025b}]{2025PDU....4801848M}
{Montani} G.,  {Carlevaro} N.,  {Escamilla} L.~A.,   {Di Valentino} E.,  2025b,
  \mn@doi [Physics of the Dark Universe] {10.1016/j.dark.2025.101848}, \href
  {https://ui.adsabs.harvard.edu/abs/2025PDU....4801848M} {48, 101848}

\bibitem[\protect\citeauthoryear{{Moresco}}{{Moresco}}{2015}]{2015MNRAS.450L..16M}
{Moresco} M.,  2015, \mn@doi [\mnras] {10.1093/mnrasl/slv037}, \href
  {https://ui.adsabs.harvard.edu/abs/2015MNRAS.450L..16M} {450, L16}

\bibitem[\protect\citeauthoryear{{Moresco} et~al.,}{{Moresco}
  et~al.}{2012}]{2012JCAP...08..006M}
{Moresco} M.,  et~al., 2012, \mn@doi [\jcap] {10.1088/1475-7516/2012/08/006},
  \href {https://ui.adsabs.harvard.edu/abs/2012JCAP...08..006M} {2012, 006}

\bibitem[\protect\citeauthoryear{{Moresco} et~al.,}{{Moresco}
  et~al.}{2016}]{2016JCAP...05..014M}
{Moresco} M.,  et~al., 2016, \mn@doi [\jcap] {10.1088/1475-7516/2016/05/014},
  \href {https://ui.adsabs.harvard.edu/abs/2016JCAP...05..014M} {2016, 014}

\bibitem[\protect\citeauthoryear{{Moresco}, {Jimenez}, {Verde}, {Cimatti}  \&
  {Pozzetti}}{{Moresco} et~al.}{2020}]{2020ApJ...898...82M}
{Moresco} M.,  {Jimenez} R.,  {Verde} L.,  {Cimatti} A.,   {Pozzetti} L.,
  2020, \mn@doi [\apj] {10.3847/1538-4357/ab9eb0}, \href
  {https://ui.adsabs.harvard.edu/abs/2020ApJ...898...82M} {898, 82}

\bibitem[\protect\citeauthoryear{{Moresco} et~al.,}{{Moresco}
  et~al.}{2022}]{2022LRR....25....6M}
{Moresco} M.,  et~al., 2022, \mn@doi [Living Reviews in Relativity]
  {10.1007/s41114-022-00040-z}, \href
  {https://ui.adsabs.harvard.edu/abs/2022LRR....25....6M} {25, 6}

\bibitem[\protect\citeauthoryear{{Muccino}, {Luongo}  \& {Jain}}{{Muccino}
  et~al.}{2023}]{2023MNRAS.523.4938M}
{Muccino} M.,  {Luongo} O.,   {Jain} D.,  2023, \mn@doi [\mnras]
  {10.1093/mnras/stad1760}, \href
  {https://ui.adsabs.harvard.edu/abs/2023MNRAS.523.4938M} {523, 4938}

\bibitem[\protect\citeauthoryear{{Myrzakulov}}{{Myrzakulov}}{2011}]{2011EPJC...71.1752M}
{Myrzakulov} R.,  2011, \mn@doi [European Physical Journal C]
  {10.1140/epjc/s10052-011-1752-9}, \href
  {https://ui.adsabs.harvard.edu/abs/2011EPJC...71.1752M} {71, 1752}

\bibitem[\protect\citeauthoryear{{Myrzakulov}, {Donmez}, {Koussour},
  {Alizhanov}, {Bekchanov}  \& {Rayimbaev}}{{Myrzakulov}
  et~al.}{2024}]{2024PDU....4601614M}
{Myrzakulov} Y.,  {Donmez} O.,  {Koussour} M.,  {Alizhanov} D.,  {Bekchanov}
  S.,   {Rayimbaev} J.,  2024, \mn@doi [Physics of the Dark Universe]
  {10.1016/j.dark.2024.101614}, \href
  {https://ui.adsabs.harvard.edu/abs/2024PDU....4601614M} {46, 101614}

\bibitem[\protect\citeauthoryear{{Nakamura} \& {Chiba}}{{Nakamura} \&
  {Chiba}}{1999}]{1999MNRAS.306..696N}
{Nakamura} T.,  {Chiba} T.,  1999, \mn@doi [\mnras]
  {10.1046/j.1365-8711.1999.02551.x}, \href
  {https://ui.adsabs.harvard.edu/abs/1999MNRAS.306..696N} {306, 696}

\bibitem[\protect\citeauthoryear{{Neveux} et~al.,}{{Neveux}
  et~al.}{2020}]{2020MNRAS.499..210N}
{Neveux} R.,  et~al., 2020, \mn@doi [\mnras] {10.1093/mnras/staa2780}, \href
  {https://ui.adsabs.harvard.edu/abs/2020MNRAS.499..210N} {499, 210}

\bibitem[\protect\citeauthoryear{{{\'O} Colg{\'a}in}, {Sheikh-Jabbari},
  {Solomon}, {Bargiacchi}, {Capozziello}, {Dainotti}  \& {Stojkovic}}{{{\'O}
  Colg{\'a}in} et~al.}{2022}]{2022PhRvD.106d1301O}
{{\'O} Colg{\'a}in} E.,  {Sheikh-Jabbari} M.~M.,  {Solomon} R.,  {Bargiacchi}
  G.,  {Capozziello} S.,  {Dainotti} M.~G.,   {Stojkovic} D.,  2022, \mn@doi
  [\prd] {10.1103/PhysRevD.106.L041301}, \href
  {https://ui.adsabs.harvard.edu/abs/2022PhRvD.106d1301O} {106, L041301}

\bibitem[\protect\citeauthoryear{{{\'O} Colg{\'a}in}, {Sheikh-Jabbari},
  {Solomon}, {Dainotti}  \& {Stojkovic}}{{{\'O} Colg{\'a}in}
  et~al.}{2024}]{2024PDU....4401464O}
{{\'O} Colg{\'a}in} E.,  {Sheikh-Jabbari} M.~M.,  {Solomon} R.,  {Dainotti}
  M.~G.,   {Stojkovic} D.,  2024, \mn@doi [Physics of the Dark Universe]
  {10.1016/j.dark.2024.101464}, \href
  {https://ui.adsabs.harvard.edu/abs/2024PDU....4401464O} {44, 101464}

\bibitem[\protect\citeauthoryear{{{\'O}. Colg{\'a}in}, {Pourojaghi},
  {Sheikh-Jabbari}  \& {Sherwin}}{{{\'O}. Colg{\'a}in}
  et~al.}{2025}]{2025EPJC...85..124O}
{{\'O}. Colg{\'a}in} E.,  {Pourojaghi} S.,  {Sheikh-Jabbari} M.~M.,   {Sherwin}
  D.,  2025, \mn@doi [European Physical Journal C]
  {10.1140/epjc/s10052-024-13727-0}, \href
  {https://ui.adsabs.harvard.edu/abs/2025EPJC...85..124O} {85, 124}

\bibitem[\protect\citeauthoryear{{Overduin} \& {Wesson}}{{Overduin} \&
  {Wesson}}{1997}]{1997PhR...283..303O}
{Overduin} J.~M.,  {Wesson} P.~S.,  1997, \mn@doi [\physrep]
  {10.1016/S0370-1573(96)00046-4}, \href
  {https://ui.adsabs.harvard.edu/abs/1997PhR...283..303O} {283, 303}

\bibitem[\protect\citeauthoryear{{Padmanabhan}}{{Padmanabhan}}{2003}]{2003PhR...380..235P}
{Padmanabhan} T.,  2003, \mn@doi [\physrep] {10.1016/S0370-1573(03)00120-0},
  \href {https://ui.adsabs.harvard.edu/abs/2003PhR...380..235P} {380, 235}

\bibitem[\protect\citeauthoryear{{Park} \& {Ratra}}{{Park} \&
  {Ratra}}{2025}]{2025arXiv250103480P}
{Park} C.-G.,  {Ratra} B.,  2025, \mn@doi [arXiv e-prints]
  {10.48550/arXiv.2501.03480}, \href
  {https://ui.adsabs.harvard.edu/abs/2025arXiv250103480P} {p. arXiv:2501.03480}

\bibitem[\protect\citeauthoryear{{Pedregosa} et~al.,}{{Pedregosa}
  et~al.}{2011}]{2011JMLR...12.2825P}
{Pedregosa} F.,  et~al., 2011, \mn@doi [Journal of Machine Learning Research]
  {10.48550/arXiv.1201.0490}, \href
  {https://ui.adsabs.harvard.edu/abs/2011JMLR...12.2825P} {12, 2825}

\bibitem[\protect\citeauthoryear{{Peebles} \& {Ratra}}{{Peebles} \&
  {Ratra}}{2003}]{2003RvMP...75..559P}
{Peebles} P.~J.,  {Ratra} B.,  2003, \mn@doi [Reviews of Modern Physics]
  {10.1103/RevModPhys.75.559}, \href
  {https://ui.adsabs.harvard.edu/abs/2003RvMP...75..559P} {75, 559}

\bibitem[\protect\citeauthoryear{{Perivolaropoulos} \&
  {Skara}}{{Perivolaropoulos} \& {Skara}}{2022}]{2022NewAR..9501659P}
{Perivolaropoulos} L.,  {Skara} F.,  2022, \mn@doi [\nar]
  {10.1016/j.newar.2022.101659}, \href
  {https://ui.adsabs.harvard.edu/abs/2022NewAR..9501659P} {95, 101659}

\bibitem[\protect\citeauthoryear{{Planck Collaboration} et~al.,}{{Planck
  Collaboration} et~al.}{2014}]{2014A&A...571A..16P}
{Planck Collaboration} et~al., 2014, \mn@doi [\aap]
  {10.1051/0004-6361/201321591}, \href
  {https://ui.adsabs.harvard.edu/abs/2014A&A...571A..16P} {571, A16}

\bibitem[\protect\citeauthoryear{{Planck Collaboration} et~al.,}{{Planck
  Collaboration} et~al.}{2020a}]{2020AA...641A...6P}
{Planck Collaboration} et~al., 2020a, \mn@doi [\aap]
  {10.1051/0004-6361/201833910}, \href
  {https://ui.adsabs.harvard.edu/abs/2020A&A...641A...6P} {641, A6}

\bibitem[\protect\citeauthoryear{{Planck Collaboration} et~al.,}{{Planck
  Collaboration} et~al.}{2020b}]{2020A&A...641A...6P}
{Planck Collaboration} et~al., 2020b, \mn@doi [\aap]
  {10.1051/0004-6361/201833910}, \href
  {https://ui.adsabs.harvard.edu/abs/2020A&A...641A...6P} {641, A6}

\bibitem[\protect\citeauthoryear{{Pourojaghi}, {Malekjani}  \&
  {Davari}}{{Pourojaghi} et~al.}{2025}]{2025MNRAS.537..436P}
{Pourojaghi} S.,  {Malekjani} M.,   {Davari} Z.,  2025, \mn@doi [\mnras]
  {10.1093/mnras/staf037}, \href
  {https://ui.adsabs.harvard.edu/abs/2025MNRAS.537..436P} {537, 436}

\bibitem[\protect\citeauthoryear{{Rahman}}{{Rahman}}{2023}]{2023GrCo...29..177R}
{Rahman} S. F.~u.,  2023, \mn@doi [Gravitation and Cosmology]
  {10.1134/S020228932302010X}, \href
  {https://ui.adsabs.harvard.edu/abs/2023GrCo...29..177R} {29, 177}

\bibitem[\protect\citeauthoryear{{Randall} \& {Sundrum}}{{Randall} \&
  {Sundrum}}{1999}]{1999PhRvL..83.4690R}
{Randall} L.,  {Sundrum} R.,  1999, \mn@doi [\prl]
  {10.1103/PhysRevLett.83.4690}, \href
  {https://ui.adsabs.harvard.edu/abs/1999PhRvL..83.4690R} {83, 4690}

\bibitem[\protect\citeauthoryear{{Rani}, {Jain}, {Mahajan}, {Mukherjee}  \&
  {Pires}}{{Rani} et~al.}{2015}]{2015JCAP...12..045R}
{Rani} N.,  {Jain} D.,  {Mahajan} S.,  {Mukherjee} A.,   {Pires} N.,  2015,
  \mn@doi [\jcap] {10.1088/1475-7516/2015/12/045}, \href
  {https://ui.adsabs.harvard.edu/abs/2015JCAP...12..045R} {2015, 045}

\bibitem[\protect\citeauthoryear{{Rasmussen} \& {Williams}}{{Rasmussen} \&
  {Williams}}{2006}]{2006gpml.book.....R}
{Rasmussen} C.~E.,  {Williams} C. K.~I.,  2006, {Gaussian Processes for Machine
  Learning}

\bibitem[\protect\citeauthoryear{{Ratsimbazafy}, {Loubser}, {Crawford},
  {Cress}, {Bassett}, {Nichol}  \& {V{\"a}is{\"a}nen}}{{Ratsimbazafy}
  et~al.}{2017}]{2017MNRAS.467.3239R}
{Ratsimbazafy} A.~L.,  {Loubser} S.~I.,  {Crawford} S.~M.,  {Cress} C.~M.,
  {Bassett} B.~A.,  {Nichol} R.~C.,   {V{\"a}is{\"a}nen} P.,  2017, \mn@doi
  [\mnras] {10.1093/mnras/stx301}, \href
  {https://ui.adsabs.harvard.edu/abs/2017MNRAS.467.3239R} {467, 3239}

\bibitem[\protect\citeauthoryear{{Riess} \& {Breuval}}{{Riess} \&
  {Breuval}}{2024}]{2024IAUS..376...15R}
{Riess} A.~G.,  {Breuval} L.,  2024, in {de Grijs} R.,  {Whitelock} P.~A.,
  {Catelan} M.,  eds,  IAU Symposium Vol. 376, IAU Symposium. pp 15--29
  (\mn@eprint {arXiv} {2308.10954}), \mn@doi{10.1017/S1743921323003034}

\bibitem[\protect\citeauthoryear{{Riess} et~al.,}{{Riess}
  et~al.}{1998}]{1998AJ....116.1009R}
{Riess} A.~G.,  et~al., 1998, \mn@doi [\aj] {10.1086/300499}, \href
  {https://ui.adsabs.harvard.edu/abs/1998AJ....116.1009R} {116, 1009}

\bibitem[\protect\citeauthoryear{{Riess} et~al.,}{{Riess}
  et~al.}{2004}]{2004ApJ...607..665R}
{Riess} A.~G.,  et~al., 2004, \mn@doi [\apj] {10.1086/383612}, \href
  {https://ui.adsabs.harvard.edu/abs/2004ApJ...607..665R} {607, 665}

\bibitem[\protect\citeauthoryear{{Riess} et~al.,}{{Riess}
  et~al.}{2007}]{2007ApJ...659...98R}
{Riess} A.~G.,  et~al., 2007, \mn@doi [\apj] {10.1086/510378}, \href
  {https://ui.adsabs.harvard.edu/abs/2007ApJ...659...98R} {659, 98}

\bibitem[\protect\citeauthoryear{{Riess} et~al.,}{{Riess}
  et~al.}{2011}]{2011ApJ...730..119R}
{Riess} A.~G.,  et~al., 2011, \mn@doi [\apj] {10.1088/0004-637X/730/2/119},
  \href {https://ui.adsabs.harvard.edu/abs/2011ApJ...730..119R} {730, 119}

\bibitem[\protect\citeauthoryear{{Riess} et~al.,}{{Riess}
  et~al.}{2022}]{2022ApJ...934L...7R}
{Riess} A.~G.,  et~al., 2022, \mn@doi [\apjl] {10.3847/2041-8213/ac5c5b}, \href
  {https://ui.adsabs.harvard.edu/abs/2022ApJ...934L...7R} {934, L7}

\bibitem[\protect\citeauthoryear{{Roos}}{{Roos}}{2012}]{2012JMPh....3.1152R}
{Roos} M.,  2012, \mn@doi [Journal of Modern Physics]
  {10.4236/jmp.2012.329150}, \href
  {https://ui.adsabs.harvard.edu/abs/2012JMPh....3.1152R} {3, 1152}

\bibitem[\protect\citeauthoryear{{Rubin} et~al.,}{{Rubin}
  et~al.}{2023}]{2023arXiv231112098R}
{Rubin} D.,  et~al., 2023, \mn@doi [arXiv e-prints]
  {10.48550/arXiv.2311.12098}, \href
  {https://ui.adsabs.harvard.edu/abs/2023arXiv231112098R} {p. arXiv:2311.12098}

\bibitem[\protect\citeauthoryear{{Sah}, {Rameez}, {Sarkar}  \& {Tsagas}}{{Sah}
  et~al.}{2024}]{2024arXiv241110838S}
{Sah} A.,  {Rameez} M.,  {Sarkar} S.,   {Tsagas} C.,  2024, \mn@doi [arXiv
  e-prints] {10.48550/arXiv.2411.10838}, \href
  {https://ui.adsabs.harvard.edu/abs/2024arXiv241110838S} {p. arXiv:2411.10838}

\bibitem[\protect\citeauthoryear{{Salehi} \& {Hatami}}{{Salehi} \&
  {Hatami}}{2022}]{2022EPJC...82.1165S}
{Salehi} A.,  {Hatami} H.,  2022, \mn@doi [European Physical Journal C]
  {10.1140/epjc/s10052-022-11105-2}, \href
  {https://ui.adsabs.harvard.edu/abs/2022EPJC...82.1165S} {82, 1165}

\bibitem[\protect\citeauthoryear{{Scherer}, {Sabogal}, {Nunes}  \& {De
  Felice}}{{Scherer} et~al.}{2025}]{2025arXiv250420664S}
{Scherer} M.,  {Sabogal} M.~A.,  {Nunes} R.~C.,   {De Felice} A.,  2025,
  \mn@doi [arXiv e-prints] {10.48550/arXiv.2504.20664}, \href
  {https://ui.adsabs.harvard.edu/abs/2025arXiv250420664S} {p. arXiv:2504.20664}

\bibitem[\protect\citeauthoryear{Schulz, Speekenbrink  \& Krause}{Schulz
  et~al.}{2018}]{SCHULZ20181}
Schulz E.,  Speekenbrink M.,   Krause A.,  2018, \mn@doi [Journal of
  Mathematical Psychology] {https://doi.org/10.1016/j.jmp.2018.03.001}, 85, 1

\bibitem[\protect\citeauthoryear{{Scolnic} et~al.,}{{Scolnic}
  et~al.}{2018}]{2018ApJ...859..101S}
{Scolnic} D.~M.,  et~al., 2018, \mn@doi [\apj] {10.3847/1538-4357/aab9bb},
  \href {https://ui.adsabs.harvard.edu/abs/2018ApJ...859..101S} {859, 101}

\bibitem[\protect\citeauthoryear{{Seikel}, {Clarkson}  \& {Smith}}{{Seikel}
  et~al.}{2012}]{2012JCAP...06..036S}
{Seikel} M.,  {Clarkson} C.,   {Smith} M.,  2012, \mn@doi [\jcap]
  {10.1088/1475-7516/2012/06/036}, \href
  {https://ui.adsabs.harvard.edu/abs/2012JCAP...06..036S} {2012, 036}

\bibitem[\protect\citeauthoryear{Silk et~al.}{Silk
  et~al.}{2010}]{Bertone:2010zza}
Silk J.,  et~al., 2010, {Particle Dark Matter: Observations, Models and
  Searches}.
Cambridge Univ. Press, Cambridge, \mn@doi{10.1017/CBO9780511770739}

\bibitem[\protect\citeauthoryear{{Simon}, {Verde}  \& {Jimenez}}{{Simon}
  et~al.}{2005}]{2005PhRvD..71l3001S}
{Simon} J.,  {Verde} L.,   {Jimenez} R.,  2005, \mn@doi [\prd]
  {10.1103/PhysRevD.71.123001}, \href
  {https://ui.adsabs.harvard.edu/abs/2005PhRvD..71l3001S} {71, 123001}

\bibitem[\protect\citeauthoryear{{Smee} et~al.,}{{Smee}
  et~al.}{2013}]{2013AJ....146...32S}
{Smee} S.~A.,  et~al., 2013, \mn@doi [\aj] {10.1088/0004-6256/146/2/32}, \href
  {https://ui.adsabs.harvard.edu/abs/2013AJ....146...32S} {146, 32}

\bibitem[\protect\citeauthoryear{{Sudharani}, {Bamba}, {Kavya}  \&
  {Venkatesha}}{{Sudharani} et~al.}{2024}]{2024PDU....4501522S}
{Sudharani} L.,  {Bamba} K.,  {Kavya} N.~S.,   {Venkatesha} V.,  2024, \mn@doi
  [Physics of the Dark Universe] {10.1016/j.dark.2024.101522}, \href
  {https://ui.adsabs.harvard.edu/abs/2024PDU....4501522S} {45, 101522}

\bibitem[\protect\citeauthoryear{{Tomasetti} et~al.,}{{Tomasetti}
  et~al.}{2023a}]{2023A&A...679A..96T}
{Tomasetti} E.,  et~al., 2023a, \mn@doi [\aap] {10.1051/0004-6361/202346992},
  \href {https://ui.adsabs.harvard.edu/abs/2023A&A...679A..96T} {679, A96}

\bibitem[\protect\citeauthoryear{{Tomasetti} et~al.,}{{Tomasetti}
  et~al.}{2023b}]{2023AA...679A..96T}
{Tomasetti} E.,  et~al., 2023b, \mn@doi [\aap] {10.1051/0004-6361/202346992},
  \href {https://ui.adsabs.harvard.edu/abs/2023A&A...679A..96T} {679, A96}

\bibitem[\protect\citeauthoryear{{Vagnozzi}}{{Vagnozzi}}{2023}]{2023Univ....9..393V}
{Vagnozzi} S.,  2023, \mn@doi [Universe] {10.3390/universe9090393}, \href
  {https://ui.adsabs.harvard.edu/abs/2023Univ....9..393V} {9, 393}

\bibitem[\protect\citeauthoryear{{Velasquez-Toribio} \&
  {Fabris}}{{Velasquez-Toribio} \& {Fabris}}{2022}]{2022BrJPh..52..115V}
{Velasquez-Toribio} A.~M.,  {Fabris} J.~C.,  2022, \mn@doi [Brazilian Journal
  of Physics] {10.1007/s13538-022-01113-8}, \href
  {https://ui.adsabs.harvard.edu/abs/2022BrJPh..52..115V} {52, 115}

\bibitem[\protect\citeauthoryear{{Velasquez-Toribio} \&
  {Magnago}}{{Velasquez-Toribio} \& {Magnago}}{2020}]{2020EPJC...80..562V}
{Velasquez-Toribio} A.~M.,  {Magnago} A. d.~R.,  2020, \mn@doi [European
  Physical Journal C] {10.1140/epjc/s10052-020-8120-6}, \href
  {https://ui.adsabs.harvard.edu/abs/2020EPJC...80..562V} {80, 562}

\bibitem[\protect\citeauthoryear{{Visser}}{{Visser}}{2004}]{2004CQGra..21.2603V}
{Visser} M.,  2004, \mn@doi [Classical and Quantum Gravity]
  {10.1088/0264-9381/21/11/006}, \href
  {https://ui.adsabs.harvard.edu/abs/2004CQGra..21.2603V} {21, 2603}

\bibitem[\protect\citeauthoryear{{Visser}}{{Visser}}{2015}]{2015CQGra..32m5007V}
{Visser} M.,  2015, \mn@doi [Classical and Quantum Gravity]
  {10.1088/0264-9381/32/13/135007}, \href
  {https://ui.adsabs.harvard.edu/abs/2015CQGra..32m5007V} {32, 135007}

\bibitem[\protect\citeauthoryear{{Volkov}}{{Volkov}}{2012}]{2012PhRvD..86f1502V}
{Volkov} M.~S.,  2012, \mn@doi [\prd] {10.1103/PhysRevD.86.061502}, \href
  {https://ui.adsabs.harvard.edu/abs/2012PhRvD..86f1502V} {86, 061502}

\bibitem[\protect\citeauthoryear{{Wang} \& {Dai}}{{Wang} \&
  {Dai}}{2006}]{2006MNRAS.368..371W}
{Wang} F.~Y.,  {Dai} Z.~G.,  2006, \mn@doi [\mnras]
  {10.1111/j.1365-2966.2006.10108.x}, \href
  {https://ui.adsabs.harvard.edu/abs/2006MNRAS.368..371W} {368, 371}

\bibitem[\protect\citeauthoryear{{Wang} \& {Dai}}{{Wang} \&
  {Dai}}{2013}]{2013NatPh...9..465W}
{Wang} F.~Y.,  {Dai} Z.~G.,  2013, \mn@doi [Nature Physics]
  {10.1038/nphys2670}, \href
  {https://ui.adsabs.harvard.edu/abs/2013NatPh...9..465W} {9, 465}

\bibitem[\protect\citeauthoryear{{Wang}, {Dai}  \& {Liang}}{{Wang}
  et~al.}{2015}]{2015NewAR..67....1W}
{Wang} F.~Y.,  {Dai} Z.~G.,   {Liang} E.~W.,  2015, \mn@doi [\nar]
  {10.1016/j.newar.2015.03.001}, \href
  {https://ui.adsabs.harvard.edu/abs/2015NewAR..67....1W} {67, 1}

\bibitem[\protect\citeauthoryear{{Wang}, {Abdalla}, {Atrio-Barandela}  \&
  {Pav{\'o}n}}{{Wang} et~al.}{2016}]{2016RPPh...79i6901W}
{Wang} B.,  {Abdalla} E.,  {Atrio-Barandela} F.,   {Pav{\'o}n} D.,  2016,
  \mn@doi [Reports on Progress in Physics] {10.1088/0034-4885/79/9/096901},
  \href {https://ui.adsabs.harvard.edu/abs/2016RPPh...79i6901W} {79, 096901}

\bibitem[\protect\citeauthoryear{{Wang} et~al.,}{{Wang}
  et~al.}{2017}]{2017MNRAS.469.3762W}
{Wang} Y.,  et~al., 2017, \mn@doi [\mnras] {10.1093/mnras/stx1090}, \href
  {https://ui.adsabs.harvard.edu/abs/2017MNRAS.469.3762W} {469, 3762}

\bibitem[\protect\citeauthoryear{{Wang}, {Hu}, {Zhang}  \& {Dai}}{{Wang}
  et~al.}{2022}]{2022ApJ...924...97W}
{Wang} F.~Y.,  {Hu} J.~P.,  {Zhang} G.~Q.,   {Dai} Z.~G.,  2022, \mn@doi [\apj]
  {10.3847/1538-4357/ac3755}, \href
  {https://ui.adsabs.harvard.edu/abs/2022ApJ...924...97W} {924, 97}

\bibitem[\protect\citeauthoryear{{Wang}, {L{\'o}pez-Corredoira}  \&
  {Wei}}{{Wang} et~al.}{2024}]{2024MNRAS.527.7692W}
{Wang} B.,  {L{\'o}pez-Corredoira} M.,   {Wei} J.-J.,  2024, \mn@doi [\mnras]
  {10.1093/mnras/stad3724}, \href
  {https://ui.adsabs.harvard.edu/abs/2024MNRAS.527.7692W} {527, 7692}

\bibitem[\protect\citeauthoryear{Weinberg}{Weinberg}{1972}]{1972gcpa.book.....W}
Weinberg S.,  1972, Gravitation and Cosmology: Principles and Applications of
  the General Theory of Relativity, by Steven Weinberg, pp. 688. ISBN
  0-471-92567-5. Wiley-VCH , July 1972.

\bibitem[\protect\citeauthoryear{{Weinberg}}{{Weinberg}}{1989}]{1989RvMP...61....1W}
{Weinberg} S.,  1989, \mn@doi [Reviews of Modern Physics]
  {10.1103/RevModPhys.61.1}, \href
  {https://ui.adsabs.harvard.edu/abs/1989RvMP...61....1W} {61, 1}

\bibitem[\protect\citeauthoryear{{Wong} et~al.,}{{Wong}
  et~al.}{2020}]{2020MNRAS.498.1420W}
{Wong} K.~C.,  et~al., 2020, \mn@doi [\mnras] {10.1093/mnras/stz3094}, \href
  {https://ui.adsabs.harvard.edu/abs/2020MNRAS.498.1420W} {498, 1420}

\bibitem[\protect\citeauthoryear{{Wu}, {Zhang}  \& {Wang}}{{Wu}
  et~al.}{2022}]{2022MNRAS.515L...1W}
{Wu} Q.,  {Zhang} G.-Q.,   {Wang} F.-Y.,  2022, \mn@doi [\mnras]
  {10.1093/mnrasl/slac022}, \href
  {https://ui.adsabs.harvard.edu/abs/2022MNRAS.515L...1W} {515, L1}

\bibitem[\protect\citeauthoryear{{Xu}, {Li}  \& {Lu}}{{Xu}
  et~al.}{2009}]{2009JCAP...07..031X}
{Xu} L.,  {Li} W.,   {Lu} J.,  2009, \mn@doi [\jcap]
  {10.1088/1475-7516/2009/07/031}, \href
  {https://ui.adsabs.harvard.edu/abs/2009JCAP...07..031X} {2009, 031}

\bibitem[\protect\citeauthoryear{{Xu}, {Xu}, {Zhang}, {Fu}  \& {Huang}}{{Xu}
  et~al.}{2024}]{2024MNRAS.530.5091X}
{Xu} B.,  {Xu} J.,  {Zhang} K.,  {Fu} X.,   {Huang} Q.,  2024, \mn@doi [\mnras]
  {10.1093/mnras/stae1135}, \href
  {https://ui.adsabs.harvard.edu/abs/2024MNRAS.530.5091X} {530, 5091}

\bibitem[\protect\citeauthoryear{{Yadav}}{{Yadav}}{2023}]{2023PDU....4201365Y}
{Yadav} V.,  2023, \mn@doi [Physics of the Dark Universe]
  {10.1016/j.dark.2023.101365}, \href
  {https://ui.adsabs.harvard.edu/abs/2023PDU....4201365Y} {42, 101365}

\bibitem[\protect\citeauthoryear{{Yadav}, {Yadav}  \& {Rajpal}}{{Yadav}
  et~al.}{2024}]{2024PDU....4601626Y}
{Yadav} V.,  {Yadav} S.~K.,   {Rajpal} 2024, \mn@doi [Physics of the Dark
  Universe] {10.1016/j.dark.2024.101626}, \href
  {https://ui.adsabs.harvard.edu/abs/2024PDU....4601626Y} {46, 101626}

\bibitem[\protect\citeauthoryear{{Yang}, {Ren}, {Wang}, {Lu}, {Zhang}, {Cai}
  \& {Saridakis}}{{Yang} et~al.}{2024}]{2024SciBu..69.2698Y}
{Yang} Y.,  {Ren} X.,  {Wang} Q.,  {Lu} Z.,  {Zhang} D.,  {Cai} Y.-F.,
  {Saridakis} E.~N.,  2024, \mn@doi [Science Bulletin]
  {10.1016/j.scib.2024.07.029}, \href
  {https://ui.adsabs.harvard.edu/abs/2024SciBu..69.2698Y} {69, 2698}

\bibitem[\protect\citeauthoryear{{Yu}, {Ratra}  \& {Wang}}{{Yu}
  et~al.}{2018}]{2018ApJ...856....3Y}
{Yu} H.,  {Ratra} B.,   {Wang} F.-Y.,  2018, \mn@doi [\apj]
  {10.3847/1538-4357/aab0a2}, \href
  {https://ui.adsabs.harvard.edu/abs/2018ApJ...856....3Y} {856, 3}

\bibitem[\protect\citeauthoryear{{Zhang}, {Zhang}, {Yuan}, {Liu}, {Zhang}  \&
  {Sun}}{{Zhang} et~al.}{2014}]{2014RAA....14.1221Z}
{Zhang} C.,  {Zhang} H.,  {Yuan} S.,  {Liu} S.,  {Zhang} T.-J.,   {Sun} Y.-C.,
  2014, \mn@doi [Research in Astronomy and Astrophysics]
  {10.1088/1674-4527/14/10/002}, \href
  {https://ui.adsabs.harvard.edu/abs/2014RAA....14.1221Z} {14, 1221}

\bibitem[\protect\citeauthoryear{{du Mas des Bourboux} et~al.,}{{du Mas des
  Bourboux} et~al.}{2017}]{2017AA...608A.130D}
{du Mas des Bourboux} H.,  et~al., 2017, \mn@doi [\aap]
  {10.1051/0004-6361/201731731}, \href
  {https://ui.adsabs.harvard.edu/abs/2017A&A...608A.130D} {608, A130}

\makeatother
\end{thebibliography}
\bsp	

\onecolumn
\begin{appendix}
\section{Additional materials} 
Tables \ref{tab:CC} and \ref{tab:BAO} show the H(z) measurements used in this paper. Table \ref{A3} shows the $z_{tr}$ constraints collected from 2004 to 2024.
\begin{table*}\footnotesize
	\caption{ $H(z)$ measurements from Cosmic Chronometer (in units of $\textrm{km}~\textrm{s}^{-1} \textrm{Mpc}^{-1}$). \label{tab:CC}}
	\centering
	\begin{tabular}{cccc|cccc}
		\hline\hline
   No.	&	$z$  & $H(z)$  & Ref. & No.	&	$z$  & $H(z)$  & Ref.    \\  
	\hline
  (1) 	&	$0.07$    & $69.0\pm19.6$ & \citet{2014RAA....14.1221Z}&(2) 	&	$0.09$    & $69.0\pm12.0$ & \citet{2005PhRvD..71l3001S}\\
  (3) 	&	$0.12$    & $68.6\pm26.2$ & \citet{2014RAA....14.1221Z}&(4)   &	$0.17$    & $83.0\pm8.0$ &  \citet{2005PhRvD..71l3001S}\\
  (5) 	&	$0.179$   & $75.0\pm4.0$ &  \citet{2012JCAP...08..006M}&(6) 	&	$0.199$   & $75.0\pm5.0$ &  \citet{2012JCAP...08..006M}\\
  (7) 	&	$0.2$     & $72.9\pm29.6$ & \citet{2014RAA....14.1221Z}&(8)	&	$0.27$    & $77.0\pm14.0$ & \citet{2005PhRvD..71l3001S}\\
  (9)	&	$0.28$    & $88.8\pm36.6$ & \citet{2014RAA....14.1221Z}&(10)	&	$0.352$   & $83.0\pm14.0$ & \citet{2012JCAP...08..006M}\\
  (11)	&	$0.3802$  & $83.0\pm13.5$ & \citet{2016JCAP...05..014M}&(12)	&	$0.4$     & $95.0\pm17.0$ & \citet{2005PhRvD..71l3001S}\\
  (13)	&	$0.4004$  & $77.0\pm10.2$ & \citet{2016JCAP...05..014M}& (14)	&	$0.4247$  & $87.1\pm11.2$ & \citet{2016JCAP...05..014M}\\
  (15)	&	$0.4497$ & $92.8\pm12.9$ &  \citet{2016JCAP...05..014M}&(16)	&	$0.47$    & $89.0\pm50$  &  \citet{2017MNRAS.467.3239R}\\
  (17)	&	$0.4783$  & $80.9\pm9.0$ &  \citet{2016JCAP...05..014M}& (18)	&	$0.48$    & $97.0\pm62.0$ &  \citet{2017MNRAS.467.3239R}\\
  (19)	&	$0.593$   & $104.0\pm13.0$ & \citet{2012JCAP...08..006M}& (20)	&	$0.68$    & $92.0\pm8.0$  &  \citet{2012JCAP...08..006M}\\
  (21)	&	$0.75$    & $98.8\pm33.6$  &  \citet{2022ApJ...928L...4B}& (22)	&	$0.75$    & $105.0\pm7.9$  &  \citet{2023JCAP...11..047J}\\
  (23)	&	$0.781$   & $105.0\pm12.0$ & \citet{2012JCAP...08..006M}& (24)	&	$0.80$    & $113.1\pm15.1$  &  \citet{2023ApJS..265...48J}\\
  (25)	&	$0.875$   & $125.0\pm17.0$ & \citet{2012JCAP...08..006M}& (26)	&	$0.88$    & $90.0\pm40.0$  & \citet{2017MNRAS.467.3239R}\\
  (27)	&	$0.9$     & $117.0\pm23.0$ & \citet{2005PhRvD..71l3001S}& (28)	&	$1.037$   & $154.0\pm20.0$ & \citet{2012JCAP...08..006M}\\
  (29)	&	$1.26$    & $135.0\pm65$  &  \citet{2023AA...679A..96T}& (30)	&	$1.3$     & $168.0\pm17.0$ & \citet{2005PhRvD..71l3001S}\\
  (31)	&	$1.363$   & $160.0\pm33.6$ & \citet{2015MNRAS.450L..16M}& (32)	&	$1.43$    & $177.0\pm18.0$ & \citet{2005PhRvD..71l3001S}\\
  (33)	&	$1.53$    & $140.0\pm14.0$ & \citet{2005PhRvD..71l3001S}& (34)	&	$1.75$    & $202.0\pm40.0$ & \citet{2005PhRvD..71l3001S}\\
  (35)	&	$1.965$   & $186.5\pm50.4$ & \citet{2015MNRAS.450L..16M}& & & & \\
		\hline\hline
	\end{tabular}
\end{table*}

\begin{table}\footnotesize
	\caption{$H(z)$ measurements from the BAO measurements including 5 DESI BAOs (in units of $\textrm{km}~\textrm{s}^{-1} \textrm{Mpc}^{-1}$). \label{tab:BAO}}
	\centering
	\begin{tabular}{cccc|cccc}
		\hline\hline
   No.	&	$z$  & $H(z)$  & Ref. & No.	&	$z$  & $H(z)$  & Ref.  \\  
  \hline
  \textbf{DESI:}	&   &   &   \\
  (1)	&	$0.51$    & $97.21\pm2.83$ &  \citet{2025JCAP...04..012A} & (2)	&	$0.71$    & $101.57\pm3.04$ &  \citet{2025JCAP...04..012A} \\
  (3)	&	$0.93$    & $114.07\pm2.24$ &  \citet{2025JCAP...04..012A} & (4)	&	$1.32$    & $147.58\pm4.49$ &  \citet{2025JCAP...04..012A} \\
  (5)	&	$2.33$    & $239.38\pm4.80$ &  \citet{2025JCAP...01..124A} & 	&    &   &    \\
  \hline \hline
  \textbf{Other:}	&   &   &   \\
  (1) 	&	$0.24$    & $79.69\pm2.99$ & \citet{2009MNRAS.399.1663G} & (2) 	&	$0.36$    & $79.93\pm3.39$ & \citet{2017MNRAS.469.3762W} \\
  (3) 	&	$0.38$    & $81.50\pm1.90$ & \citet{2017MNRAS.470.2617A} & (4)  &	$0.40$    & $82.04\pm2.03$ &  \citet{2017MNRAS.469.3762W} \\
  (5) 	&	$0.43$    & $86.45\pm3.68$ &  \citet{2009MNRAS.399.1663G} & (6)  &	$0.44$    & $84.81\pm1.83$ &  \citet{2017MNRAS.469.3762W} \\
  (7) 	&	$0.48$    & $87.79\pm2.03$ & \citet{2017MNRAS.469.3762W} & (8)  	&	$0.51$    & $90.40\pm1.90$ & \citet{2017MNRAS.470.2617A} \\
  (9) 	&	$0.56$    & $93.33\pm2.32$ & \citet{2017MNRAS.469.3762W} & (10)	&	$0.57$    & $87.60\pm7.80$ & \citet{2013MNRAS.433.3559C} \\
  (11)	&	$0.57$    & $96.80\pm3.40$ & \citet{2014MNRAS.441...24A} & (12)	&	$0.59$    & $98.48\pm3.19$ & \citet{2017MNRAS.469.3762W} \\
  (13)	&	$0.60$    & $87.90\pm6.10$ & \citet{2012MNRAS.425..405B} & (14)	&	$0.61$    & $97.30\pm2.10$ & \citet{2017MNRAS.470.2617A} \\
  (15)	&	$0.64$    & $98.82\pm2.99$ &  \citet{2017MNRAS.469.3762W} & (16)	&	$1.48$    & $153.81\pm6.39$ &  \citet{2020MNRAS.499..210N} \\
  (17)	&	$2.30$    & $224.0\pm8.0$ &  \citet{2013AA...552A..96B} & (18)	&	$2.34$    & $223.0\pm7.0$ &  \citet{2015AA...574A..59D} \\
  (19)	&	$2.36$    & $226.0\pm8.0$ &  \citet{2014JCAP...05..027F} & (20)	&	$2.40$    & $227.8\pm5.61$ &  \citet{2017AA...608A.130D} \\
  \hline\hline
	\end{tabular}
\end{table}

\begin{table*}
\tiny
\centering
	\caption{ Constraints of the transition redshift $z_{tr}$ obtained by different methods and different datasets. \label{A3}}
		\begin{tabular}{cccc|cccc}
			\hline\hline
	Year & $z_{tr}$ & method & Ref.  & Year & $z_{tr}$ & method & Ref. \\ \hline	
    2004 & 0.46$_{-0.13}^{+0.13}$  & SNe Ia (a)  &  \citet{2004ApJ...607..665R}  &  2018  &  0.81$_{-0.09}^{+0.09}$  & H(z) + SNe Ia (e)  &  \citet{2018JCAP...05..073J}   \\ 
    2005 & 0.73$_{-0.09}^{+0.09}$  & GRBs + SNe Ia (b)  &  \citet{2005MNRAS.360L...1F}  &  2018  &  0.87$_{-0.06}^{+0.06}$  & H(z) + SNe Ia (e) &  \citet{2018JCAP...05..073J}  \\ 
    2005 & 0.40  & GRBs + SNe Ia (c)  &  \citet{2005MNRAS.360L...1F}  &  2018  &  0.97$_{-0.06}^{+0.06}$  & H(z) + SNe Ia (e) &  \citet{2018JCAP...05..073J}  \\ 
    2005 & 0.55  & GRBs + SNe Ia (d)  &  \citet{2005MNRAS.360L...1F}  &  2018  &  0.57$_{-0.23}^{+0.27}$  & CCs + two ly$\alpha$ (j)  &  \citet{2018ApJ...856....3Y}  \\  
    2005 & 0.78$_{-0.23}^{+0.32}$  & GRBs (b) &  \citet{2005ApJ...633..611L}  &  2018  &  0.46$_{-0.12}^{+0.32}$  & H(z) (j)  &  \citet{2018ApJ...856....3Y}  \\ 
    2006 & 0.69$_{-0.12}^{+0.11}$   & GRBs (b)  &  \citet{2006MNRAS.368..371W}  &  2018  &  0.55$_{-0.20}^{+0.23}$  & CCs (j)  &  \citet{2018ApJ...856....3Y}  \\ 
    2006 & 0.61$_{-0.06}^{+0.05}$   & GRBs + SNe Ia (b)  &  \citet{2006MNRAS.368..371W}  &  2018  &  0.44$_{-0.11}^{+0.56}$  & H(z) (j)  &  \citet{2018ApJ...856....3Y} \\ 
    2007 & 0.43$_{-0.07}^{+0.07}$  & SNe Ia (a)  &  \citet{2007ApJ...659...98R}  &  2018  &  0.64$_{-0.09}^{+0.12}$  & H(z) + SNe Ia (j)  &  \citet{2018JCAP...10..015H}   \\ 
    2008 & 0.61$_{-0.21}^{+3.68}$  &  SN LS (e)  &  \citet{2008MNRAS.390..210C}  &  2020  &  0.59$_{-0.11}^{+0.12}$  & H(z) (j)  &  \citet{2020JCAP...04..053J}   \\ 
    2008 & 0.43$_{-0.05}^{+0.09}$  & SNe Ia (e)  &  \citet{2008MNRAS.390..210C}  &  2020  &  0.68$_{-0.08}^{+0.11}$  & SNe Ia (j)  &  \citet{2020JCAP...04..053J}  \\ 
    2008 & 0.60$_{-0.09}^{+0.28}$  & SNe Ia (e)  &  \citet{2008MNRAS.390..210C}  &  2020  &  0.63$_{-0.02}^{+0.02}$  & Planck (b) &  \citet{2020AA...641A...6P} \\ 
    2009 & 0.71$_{-0.08}^{+0.08}$  &  SNe Ia (b)  &  \citet{2009JCAP...10..010G}  &  2020  &  0.70$_{-0.30}^{+0.30}$  & H(z) + SNe Ia (b)  &  \citet{2020EPJC...80..562V}  \\ 
    2009 & 0.49$_{-0.09}^{+0.27}$  & SNe Ia (a) &  \citet{2009JCAP...10..010G}  &   2021  &  0.65$_{-0.16}^{+0.03}$  & H(z) + BAOs + SNe Ia (e)  &  \citet{2022MNRAS.509.5399C}   \\ 
    2009 & 0.46$_{-0.28}^{+0.40}$  & SNe Ia (a)  &  \citet{2009JCAP...10..010G}  &   2021  &  0.57$_{-0.12}^{+0.37}$  & H(z) (e)  &  \citet{2022MNRAS.509.5399C}   \\ 
    2009 & 0.48$_{-0.11}^{+0.36}$  & SNe Ia (a)  &  \citet{2009JCAP...10..010G}  &  2021  &  1.02$_{-0.17}^{+0.01}$  & H(z) + BAOs + SNe Ia (e) &  \citet{2022MNRAS.509.5399C}   \\ 
    2009 & 0.52$_{-0.08}^{+0.21}$  & SNe Ia (a)  &  \citet{2009JCAP...10..010G}  &  2021  &  0.60$_{-0.26}^{+0.12}$  & H(z) (e) &  \citet{2022MNRAS.509.5399C}  \\ 
    2009 & 0.61$_{-0.07}^{+0.11}$  & H(z) + BAOs + SNe Ia (a)  &  \citet{2009JCAP...07..031X}  &  2022  &  0.43$_{-0.04}^{+0.04}$  & H(z) + BAOs + SNe Ia (k) &  \citet{2022EPJC...82.1165S} \\ 
    2009 & 0.59$_{-0.06}^{+0.09}$  & H(z) + BAOs + SNe Ia (a)  &  \citet{2009JCAP...07..031X}  &  2023  &  0.59$_{-0.33}^{+0.33}$  & H(z) (l)  &  \citet{2023IJMPD..3250039K}  \\
    2009 & 0.95$_{-0.08}^{+0.11}$  & H(z) + BAOs + SNe Ia (a)  &  \citet{2009JCAP...07..031X}  &  2023   &  0.62$_{-0.76}^{+0.58}$  & H(z) (m)  &  \citet{2023IJMPD..3250039K} \\ 
    2013 & 0.82$_{-0.08}^{+0.08}$  & BAOs (b)   &  \citet{2013AA...552A..96B}  &  2023   &  0.85$_{-0.12}^{+0.12}$  & SNe Ia (l) &  \citet{2023IJMPD..3250039K}  \\ 
    2013 & 0.74$_{-0.05}^{+0.05}$  & H(z) (*)  &  \citet{2013ApJ...766L...7F}  &  2023  &  0.72$_{-0.05}^{+0.05}$  & CCs + SNe Ia +  ISW (b)  &  \citet{2023GrCo...29..177R}   \\  
    2013 & 0.74$_{-0.04}^{+0.04}$  & H(z) (*)  &  \citet{2013PhLB..726...72F}  &  2023  &  0.62$_{-0.13}^{+0.08}$  &  H(z) (n)  &  \citet{2023MNRAS.523.4938M}  \\ 
    2014 & 0.77$_{-0.18}^{+0.18}$  & H(z) + SNe Ia (f)  &  \citet{2014PhRvD..90d4016C}  &  2023  &  0.70$_{-0.04}^{+0.09}$  &   H(z) (e)  &  \citet{2023MNRAS.523.4938M} \\  
    2015 & 0.64$_{-0.03}^{+0.03}$  & SNe Ia (g)  &  \citet{2015PhRvD..91l4037C}  &  2023  &  0.81$_{-0.05}^{+0.08}$  & H(z) (e)  &  \citet{2023MNRAS.523.4938M}  \\ 
    2015 & 0.25$_{-0.27}^{+0.35}$  & SNe Ia (g)  &  \citet{2015PhRvD..91l4037C}  &  2023  &  0.70$_{-0.06}^{+0.08}$  & GRBs (e)  &  \citet{2023MNRAS.523.4938M}  \\
    2015 & 0.98  & AG + SL + SNe Ia (e)  &  \citet{2015JCAP...12..045R}  &  2023  &  0.71$_{-0.07}^{+0.07}$  & GRBs (e)  &  \citet{2023MNRAS.523.4938M}  \\
    2015 & 0.96  & AG + SL + SNe Ia (e)  &  \citet{2015JCAP...12..045R}  &  2024  &  0.60  & H(z) + BAOs + SNe Ia (o)  &  \citet{2024PDU....4601614M}   \\
    2015 & 0.60  & AG + SL + SNe Ia (e)  &  \citet{2015JCAP...12..045R}  &  2024  &  0.78  & H(z) + BAOs + SNe Ia (o)  &  \citet{2024PDU....4601614M}  \\
    2016  &  0.64$_{-0.06}^{+0.11}$  & H(z) (h)  &  \citet{2016JCAP...05..014M}  &  2024  &  0.66  & H(z) + BAOs + SNe Ia (o)  &  \citet{2024PDU....4601614M}  \\
    2016   &  0.40$_{-0.10}^{+0.10}$  & H(z) (i)  &  \citet{2016JCAP...05..014M}  &  -  &  -  & -  &  -  \\
    2017   &  0.72$_{-0.05}^{+0.05}$  & H(z) (*)   &  \citet{2017ApJ...835...26F}  &  -  &  -  & -  &  -  \\
    2017  &  0.84$_{-0.03}^{+0.03}$  & H(z) (*)   &  \citet{2017ApJ...835...26F}  &  -  &  -  & -  &  -  \\
    2017  &  0.72$_{-0.09}^{+0.09}$  & H(z) (b)  &  \citet{2017ApJ...835...26F}  &  -  &  -  & -  &  -  \\
    2017  &  0.83$_{-0.06}^{+0.06}$  & H(z) (b)  &  \citet{2017ApJ...835...26F}  &  -  &  -  & -  &  -  \\
	\hline\hline
	\end{tabular}
	\begin{itemize}	
	\tiny
	\item[Note:] (a) Cosmography; (b) $\Lambda$CDM; (c) $w_{0}$CDM; (d) $w_{z}$CDM;
 (e) Linear parametrization; (f) f(R) gravity; (g) f(T) gravity; (h) o$\Lambda$CDM; (i) Piecewise linear fit; (j) Gaussian Process; (k) Chameleon model; (l) Phase space portrait; (m) Cosmic triangle; (n) Bezier curve; (o) f(Q + $L_{m}$) gravity; (*) Statistical results.  
\item[Abb.:] SN LS: Superluminous supernova; AG: Age of galaxies; SL: Strong lensing; ISW: Integrated Sachs-Wolfe effect.    
	\end{itemize}
\end{table*}
\end{appendix}
\label{lastpage}
\end{document}